\documentclass{fundam}

\publyear{22}
\papernumber{2102}
\volume{185}
\issue{1}

\usepackage{url} % takes care of hyperlinks, preferred over hyperref
\usepackage[ruled,lined]{algorithm2e}% provides Algorithm environment
\usepackage{graphicx}% allows for inclusion of graphic files (figures)
\usepackage{tikz}
\usetikzlibrary{automata,positioning,%
patterns,decorations,decorations.pathmorphing,%
arrows,shapes,automata,backgrounds,fit,calc,petri}
\usepackage{latexsym}
\usepackage{amsfonts}
\usepackage{amssymb}
\usepackage{amsmath} 
\usepackage{color} 
\usepackage{textcomp} 
\usepackage{lineno}   
\usepackage[T1]{fontenc}
\usepackage{mathrsfs}
\usepackage{mathptmx}
\usepackage{helvet}
\usepackage{multirow} 
\usepackage{courier}
\usepackage{type1cm}
\usepackage{makeidx} 
\usepackage{multicol} 
\usepackage{verbatim}
\usepackage{stmaryrd} 
\usepackage{fancyhdr}
\usepackage{fancybox} 
\usepackage[fleqn,tbtags]{mathtools}
\usepackage{longtable}
\usepackage{wrapfig} 
\usepackage{algorithmic} 
\usepackage{rotating}
\usepackage{float}
\floatstyle{boxed}
\restylefloat{figure} 

\def\eod{\penalty 1000\hfill\penalty 1000$\diamond$\hskip 0pt\penalty-1000\par} 
\newcommand{\eg} {e.g., }
\newcommand{\ie} {i.e., } 

\newcommand{\pref}{\textit{pref}}
\newcommand{\reach}{\textit{reach}}
\newcommand{\fseq}{\textit{fseq}}
\newcommand{\fireable}{\textit{fireable}} 
\newcommand{\DP}{\textit{dplaces}}
\newcommand{\BOX}{\textit{net}}
\newcommand{\maxCLIQUE}{\textit{maxclq}}
\newcommand{\outseq}{\textit{outseq}}
\newcommand{\inseq}{\textit{inseq}}
\newcommand{\disables}{\textit{disables}}
\newcommand{\enables}{\textit{enables}}
\newcommand{\comploutseq}{\textit{coutseq}}
\newcommand{\complinseq}{\textit{cinseq}}
\newcommand{\In}{\textit{ins}}
\newcommand{\Out}{\textit{rem}}
\newcommand{\Read}{\textit{read}}
\newcommand{\PREE}{\textit{pre}}
\newcommand{\POSTT}{\textit{post}} 
\newcommand{\CG}{\textit{cgraph}}
\newcommand{\PLACES}{\textit{places}}

\newcommand{\init}{\textit{init}}

\newcommand{\ITER}[3]{[ {#1} *{#2}*{#3}]} 
\newcommand{\FA}[1]{\langle #1\rangle} 
\newcommand{\join}{\boldsymbol{-\!\!\!\!\!\!\!-}}
\newcommand{\es}{\varnothing} 
\newcommand{\PC}{\mathcal{P}} 
\newcommand{\clq}{\textit{clq}} 
\newcommand{\Flow}{\mathit{Fl}} 

\newcommand{\PREPOST}[1] {\PRE{\POST{#1}}} 
\newcommand{\POST}[1] {#1^\bullet }
\newcommand{\PRE}[1] {{}^\bullet #1} 
\newcommand{\cPOST}[1] {#1^\blacktriangleright }
\newcommand{\cPRE}[1] {{}^\blacktriangleright #1}
\newcommand{\SEQ} {\,{;}\,} 
\newcommand{\CHOICE} {\,{\Box}\,}
\newcommand{\PAR} {\,{\|}\,} 
 
\newcommand{\IN}{\textit{in}}
\newcommand{\OUT}{\textit{out}} 
\newcommand{\NN} {\mathit{N}}
\newcommand{\OUTT}{\mathcal{O}}
\newcommand{\INN}{\mathcal{I}}
\newcommand{\SSS}{\mathcal{S}} 
\newcommand{\RG} {\textit{RG}} 
\newcommand{\PLUS} {\sqcup}

\begin{document}

\pgfdeclarelayer{background}
\pgfdeclarelayer{foreground}
\pgfsetlayers{background,main,foreground}

% Commands for Petri nets and transtion systems
% StandardNet and StandardTS commands open tikzpicture environment
%     which has to be closed (using \end{tikzpicture})

\newcommand{\DiagramBig}[1][0.5]
     {
     \begin{tikzpicture}[node distance=1.3cm,>=arrow30,%
     line  width=0.3mm,scale=#1,bend angle=45]
     \tikzstyle{box}=[draw,regular polygon,thick,%
     regular polygon sides=4,minimum size=22mm, inner sep = -3pt]
     \tikzstyle{ybox}=[draw,regular polygon,thick,%
     regular polygon sides=4,minimum size=22mm, inner sep = -3pt,fill=yellow]
     \tikzstyle{rbox}=[draw,regular polygon,thick,%
     regular polygon sides=4,minimum size=22mm, inner sep = -3pt,fill=red]
     \tikzstyle{gbox}=[draw,regular polygon,thick,%
     regular polygon sides=4,minimum size=22mm, inner sep = -3pt,fill=green]

     \tikzstyle{dybox}=[draw,regular polygon,thick,style=dashed,%
     regular polygon sides=4,minimum size=22mm, inner sep = -3pt,fill=yellow]
     \tikzstyle{drbox}=[draw,regular polygon,thick,style=dashed,%
     regular polygon sides=4,minimum size=22mm, inner sep = -3pt,fill=red]
     \tikzstyle{dgbox}=[draw,regular polygon,thick,style=dashed,%
     regular polygon sides=4,minimum size=22mm, inner sep = -3pt,fill=green]
     }

\newcommand{\Diagram}[1][0.5]
     {
     \begin{tikzpicture}[node distance=1.3cm,>=arrow30,%
     line  width=0.3mm,scale=#1,bend angle=45]
     \tikzstyle{box}=[draw,regular polygon,thick,rounded corners,%
     regular polygon sides=4,minimum size=14mm, inner sep = -3pt]
     \tikzstyle{ybox}=[draw,regular polygon,thick,rounded corners,%
     regular polygon sides=4,minimum size=14mm, inner sep = -3pt,fill=yellow!40]
     \tikzstyle{rbox}=[draw,regular polygon,thick,rounded corners,%
     regular polygon sides=4,minimum size=14mm, inner sep = -3pt,fill=red!40]
     \tikzstyle{gbox}=[draw,regular polygon,thick,rounded corners,%
     regular polygon sides=4,minimum size=14mm, inner sep = -3pt,fill=green!40]
     \tikzstyle{bbox}=[draw,regular polygon,thick,rounded corners,%
     regular polygon sides=4,minimum size=14mm, inner sep = -3pt,fill=cyan!40]

     \tikzstyle{dybox}=[draw,regular polygon,thick,style=dashed,rounded corners,%
     regular polygon sides=4,minimum size=14mm, inner sep = -3pt,fill=yellow!40]
     \tikzstyle{drbox}=[draw,regular polygon,thick,style=dashed,rounded corners,%
     regular polygon sides=4,minimum size=14mm, inner sep = -3pt,fill=red!40]
     \tikzstyle{dgbox}=[draw,regular polygon,thick,style=dashed,rounded corners,%
     regular polygon sides=4,minimum size=14mm, inner sep = -3pt,fill=green!40]
     \tikzstyle{dbbox}=[draw,regular polygon,thick,style=dashed,rounded corners,%
     regular polygon sides=4,minimum size=14mm, inner sep = -3pt,fill=cyan!40]
     }

\newcommand{\Relation}[2]
     {
     \begin{tikzpicture}[baseline=-0.5cm, node distance=.8cm,>=arrow30,line width=0.3mm]
     \tikzstyle{tlcorner}=[xshift=#2,yshift=#1]
     \tikzstyle{brcorner}=[xshift=#1,yshift=#2]
     \tikzstyle{bgcolor}=[fill=brown!20]
     }

\newcommand{\StandardNet}[1][0.5]
     {
     \begin{tikzpicture}[node distance=1.3cm,>=latex',line  width=0.3mm,scale=#1,auto,bend angle=45]
     \tikzstyle{place}=[draw,circle,thick,minimum size=3mm]
     \tikzstyle{transition}=
                [draw,regular polygon,thick,
                 regular polygon sides=4,minimum size=6mm, inner sep = -2pt]
     }

\newcommand{\StandardTS}[1][0.5]
     {
     \begin{tikzpicture}[node distance=0.5cm,>=stealth',bend angle=45,scale=#1]
     }

% edges and loops
% edgetext and loopedgetext draw edges!
\newcommand{\bdiredge}    [2]{\path (#1) edge [->,ultra thick] (#2);}
\newcommand{\dirDOTS}     [2]{\path (#1) edge [dotted](#2);}
\newcommand{\dirDASH}     [2]{\path (#1) edge [dashed,->](#2);}
\newcommand{\edgeDASH}    [2]{\path (#1) edge [dashed](#2);}
\newcommand{\diredge}     [2]{\path (#1) edge [->] (#2);}
\newcommand{\plainedge}   [2]{\path (#1) edge (#2);}
\newcommand{\inhedge}     [2]{\path (#1) edge [-o] (#2);}
\newcommand{\actedge}     [2]{\path (#1) edge [-*] (#2);}
\newcommand{\edgetext}    [3]{\path [->] (#1) edge node [fill=white, inner sep=1pt] {#3} (#2);}
\newcommand{\dirleftbow}  [5]{\path [->] (#1) edge [bend left,out=#3,in=#4,min distance=#5] (#2);}
\newcommand{\dirrightbow} [5]{\path [->] (#1) edge [bend right,out=#3,in=#4,min distance=#5] (#2);}
\newcommand{\dirleftbowDASH}  [5]{\path [dashed,->] (#1) edge [bend left,out=#3,in=#4,min distance=#5] (#2);}
\newcommand{\dirrightbowDASH} [5]{\path [dashed,->] (#1) edge [bend right,out=#3,in=#4,min distance=#5] (#2);}
\newcommand{\leftbowtext} [6]{\path [->] (#1) edge [bend left,out=#3,in=#4,min distance=#5]
                              node[fill=white, inner sep=1pt]{#6}(#2);}
\newcommand{\rightbowtext}[6]{\path [->] (#1) edge [bend right,out=#3,in=#4,min distance=#5]
                              node[fill=white, inner sep=1pt, swap]{#6}(#2);}
\newcommand{\dirloopedge} [4]{\path[->](#1)edge[loop,out=#2,in=#3,min distance=#4] (#1);}
\newcommand{\loopedgetext}[6]{\path[->](#1)edge[loop,out=#2,in=#3,min distance=#4] node[#6]{#5}(#1);}
\newcommand{\Rightbowtext}[6]{\path [<-] (#1) edge [bend right,out=#3,in=#4,min distance=#5]
                              node[fill=white, inner sep=1pt, swap]{#6}(#2);}

% Nodes in graphs:
\newcommand{\DUMMY}[3]{\fill(#2,#3)circle(0.00003pt)node(#1) {};}
\newcommand{\DUMMYwest}[4]{\fill(#2,#3)circle(0.000003pt)node[label=left :$#4$]        (#1) {};}

\newcommand{\Node}[3]{\fill(#2,#3)circle(3pt)node(#1) {};}
\newcommand{\nodE}[4]{\fill(#2,#3)circle(3pt)node[label=right:$#4$]        (#1) {};}
\newcommand{\nodW}[4]{\fill(#2,#3)circle(3pt)node[label=left :$#4$]        (#1) {};}
\newcommand{\nodN}[4]{\fill(#2,#3)circle(3pt)node[label=above:$#4$]        (#1) {};}
\newcommand{\nodS}[4]{\fill(#2,#3)circle(3pt)node[label=below:$#4$]        (#1) {};}
\newcommand{\noSE}[4]{\fill(#2,#3)circle(3pt)node[label=below right:$#4$]  (#1) {};}
\newcommand{\noNE}[4]{\fill(#2,#3)circle(3pt)node[label=above right:$#4$]  (#1) {};}
\newcommand{\noSW}[4]{\fill(#2,#3)circle(3pt)node[label=below left :$#4$]  (#1) {};}
\newcommand{\noNW}[4]{\fill(#2,#3)circle(3pt)node[label=above left :$#4$]  (#1) {};}

\newcommand{\noNS}[5]{\fill(#2,#3)circle(3pt)node[label=above:$#4$,label=below:$#5$]  (#1) {};}
\newcommand{\noSN}[5]{\fill(#2,#3)circle(3pt)node[label=above:$#5$,label=below:$#4$]  (#1) {};}

% Places in nets:
\newcommand{\place}[4]{\node(#1)at(#2,#3)[place,tokens=#4]                        {};}
\newcommand{\placE}[5]{\node(#1)at(#2,#3)[place,tokens=#4,label=right:$#5$]       {};}
\newcommand{\placW}[5]{\node(#1)at(#2,#3)[place,tokens=#4,label=left :$#5$]       {};}
\newcommand{\placN}[5]{\node(#1)at(#2,#3)[place,tokens=#4,label=above:$#5$]       {};}
\newcommand{\placS}[5]{\node(#1)at(#2,#3)[place,tokens=#4,label=below:$#5$]       {};}
\newcommand{\plaSE}[5]{\node(#1)at(#2,#3)[place,tokens=#4,label=below right:$#5$] {};}
\newcommand{\plaNE}[5]{\node(#1)at(#2,#3)[place,tokens=#4,label=above right:$#5$] {};}
\newcommand{\plaSW}[5]{\node(#1)at(#2,#3)[place,tokens=#4,label=below left :$#5$] {};}
\newcommand{\plaNW}[5]{\node(#1)at(#2,#3)[place,tokens=#4,label=above left :$#5$] {};}

\newcommand{\PlacN}[5]{\node(#1)at(#2,#3)[place,tokens=0,label=above:$#5$]      {$#4$};}
\newcommand{\PlacS}[5]{\node(#1)at(#2,#3)[place,tokens=0,label=below:$#5$]      {$#4$};}
\newcommand{\PlacE}[5]{\node(#1)at(#2,#3)[place,tokens=0,label=right:$#5$]      {$#4$};}
\newcommand{\PlacW}[5]{\node(#1)at(#2,#3)[place,tokens=0,label=left :$#5$]      {$#4$};}

\newcommand{\bPlacN}[5]{\node(#1)at(#2,#3)[place,tokens=0,label=above:$#5$,ultra thick]      {$#4$};}
\newcommand{\bPlacS}[5]{\node(#1)at(#2,#3)[place,tokens=0,label=below:$#5$,ultra thick]      {$#4$};}
\newcommand{\bPlacE}[5]{\node(#1)at(#2,#3)[place,tokens=0,label=right:$#5$,ultra thick]      {$#4$};}
\newcommand{\bPlacW}[5]{\node(#1)at(#2,#3)[place,tokens=0,label=left :$#5$,ultra thick]      {$#4$};}

\newcommand{\WhitetraN}[5]{\node (#1) at (#2,#3)[transition,label=above:$#5$] {$#4$};}
\newcommand{\WhitetraS}[5]{\node (#1) at (#2,#3)[transition,label=below:$#5$] {$#4$};}
\newcommand{\WhitetraE}[5]{\node (#1) at (#2,#3)[transition,label=right:$#5$] {$#4$};}
\newcommand{\WhitetraW}[5]{\node (#1) at (#2,#3)[transition,label=left :$#5$] {$#4$};}

\newcommand{\Whitetran}[4]{\node (#1) at (#2,#3)[transition] {$#4$};}

\newcommand{\colorarc}[4]
    {\path [color=#1, -stealth, line width=#2,
     postaction={draw, line width=#3, shorten >=#4, -}]}
\newcommand{\coloredge}[2]{\path [color=#1, line width=#2]}

\newcommand{\Whitebox}[4]{\node (#1) at (#2,#3) [box] {$#4$};}
\newcommand{\Ybox}    [4]{\node (#1) at (#2,#3) [ybox]{$#4$};}
\newcommand{\Rbox}    [4]{\node (#1) at (#2,#3) [rbox]{$#4$};}
\newcommand{\Gbox}    [4]{\node (#1) at (#2,#3) [gbox]{$#4$};}
\newcommand{\Bbox}    [4]{\node (#1) at (#2,#3) [bbox]{$#4$};}
\newcommand{\dYbox}    [4]{\node (#1) at (#2,#3) [dybox]{$#4$};}
\newcommand{\dRbox}    [4]{\node (#1) at (#2,#3) [drbox]{$#4$};}
\newcommand{\dGbox}    [4]{\node (#1) at (#2,#3) [dgbox]{$#4$};}
\newcommand{\dBbox}    [4]{\node (#1) at (#2,#3) [dbbox]{$#4$};}

% command to put free text on net/diagram
%\renewcommand{\put}[3]{\node at (#1,#2) {#3};}

\newcommand{\Put}[3]{
	\node at (#1,#2) {#3};
}

\pgfarrowsdeclare{arrow30}{arrow30}
{
  \arrowsize=0.2pt
  \advance\arrowsize by.275\pgflinewidth%
  \pgfarrowsleftextend{+-\arrowsize}
  \advance\arrowsize by.5\pgflinewidth
  \pgfarrowsrightextend{+\arrowsize}
}
{
  \arrowsize=0.2pt
  \advance\arrowsize by.275\pgflinewidth%
  \pgfsetdash{}{+0pt}
  \pgfsetroundjoin
  \pgfpathmoveto{\pgfqpoint{1\arrowsize}{0\arrowsize}}
  \pgfpathlineto{\pgfqpoint{-12\arrowsize}{2.5\arrowsize}}
  \pgfpathlineto{\pgfqpoint{-12\arrowsize}{-2.5\arrowsize}}
  \pgfpathclose
  \pgfusepathqfillstroke
}

\title{Distributed Places and Safe Net Reduction} 

\address{maciej.koutny@ncl.ac.uk}

\author{Victor Khomenko\\
Renesas Electronics\\
Delta 200 Office Park, Welton Rd, Swindon
SN5 7XB, United Kingdom \\
victor.khomenko.xh@renesas.com
\and
Maciej Koutny\\
School of Computing,
Newcastle University\\
1 Science Square, Newcastle upon Tyne, NE4 5TG, United Kingdom\\
maciej.koutny@ncl.ac.uk
\and 
Alex Yakovlev\\
School of Engineering,
Newcastle University\\ 
Merz Court, Newcastle upon Tyne, NE1 7RU, United Kingdom\\
alex.yakovlev@ncl.ac.uk 
} 

\maketitle

\runninghead{V. Khomenko, M. Koutny, and A. Yakovlev}{Distributed Places}

\begin{abstract}

Being able to find small 
Petri nets with the same behaviour as 
formal specifications of concurrent 
systems benefits both 
effective verification and 
practical implementation of such systems.

For specifications given 
in the form of compositionally defined safe nets,
this paper discusses a novel concept 
of `distributed place' which implements the behaviour 
of an individual net place. 
It is shown that if distributed places cover 
a safe Petri net, then it is possible to delete some places 
without changing the behaviour.
Crucially, the reduction is both static and local,
making it computationally feasible in practice.

The resulting reduction technique is then applied to 
an algebra of safe Petri nets (boxes) 
derived compositionally from process (box) expressions. 
Though the original derivation can yield
exponentially large nets, prior research demonstrated that if a
box expression does not involve cyclic 
behaviours, the exponential number of places 
can be reduced down to polynomial (quadratic).
Using distributed places, 
it is shown that similar optimisation can also be achieved 
in the case of process expressions with iteration.
\end{abstract}


\section{Introduction}

\begin{figure}[t!]
\begin{center}
\begin{tabular}{c@{~~~~~}c}
 
\StandardNet[0.45] 

 \placW{p111}{ 0}{8}{1}{} 
 \placW{p110}{ 2}{8}{1}{} 
 \placW{p101}{ 4}{8}{1}{} 
 \placW{p100}{ 6}{8}{1}{} 
 \placW{p011}{ 8}{8}{1}{} 
 \placW{p010}{10}{8}{1}{} 
 \placW{p001}{12}{8}{1}{} 
 \placW{p000}{14}{8}{1}{} 

 \Whitetran{a1}{0}{1}{a_1}
 \Whitetran{b1}{2}{1}{b_1}
 \Whitetran{a2}{6}{1}{a_2} 
 \Whitetran{b2}{8}{1}{b_2} 
 \Whitetran{a3}{12}{1}{a_3} 
 \Whitetran{b3}{14}{1}{b_3}

 \diredge{p111}{b1}\diredge{p111}{b2}\diredge{p111}{b3} 
 \diredge{p110}{b1}\diredge{p110}{b2}\diredge{p110}{a3}
 \diredge{p101}{b1}\diredge{p101}{a2}\diredge{p101}{b3}
 \diredge{p100}{b1}\diredge{p100}{a2}\diredge{p100}{a3}
 \diredge{p011}{a1}\diredge{p011}{b2}\diredge{p011}{b3} 
 \diredge{p010}{a1}\diredge{p010}{b2}\diredge{p010}{a3}
 \diredge{p001}{a1}\diredge{p001}{a2}\diredge{p001}{b3}
 \diredge{p000}{a1}\diredge{p000}{a2}\diredge{p000}{a3}
\end{tikzpicture} 
&
\StandardNet[0.45] 
 \placW{p110}{ 2}{8}{1}{} 
 \placW{p101}{ 4}{8}{1}{}
 \placW{p011}{ 8}{8}{1}{}
 \placW{p000}{14}{8}{1}{} 

 \Whitetran{a1}{0}{1}{a_1}
 \Whitetran{b1}{2}{1}{b_1}
 \Whitetran{a2}{6}{1}{a_2} 
 \Whitetran{b2}{8}{1}{b_2} 
 \Whitetran{a3}{12}{1}{a_3} 
 \Whitetran{b3}{14}{1}{b_3} 

 \diredge{p110}{b1}\diredge{p110}{b2}\diredge{p110}{a3}
 \diredge{p101}{b1}\diredge{p101}{a2}\diredge{p101}{b3}
 \diredge{p011}{a1}\diredge{p011}{b2}\diredge{p011}{b3}
 \diredge{p000}{a1}\diredge{p000}{a2}\diredge{p000}{a3}
\end{tikzpicture} 
\\
($a$)&($b$)
\end{tabular}
\\ [2mm]
\StandardNet[0.5] 

 \placW{p1}{ 3}{0}{1}{} 
 \placW{p2}{ 7}{0}{1}{} 
 \placW{p3}{11}{0}{1}{} 
 \placW{p4}{15}{0}{1}{} 
 \placW{p5}{19}{0}{1}{} 
 \placW{p6}{23}{0}{1}{} 

 \Whitetran{a135}{0}{8}{a_1}
 \Whitetran{a246}{2}{8}{b_1}
 \Whitetran{a136}{6}{8}{a_2}
 \Whitetran{a245}{8}{8}{b_2}
 \Whitetran{a145}{12}{8}{a_3}
 \Whitetran{a236}{14}{8}{b_3}
 \Whitetran{a146}{18}{8}{a_4}
 \Whitetran{a235}{20}{8}{b_4}
 \Whitetran{a156}{24}{8}{a_5}
 \Whitetran{a234}{26}{8}{b_5}

\diredge{p1}{a135}\diredge{p3}{a135}\diredge{p5}{a135}
\diredge{p2}{a246}\diredge{p4}{a246}\diredge{p6}{a246}
\diredge{p1}{a136}\diredge{p3}{a136}\diredge{p6}{a136}
\diredge{p2}{a245}\diredge{p4}{a245}\diredge{p5}{a245}
\diredge{p1}{a145}\diredge{p4}{a145}\diredge{p5}{a145}
\diredge{p2}{a236}\diredge{p3}{a236}\diredge{p6}{a236}
\diredge{p1}{a146}\diredge{p4}{a146}\diredge{p6}{a146}
\diredge{p2}{a235}\diredge{p3}{a235}\diredge{p5}{a235}
\diredge{p1}{a156}\diredge{p5}{a156}\diredge{p6}{a156}
\diredge{p2}{a234}\diredge{p3}{a234}\diredge{p4}{a234}

 \Whitetran{b123}{0}{-8}{a_6}
 \Whitetran{b456}{2}{-8}{b_6}
 \Whitetran{b124}{6}{-8}{a_7}
 \Whitetran{b356}{8}{-8}{b_7}
 \Whitetran{b125}{12}{-8}{a_8}
 \Whitetran{b346}{14}{-8}{b_8}
 \Whitetran{b126}{18}{-8}{a_9}
 \Whitetran{b345}{20}{-8}{b_9}
 \Whitetran{b134}{24}{-8}{a_{10}}
 \Whitetran{b256}{26}{-8}{b_{10}}

\diredge{p1}{b123}\diredge{p2}{b123}\diredge{p3}{b123}
\diredge{p4}{b456}\diredge{p5}{b456}\diredge{p6}{b456}
\diredge{p1}{b124}\diredge{p2}{b124}\diredge{p4}{b124}
\diredge{p3}{b356}\diredge{p5}{b356}\diredge{p6}{b356}
\diredge{p1}{b125}\diredge{p2}{b125}\diredge{p5}{b125}
\diredge{p3}{b346}\diredge{p4}{b346}\diredge{p6}{b346}
\diredge{p1}{b126}\diredge{p2}{b126}\diredge{p6}{b126}
\diredge{p3}{b345}\diredge{p4}{b345}\diredge{p5}{b345}
\diredge{p1}{b134}\diredge{p3}{b134}\diredge{p4}{b134}
\diredge{p2}{b256}\diredge{p5}{b256}\diredge{p6}{b256}
\end{tikzpicture} 
\\
($c$)
\end{center}
\caption {\label{fi-ee3} 
($a$) A net expressing choice between 
three pairs of concurrent actions ($a_i$ and $b_i$, for $i=1,2,3$) 
and ($b$) its reduced version exhibiting the same behaviour. 
($c$) A reduced version of a net 
expressing choice between 10 pairs of concurrent actions ($a_i$ and $b_i$, for $i=1,\dots,10$).
}
\end{figure}

Petri nets are a formal modelling technique 
designed to deal with concurrent and 
distributed computing systems.
They support a simple yet expressive semantics, 
intuitive graphical notation, and the possibility of 
capturing behaviours concisely without making subsequent 
formal verification or synthesis intractable. 
There are many software tools for Petri nets, 
and Petri nets have been widely used both as a modelling 
formalism, and as an intermediate representation to which 
designs initially expressed in other formalisms are translated. 
In fact, developing translations from, 
\eg process algebras, concurrent programming languages,
and other formalisms to Petri nets has 
been extensively pursued for the past 
four decades, see \eg \cite{BDK-01,gm,kmh,old}.

The possibility to create concise system models is 
often the key advantage of Petri nets over simpler 
formalisms like Finite State Machines (\textsc{fsm}s), 
where one often encounters the exponential 
\emph{state space explosion}~\cite{V-98} already during the modelling stage.
Unfortunately, as we observed in~\cite{KKY-22}, translating even 
simple control flows to Petri nets may also lead to an exponential 
explosion in the Petri net size (in terms of places).
A typical situation in which this occurs is when   
groups (`bursts') of two or more concurrent and simultaneously 
enabled actions are put in a mutually exclusive choice. 
Consider, for instance the net in Figure~\ref{fi-ee3}($a$) which captures mutually exclusive
choice between concurrent executions of $a_1$ and $b_1$, of 
$a_2$ and $b_2$, and of $a_3$ and $b_3$.
The standard construction in Figure~\ref{fi-ee3}($a$) uses $2^3=8$ places, but 
the same behaviour can be realised using only $4$ places, as shown
in Figure~\ref{fi-ee3}($b$). In general, to express choice between $n$ pairs
of concurrent actions $a_i$ and $b_i$, 
the standard construction would use $2^n$ places. 
However, an optimised construction only requires a small fraction of these places. 
For instance, Figure~\ref{fi-ee3}($c$) shows a reduced 
solution for $n=10$ which uses 6 rather than $2^{10}=1024$ places. 
 
\paragraph{Prior works.}
As a motivating example, \cite{KKY-22} 
considered Burst Automata~\cite{CSKLY-21} 
used in the area of asynchronous circuits design. 
They are like \textsc{fsm}s with arcs labelled 
by sets of concurrently executed actions (bursts). 
In~\cite{CSKLY-21}, a language-preserving linear size translation 
was proposed, that prefixes each burst with a silent `fork' 
transition and then uses another silent `join' transition after the 
burst to detect completion. Unfortunately, there are situations 
when silent transitions are unacceptable~\cite{CKKLY-02}. 
First, language equivalence may be too weak 
(\eg it does not preserve branching time temporal properties 
or even deadlocks), and prefixing bursts with silent transitions 
breaks not only strong but also weak bisimulation~\cite{ccs}.

To preserve strong bisimulation, the \emph{cross-product} 
was often used, see \eg 
\cite{BDK-01,gm,DBLP:conf/apn/BestDH92,DBLP:conf/rex/GlabbeekG89}. 
To express a choice between several bursts (\ie sets of concurrent transitions) 
$B_1, B_2,\ldots,B_n$, this construct would create a 
set of places corresponding to tuples in the Cartesian product
$B_1\times B_2\times\cdots\times B_n$, resulting in 
the Petri net size exponential in the number of bursts. 
For example, \cite{CSKLY-21} developed translations 
from Burst Automata to Petri nets based on cross-product 
preserving weak or strong bisimulation. 
 
The paper~\cite{KKY-22} proposed an alternative to cross-product, 
that uses at most quadratic (in the total size of all bursts) 
number of places to express choice between bursts, thereby 
reducing the size of Burst Automata to Petri net translation 
from exponential~\cite{CSKLY-21} down to polynomial. 
Furthermore, in some cases a logarithmic number of places is sufficient, 
yielding a double-exponential reduction compared with the cross-product approach. 
The latter case is illustrated in Figure~\ref{fi-ee3}, as for 
a choice between $n$ binary bursts we get (asymptotically) double-exponential 
reduction from $2^n$ to $O(log~n)$.
The technique was based on showing the equivalence between the 
modelling problem of expressing choice between bursts of concurrent 
events and the problem of finding an edge clique cover of a complete multipartite graph.

For specifications of concurrent systems in the form of process expressions,
\cite{DBLP:conf/concur/KhomenkoKY22} generalised the above technique to 
box expressions involving choice, concurrency, and sequencing operators. 
We proposed a polynomial translation of such box expressions 
that preserves strong bisimulation (in fact, it guarantees the isomorphism of 
reachability graphs, which is a stronger equivalence). 
The developed translation was compositional --- 
this was ensured by augmenting Box Algebra~\cite{BDK-01} with 
the notion of \emph{interface graphs} in a 
way that allowed importing many results from Box Algebra.
 
The construction based on finding an \emph{edge clique cover} of a certain 
\emph{complement-reducible graph (cograph~\cite{L-71})}, where some of the 
edges may already be considered as `covered', yields a translation with 
the number of created places corresponding to the number of cliques in the cover.
The basic idea is that each semantically essential 
sequence and conflict relationship 
between two actions is represented by an edge.
Then, a clique corresponds to a net realisation 
of several individual constraints using a single place.
It is then easy to see that at most polynomial (quadratic) 
number of cliques are always sufficient (because the relevant 
cograph has a linear number of vertices), which yields a polynomial Petri net.
Hence, the results of~\cite{KKY-22,DBLP:conf/concur/KhomenkoKY22} 
demonstrated that if a box expression does not involve cyclic 
behaviours, the exponential number of places created by the cross-product 
construction can be reduced 
from exponential down to polynomial (quadratic) even in the worst case, 
and to logarithmic in the best case.

In this paper, we extend the 
results of~\cite{DBLP:conf/concur/KhomenkoKY22} 
to the case of process expressions 
with iteration. 
To this end, we introduce a novel notion of \emph{distributed place}
which provides a modelling 
representation for the individual places of Petri nets
(and so each distributed place can be abstracted to a single net place).

\paragraph{The approach followed in this paper.}
In general terms, the problem addressed in this paper can be 
formulated in the following way:
\begin{quote}
\emph{Given a formal specification of a concurrent 
system expressed in some formalism, find possibly smallest 
Petri net with the same set of firing sequences or branching behaviour 
represented by its reachability graph.}
\end{quote}
This means that we are interested 
in reducing the number of places, and the set of transitions is fixed 
by the given specification. 

Two particular instances of such a broad problem 
concern specifications formulated using Petri nets and process algebras:
\begin{quote}
 \emph{Given a safe Petri net $\NN$, find possibly smallest 
 safe Petri net $\NN'$ with the same behaviour as $\NN$ 
 (\ie with the same transition set and an isomorphic 
 reachability graph). \hfill\textbf{\textsc{Prob~i}}}
\end{quote}
\begin{quote}
 \emph{Given a process algebra expression, find possibly smallest 
 safe Petri net with the same behaviour
 (\ie with the same transition set and an isomorphic 
 reachability graph). \hfill\textbf{\textsc{Prob~ii}}}
\end{quote}
Aiming to solve \textsc{Prob~i}, a naive way of finding $\NN'$ might be to derive the 
sequential reachability graph $\RG_\NN$ of $\NN$ and then apply 
Petri net synthesis techniques for $\RG_\NN$ specifically aimed
at generating a minimal net solution like 
those in~\cite{bbd,BBD-book,DBLP:conf/concur/Pietkiewicz-Koutny98}.
Though theoretically sound, such a method is hardly practical 
as the size of $\RG_\NN$ is too often exponential in the size of $\NN$. 

With a rather bleak outlook at solving \textsc{Prob~i} in full generality, 
one might seek limited but still practically relevant solutions.
In particular, one can aim at developing a compositional solution,
along the following lines:
\begin{quote}
 \emph{Given a safe Petri net $\NN$ constructed from components
 $\mathcal C_1,\dots,\mathcal C_k$,
 find possibly smallest behaviourally equivalent
 safe Petri net $\NN'$ constructed from optimised 
 versions of (some of) the components. In addition, 
 all the decisions should be made 
 as `locally' as possible
 (intuitively, coming up with each such decision should avoid investigating 
 a significant part of the behaviour and/or the structure of $\NN$).}
\end{quote}
As an example of such an approach, 
consider $\NN$ derived from sequential Petri nets 
$\mathcal{S}_1,\dots,\mathcal{S}_k$ which are 
composed by gluing (synchronising on)
common transitions. 
In such a case, deleting from each $\mathcal{S}_i$ some places leading to
$\mathcal{S}'_i$ generating the same firing sequences and/or deleting some 
$\mathcal{S}_i$'s would lead to $\NN'$ with the same behaviour as $\NN$. 

\begin{figure}[t!]
\begin{center}
\begin{tabular}{cc}
 
\StandardNet[0.55] 
 \placN{Aa}{ 4}{7.5}{1}{p_1}
 \placN{Ab}{10}{7.5}{1}{p_2}
 
 \placW{cea}{ 0}{4.5}{0}{p_3\!\!\!} 
 \placW{ceb}{ 2}{4.5}{0}{p_4\!\!\!} 
 \placW{cfa}{ 4}{4.5}{0}{p_5\!\!\!} 
 \placW{cfb}{ 6}{4.5}{0}{p_6\!\!\!} 
 \placE{dea}{ 8}{4.5}{0}{\!\!\!p_7} 
 \placE{deb}{10}{4.5}{0}{\!\!\!p_8} 
 \placE{dfa}{12}{4.5}{0}{\!\!\!p_9} 
 \placE{dfb}{14}{4.5}{0}{\!\!\!p_{10}} 
 
 \Whitetran{a}{ 4}{6}{a}
 \Whitetran{b}{10}{6}{b}
 \Whitetran{c}{ 2}{1.5}{c} 
 \Whitetran{d}{ 4}{1.5}{d}
 \Whitetran{e}{10}{1.5}{e}
 \Whitetran{f}{12}{1.5}{f}

 \diredge{Aa}{a}\diredge{Ab}{b} 
 
 \diredge{a}{cea}\diredge{a}{cfa}\diredge{a}{dea}\diredge{a}{dfa}
 \diredge{b}{ceb}\diredge{b}{cfb}\diredge{b}{deb}\diredge{b}{dfb}
 \diredge{ceb}{c}\diredge{cfb}{c}\diredge{cea}{c}\diredge{cfa}{c}
 \diredge{deb}{d}\diredge{dfb}{d}\diredge{dea}{d}\diredge{dfa}{d}
 \diredge{ceb}{e}\diredge{cfb}{f}\diredge{cea}{e}\diredge{cfa}{f}
 \diredge{deb}{e}\diredge{dfb}{f}\diredge{dea}{e}\diredge{dfa}{f}
 
\end{tikzpicture} 
&
\StandardTS [0.55]
\nodN{a}{3}{6}{M_\init}
\nodE{b}{0}{5}{}
\nodE{c}{6}{5}{}
\nodE{d}{3}{4}{}
\nodE{e}{0}{2}{}
\nodE{f}{2}{2}{}
\nodE{g}{3}{0}{}
\nodE{h}{4}{2}{}
\nodE{i}{6}{2}{}

\edgetext{a}{b}{$a$};\edgetext{b}{d}{$b$};
\edgetext{a}{c}{$b$};\edgetext{c}{d}{$a$};
\edgetext{d}{e}{$c$};\edgetext{d}{f}{$d$};
\edgetext{e}{g}{$d$};\edgetext{f}{g}{$c$};

\edgetext{d}{h}{$e$};\edgetext{d}{i}{$f$};
\edgetext{h}{g}{$f$};\edgetext{i}{g}{$e$};
 
\end{tikzpicture}
\\
($a$) $\NN_0$&($c$)
\\[3mm] 
\StandardNet[0.55]
 \placN{Aa}{ 4}{7.5}{1}{p_1}
 \placN{Ab}{10}{7.5}{1}{p_2}
 
 \placW{ceb}{ 2}{4}{0}{p_4} 
 \placE{cfa}{ 4}{4}{0}{p_5} 
 \placW{deb}{10}{4}{0}{p_8} 
 \placE{dfa}{12}{4}{0}{p_9} 
 \Whitetran{a}{ 4}{6}{a}
 \Whitetran{b}{10}{6}{b}
 \Whitetran{c}{ 2}{2}{c} 
 \Whitetran{d}{ 4}{2}{d}
 \Whitetran{e}{10}{2}{e}
 \Whitetran{f}{12}{2}{f} 

 \diredge{Aa}{a}\diredge{Ab}{b} 
 \diredge{a}{cfa} \diredge{a}{dfa}
 \diredge{b}{deb}\diredge{b}{ceb}
 \diredge{ceb}{c} \diredge{cfa}{c}
 \diredge{deb}{d} \diredge{dfa}{d} 
 \diredge{ceb}{e} \diredge{cfa}{f}
 \diredge{deb}{e} \diredge{dfa}{f} 
\end{tikzpicture}
&
\StandardNet[0.55] 
 \placW{Aa}{1}{7.5}{1}{}
 \placW{cfa}{1}{4}{0}{p_i} 
 
 \Whitetran{a}{1}{6}{} 
 \Whitetran{c}{0}{2}{} 
 \Whitetran{f}{2}{2}{}

 \diredge{Aa}{a} 
 \diredge{a}{cfa} \diredge{cfa}{c} \diredge{cfa}{f}
\end{tikzpicture} 
\\
($b$) $\NN'_0$&($d$)
\end{tabular}
\end{center}
\caption {\label{fi-1}
($a$) A safe Petri net with ($b$) its reduced behaviourally equivalent
net and ($c$) its reachability graph.
($d$) A sequential component $\SSS_i$
of $\NN_0$ ($i=3,4,\dots,10$).
}
\end{figure}

Consider, for example, the Petri net $\NN_0$ shown in Figure~\ref{fi-1}($a$)
which specifies a system which first executes 
transitions $a$ and $b$ (concurrently), and then 
either $c$ and $d$ (concurrently) or $e$ and $f$ (concurrently). 
$\NN_0$ can be decomposed uniquely onto eight sequential subnets
$\mathcal{S}_3,\mathcal{S}_4,\dots,\mathcal{S}_{10}$ as depicted 
in Figure~\ref{fi-1}($d$).
None of these can be reduced in the way described above. However, 
one can remove four of these sequential nets
leaving $\mathcal{S}_4,\mathcal{S}_5,\mathcal{S}_8,\mathcal{S}_9$
in the composition.
The result is $\NN'_0$ shown in Figure~\ref{fi-1}($b$)
with an isomorphic reachability graph
depicted in Figure~\ref{fi-1}($c$). 
Notice also that removing 
$\mathcal{S}_3,\mathcal{S}_6,\mathcal{S}_7,\mathcal{S}_9$ 
could not be done locally as one needs to take into account other 
sequential subnets.

In this paper, we will show that there is an alternative 
compositional solution to the problem at hand.
To this end, we will consider a covering of the places $P$ of a safe
Petri net 
by \emph{distributed places} $\PC_1,\dots,\PC_k$. Each distributed 
place $\PC_i$ is a set of places satisfying properties which depend 
purely on the arcs between $\PC_i$ and the transitions 
in $\PRE{\PC_i}\cup\POST{\PC_i}$. What is important is that other arcs
incident to these transitions are irrelevant, and so the 
definition of a distributed place is purely \emph{local}.
We then demonstrate that it is possible to delete from the 
$\PC_i$'s some of the places, leading to 
$\PC'_1,\dots,\PC'_k$ forming places of a behaviourally equivalent 
safe Petri net.
And, crucially, the reduction from $\PC_i$ 
to $\PC'_i$ is both \emph{static} (as the overall behaviour of the net 
is not needed) and \emph{local} (as the requirements deal 
separately with distributed places and transitions linked to them,
but disregard links of such transitions to places outside the 
distributed place).

As an example, the places of $\NN_0$ shown in Figure~\ref{fi-1}($a$) 
can be partitioned into two distributed places:
$\PC_1=\{p_1,p_2\}$ and $\PC_2=\{p_3,\dots,p_{10}\}$.
In this case, $\PC_1$ cannot be reduced and so $\PC'_1=\PC_1$,
but $\PC_2$ can be reduced to $\PC'_2=\{p_4,p_5,p_8,p_9\}$.
The result is a net $\NN'_0$ shown in Figure~\ref{fi-1}($b$).
The two nets are behaviourally 
equivalent, as they have (up to isomorphism)
the same reachability graph portrayed in Figure~\ref{fi-1}($c$).
 
In this paper, we introduce a novel notion of `distributed place'
which intuitively provides a concurrent implementation of 
the individual places of Petri nets (and so the latter provide abstractions
of distributed places). 
We also formulate and prove correct the reduction procedure
based on distributed places.

Box Algebra~\cite{BDK-01,DBLP:journals/iandc/BestDK02} 
provides a generic process-algebraic framework for 
Petri nets called \emph{(Petri) boxes}.
Compositionally defined boxes can be derived from 
\emph{box expressions} which 
are a process algebra in the usual language-theoretic sense. 
Box Algebra has several
concrete incarnations, including \textsc{ccs}~\cite{ccs} 
and \textsc{tcsp}~\cite{csp}.
In general, being able to find small 
Petri nets with the same behaviour as 
formal specifications of concurrent 
systems benefits both 
effective verification and 
practical implementation of such systems.
This paper considers specifications given 
in the form of safe Petri nets and box expressions.

Moving on to \textsc{Prob~ii}, we will consider 
safe Petri nets derived from 
process expressions supporting 
choice ($E\CHOICE F$), sequence ($E\SEQ F$), parallel ($E\PAR F$),
and iteration ($\ITER{E}{F}{G}$) compositions. 
For example, the net in Figure~\ref{fi-ee3}($a$)
corresponds to 
the box expression $(a_1\PAR b_1)\CHOICE(a_2\PAR b_2)\CHOICE(a_3\PAR b_3)$.
For each such expression $E$, there is a standard translation 
yielding a safe Petri net $\BOX(E)$, called a \emph{box}.
For compositionally 
derived nets, distributed places emerge `by construction'. What is more,
to reduce $\BOX(E)$ one does not need to explicitly 
derive $\BOX(E)$ that can be exponential in the size of $E$.
Instead, we construct an undirected 
`connection graph' $\CG(E)$ 
which happens to be a complement-reducible graph (cograph) 
whose representation is at most linear in the size of $E$, and derive the $\PC'_i$'s
directly from an edge clique covering of $\CG(E)$. 
Note that edge clique cover on cographs is in 
\textsc{np} and is suspected to be \textsc{np}-complete, 
but there are polynomial heuristic algorithms, and even 
the trivial cover that covers each edge by a separate clique yields 
at most quadratic solution, which is much better than the 
exponential solution resulting from the cross-product construction.

\paragraph{Related work.}

There are several works aimed at structural property-preserving Petri net transformations.
Such rules allow one to contract a large model (reduction),
and to construct a fuller model of the system (refinement).
In the domain of PT-nets, \cite{DBLP:conf/mfcs/BerthelotR76}
proposed three place reduction rules
preserving liveness and 
boundedness.
The papers \cite{DBLP:journals/jcss/SuzukiM83,DBLP:conf/ac/Berthelot86,murata1989petri} 
investigated ways for transforming Petri net representation 
refinement of transitions and places (or abstraction of subnets to transitions),
aiming at preserving properties such as liveness, deadlock-freeness and boundedness. 
The paper \cite{DBLP:conf/spin/BerthomieuBD18} 
developed an approach aimed at discovering redundant
places and transitions in PT-nets
using integer linear programming, while \cite{Lee1985} proposed
a hierarchical reduction approach for Petri nets aimed at preserving
liveness and boundedness. 
In the domain of safe Petri nets, \cite{DBLP:journals/topnoc/BernardinelloLNP22}
investigated transformation rules (in particular, deadlock-preserving)
for elementary net systems with the help of 
morphisms, and discussed how they can be used in workflow net compositions.
The paper 
\cite{DBLP:conf/concur/Desel90} used
vicinity respecting net morphisms to formalise contractions of nets.
By restricting the interest to free choice Petri nets,
\cite{DBLP:conf/apn/EsparzaS90,Desel_Esparza_1995,DBLP:journals/iandc/Esparza94}
developed reduction rules that allow 
one to construct every live and bounded free choice net. 
Structural transformations are not restricted to the basic 
Petri net frameworks. For example, \cite{DBLP:conf/fase/EsparzaH16}
proposed reduction rules for a model based on workflow nets extended
with data. 
To deal with reconfigurable systems
\cite{ehrig2006transformations,DBLP:conf/dfg/PadbergU03} 
developed graph transformations 
in categorical setting.
To cope with nets enriched with reset and inhibitor arcs,
and after
discovering that many rules defined for basic net models are invalid,
\cite{DBLP:journals/jcss/VerbeekWAH10} 
introduced new liveness and boundedness preserving transformations.
Aiming at coping with system equivalences based on bisimilarity,
\cite{DBLP:conf/apn/SchnoebelenS00} investigated structural 
equivalences on places of P/T nets that allow reductions compatible with bisimilarity. 

When compared to the existing works on net reduction, 
this paper aims at preserving isomorphism of reachability graphs 
which is arguably the strongest behavioural
requirement expected of a reduction scheme
(note that two Petri nets with isomorphic reachability 
graphs are, in particular, equivalent with respect to deadlock-freeness, liveness, and 
proper termination). 
As a result, we focus on identifying places (rather than transitions) 
which can be deleted without 
changing the behaviour.
Second, when dealing with transitions, we only consider
a subset of their connections, viz.\ those which link to the 
distributed place.
Third, before applying the reduction rules 
to nets corresponding to box expressions, we never explicitly generate places 
using the exponential cross-product construction as in the standard semantics~\cite{BDK-01}.
In this way, we contribute techniques aimed at efficient translation
of process expressions into Petri nets.

\paragraph{Structure of this paper.}
This paper has three parts. Following a preliminary
Section~\ref{sect-prelim}, Part~I
(Section~\ref{sect-cluster}) introduces distributed places.
In particular, Theorem~\ref{th-main} provides a formal justification
of the suitability of distributed places as a means 
of behaviour-preserving reduction of safe Petri nets.
Part~II (Sections~\ref{sec-constrdp} and~\ref{sec-balg}) 
first identifies two operations for composing
distributed places, and proves 
some basic properties of distributed places. It then demonstrates how 
these operations are used in the standard translation of box expressions to 
nets. 
 Part~III (Sections~\ref{sec-cg} and \ref{sec-cgss}) 
presents basic facts concerning connection graphs and 
shows how they can be employed in an optimised translation 
of box expressions to behaviourally equivalent nets. 

This paper is a revised and extended version of the 
conference publication~\cite{DBLP:conf/apn/KhomenkoKY25} which
was used in~\cite{DBLP:conf/apn/ChanTSY26} in 
the context of reducing the size 
of safe nets modelling Tsetlin Machines (a machine learning model).
 
\section{Basic notions and notations}
\label{sect-prelim}

\paragraph{Nets.}
A \emph{net} is a tuple 
$\FA{P,T,\Flow}$, 
where $P$ is a finite set of \emph{places},
$T$ is a disjoint finite set of \emph{transitions}, and 
$\Flow\subseteq (T\times P)\cup (P\times T)$ 
is the \emph{flow relation}. 
For all $x\in P\cup T$ and $X\subseteq P\cup T$, we denote:
\[
\begin{array}{lllll}
\PRE{x}=\{y\mid \FA{y,x}\in \Flow \}& 
\POST{x}=\{y\mid \FA{x,y}\in \Flow\}& 
\PREPOST{x}=\PRE{x}\cup\POST{x}&
\cPRE{x}=\PRE{x}\setminus\POST{x}&
\cPOST{x}=\POST{x} \setminus\PRE{x}
\\
\PRE{X}=\bigcup_{x\in X} \PRE{x} &
\POST{X}=\bigcup_{x\in X} \POST{x} &
\PREPOST{X}=\bigcup_{x\in X}\PREPOST{x}&
\cPRE{X}=\bigcup_{x\in X} \cPRE{x} &
\cPOST{X}=\bigcup_{x\in X} \cPOST{x} \;.
\end{array}
\] 
It is assumed that no place $p$ is \emph{isolated}, 
\ie it has an \emph{input} transition $t\in\PRE{p}$, or an \emph{output} transition 
$t\in\POST{p}$. 
Moreover, each transition $t$ has at least one \emph{pre-place} 
$p\in\PRE{t}\neq\es$. 
We use the standard way of representing nets as directed graphs,
\eg transitions are depicted as small squares and places as circles.

\paragraph{Sequences of transitions.}
Let 
$\sigma$ and $\omega$ be finite sequences of transitions, and 
$V$ and $W$ be disjoint sets of transitions. Then:
(i) 
$\pref(\sigma)$ is the set of all the prefixes of $\sigma$;
(ii)
$\sigma|_V$ is obtained from $\sigma$ by deleting all 
the transitions outside $V$; 
(iii) 
$\sigma\circ \omega$ is the concatenation 
of $\sigma$ and $\omega$;
(iv) 
$\gamma|||\delta$ is set of all sequences $\kappa$
such that $\kappa|_V=\gamma$ and $\kappa|_W=\delta$,
for all $\gamma\in V^*$ and $\delta\in W^*$; 
and
(v) $\Gamma|||\Delta=\bigcup\{\gamma|||\delta\mid \gamma\in\Gamma \wedge
\delta\in\Delta\}$, for all $\Gamma\subseteq V^*$ and $\Delta\subseteq W^*$. 
The empty sequence is denoted by $\lambda$, and 
it is assumed that $\es^*=\{\lambda\}$.

\paragraph{Marked nets.}
A \emph{marked net} is a tuple 
$\NN=\FA{P,T,\Flow,M_\init}$ such that $\FA{P,T,\Flow}$
is a net and 
$M_\init\subseteq P$ is the \emph{initial marking}. 
In general, any set of places $M$ is a \emph{marking}. 
In the diagrams, a marking $M$ is depicted
by tokens drawn inside circles representing places $p$ belonging to $M$.
We also say that $p\in M$ is \emph{filled} and otherwise \emph{empty}.
 
A transition $t\in T$ is \emph{fireable} at a 
marking $M$ if $\PRE{t}\subseteq M$. Such 
a transition can be \emph{fired} leading to marking 
$M'=(M\setminus\PRE{t})\cup\POST{t}=(M\setminus\cPRE{t})\cup\cPOST{t} $.
We denote this by 
\[
t\in\fireable_{\NN}(M) 
~~\mbox{and}~~
M\xrightarrow[\NN]{t} M'\;, 
\]
respectively.
Note that firing transition $t$ \emph{removes} tokens from 
the places in $\cPRE{t}$ making them empty, and 
\emph{inserts} tokens into places in $\cPOST{t}$ making them filled.
The empty/filled status of other places remains the same.

A marking $M'$ is \emph{reachable} from marking $M$ if $M=M'$ or
$M\xrightarrow[\NN]{t_1} M_1\xrightarrow[\NN]{t_2}
M_2\dots M_{k-1}\xrightarrow[\NN]{t_k}M'$,
for some markings $M_1,\dots,M_{k-1}$ and transitions $t_1,\dots,t_k$ ($k\geq 1$).
We then denote 
$M\xrightarrow[\NN]{\lambda} M$ and
$M\xrightarrow[\NN]{t_1\dots t_k} M'$ as well as 
$M\xrightarrow[\NN]{ } M'$.

A \emph{firing sequence} of $\NN$ is a 
sequence of transitions $\sigma$ 
such that there is a marking $M$ satisfying 
$M_\init\xrightarrow[\NN]{\sigma} M$. 
Then $M$ is \emph{reachable} in $\NN$. 
The firing sequences and reachable markings are 
denoted by $\fseq(\NN)$ and 
$\reach(\NN)$, respectively.
Note that 
$\reach(\NN)=\{M\mid \exists \sigma \in \fseq(\NN):~M_\init\xrightarrow[\NN]{\sigma} M\}$. 

A marked net $\NN$ is \emph{safe} 
if $\POST{t}\cap(M\setminus \PRE{t}) =\es$ (or, alternatively, $\cPOST{t}\cap M =\es$), 
for each marking $M\in\reach(\NN)$ and each transition $t\in\fireable_{\NN}(M)$.

\paragraph{Reachability graphs.}
The tuple $\RG_\NN=\FA{\reach(\NN),A,M_\init}$,
where $\reach(\NN)$ are the nodes, 
$A=\{\FA{M,t,M'} \mid M\in\reach(\NN) \wedge M\xrightarrow[\NN]{t} M'\}$
are the labelled directed arcs, and $M_\init$ is the initial 
node, is the \emph{reachability graph} of a marked net $\NN$.
Note that $\RG_\NN$ is \emph{deterministic}, which means that 
if $\FA{M,t,M'}$ and $\FA{M,t,M''}$ belong to $A$, then $M'=M''$.

Note that the firing sequences in $\fseq(\NN)$ label the directed paths 
of $\RG_\NN$ originating at the initial node $M_\init$.
 
Two reachability graphs, $\RG_{\NN_i}=\FA{\reach(\NN_i),A_i,M^i_\init}$
(for $i=0,1$), 
are \emph{isomorphic} if there is a bijection
$\imath:\reach(\NN_0)\to\reach(\NN_1)$ such that 
$\imath(M^0_\init)=M^1_\init$ and 
$\FA{M,t,M'}\in A_0 \iff \FA{\imath(M),t,\imath(M')}\in A_1$,
for all $M,M'\in\reach(\NN_0)$.
We denote this by $\RG_{\NN_0}\cong\RG_{\NN_1}$.

\begin{proposition}
\label{prop-refrf}
Let $\NN_i=\FA{P_i,T_i,\Flow_i,M^i_\init}$ (for $i=0,1$) be 
marked nets such that $\fseq(\NN_0)=\fseq(\NN_1)$.
Then $\RG_{\NN_0}\cong\RG_{\NN_1}$ provided that, for all 
$\sigma,\sigma'\in\fseq(\NN_0)=\fseq(\NN_1)$:
\begin{equation}
\label{eq-ufufgg}
(\exists M: M^0_\init\xrightarrow[\NN_0]{\sigma}M ~\wedge~ 
 M^0_\init\xrightarrow[\NN_0]{\sigma'}M )
 \iff
 (\exists M':~M^1_\init\xrightarrow[\NN_1]{\sigma}M' ~\wedge~
 M^1_\init\xrightarrow[\NN_1]{\sigma'}M') \;,
\end{equation} 
\ie $\sigma$ and $\sigma'$ lead to the same 
marking in $\NN_0$ iff they lead to the same 
marking in $\NN_1$.
\end{proposition}
\begin{proof} 
Let $\RG_{\NN_i}=\FA{\reach(\NN_i),A_i,M^i_\init}$ (for $i=0,1$).
For every $M\in\reach(\NN_i)$ (for $i=0,1$), let 
$\rho_i(M)$ be the set of all label sequences of 
the directed paths in $\RG_{\NN_i}$ from $M^i_\init$ to $M$.

Let $i\in\{0,1\}$. We observe that since $\RG_{\NN_i}$ is deterministic 
and all the nodes of $\RG_{\NN_i}$ are reachable from $M^i_\init$, 
$\pi_i=\{\rho_i(M)\mid M\in\reach(\NN_i)\}$ 
is a partition of $\fseq(\NN_i)$.
That is, $\fseq(\NN_i)=\bigcup\{\rho_i(M)\mid M\in\reach(\NN_i)\}$
as well as
$\rho_i(M)\neq\es$ and $M\neq M'\implies \rho_i(M)\cap\rho_i(M')=\es$,
for all $M,M'\in\reach(\NN_i)$.

It then follows, by $\fseq(\NN_0)=\fseq(\NN_1)$ and Eq.\eqref{eq-ufufgg}, 
that $\pi_0=\pi_1$.

Let 
$\imath:\reach(\NN_0)\to\reach(\NN_1)$ be such that
$\imath(M)=M'$ if $\rho_0(M)=\rho_1(M')$,
for every $M\in\reach(\NN_0)$
(\ie $\iota(M)=\rho_1^{-1}(\rho_0(M))$).
It is straightforward to show that $\imath$
is a bijection establishing isomorphism of $\RG_{\NN_0}$ and 
$\RG_{\NN_1}$.
\end{proof}

\paragraph{Marked net size.}
The total number of tokens in the initial marking cannot 
exceed the number of places, so one can define $\NN$'s 
size as the total number of places, transitions, and arcs. 
In practice, the size of $\NN$ is dominated by its arcs.
When translating process expressions to marked nets, the set of 
transitions in general coincides with the atomic actions of the 
expression, and the aim will be to 
use a small number of places and arcs.

\section{Distributed places}
\label{sect-cluster} 

In this section, we aim at lifting various notions 
concerning the individual places of a safe marked net to the level 
of sets of places. Basically, we would like to impose 
constraints on a set of places to ensure that (at least in 
some contexts) its dynamic behaviour could be regarded 
as that of a single place. 

Let $\PC$ be a non-empty set of places of a marked net.
Then, for every transition $t$, we denote
$\POSTT_\PC(t)=\PC\cap\POST{t}$
and
$\PREE_\PC(t)=\PC\cap\PRE{t}$.
We also define three disjoint
sets of transitions:
\begin{itemize}
\item 
$\In_\PC=\{t\in\PRE{\PC}\mid \PREE_\PC(t) =\es \} $
are transitions 
which only insert tokens into $\PC$. 
\item 
$\Out_\PC=\{t\in\POST{\PC}\mid\POSTT_\PC(t)=\es \} $
are transitions 
which only remove tokens from $\PC$.
\item 
$\Read_\PC=\{t\in\PREPOST{\PC}\mid \POSTT_\PC(t) =\PREE_\PC(t) \}$
are transitions 
which only `read' tokens in $\PC$. 
\end{itemize} 
It is worth mentioning that transitions connected to the places in $\PC$ which
are \emph{not included} in any of the above three sets are
those which combine two or more characteristics of these sets;
\eg any transition $t$ such that $\PC\cap\cPOST{t}\neq\es$ 
and $\PC\cap\cPRE{t}\neq\es$.
It is also worth observing that for a singleton set $\PC=\{p\}$, 
we have 
$\In_\PC=\PRE{p}\setminus\POST{p}$,
$\Out_\PC=\POST{p}\setminus\PRE{p}$,
and
$\Read_\PC=\POST{p}\cap\PRE{p}$.

We also identify the following two kinds of 
pairs of transitions adjacent to places in $\PC$ in 
ways which influence their fireability:
\begin{itemize}
\item 
$\enables(\PC)=\{\FA{t,u}\in T\times T\mid 
 \PC\cap(\cPOST{t}\cap\PRE{u})\neq\es \}$.
 \\
Intuitively, if a place $p\in\PC\cap\PRE{u}$ is empty, then
$u$ can only become fireable if a transition $t$ satisfying 
$p\in\cPOST{t}$ is fired.
\item 
$\disables(\PC)=
 \{\FA{t,u}\in T\times T
 \mid 
 \PC\cap(\cPRE{t}\cap\PRE{u})\neq\es \}$.
 \\
 Intuitively, if $\FA{t,u}\in \disables(\PC)$, then firing $t$ empties 
 a place in $\PRE{u}$ making $u$ non-fireable.
\end{itemize}
\begin{example}
\label{ex-irvui}
For $\NN_0$ in Figure~\ref{fi-1}($a$), we have:
\[
\begin{array}{rcl}
\In_{\{p_3,p_6\}}
 &=&
 \{a,b\}
 \\
\Out_{\{p_3,p_6\}}
 &=&
 \{c,e,f\}
 \\
\Read_{\{p_3,p_6\}}
 &=&
 \es
 \\
 \enables(\{p_3,p_4,\dots,p_{10}\})
 &=&
 \{a,b\}\times \{c,d,e,f\}
 \\
 \disables(\{p_3,p_4,\dots,p_{10}\})
 &=&
 (\{c,d\}\times \{e,f\}) \cup (\{e,f\}\times\{c,d\})
 \\
 \enables(\{p_6,p_7,p_{10}\})
 &=&
 \{\FA{b,c}, \FA{b,f}, \FA{b,d}, \FA{a,d}, \FA{a,e}\}
 \\
 \disables(\{p_6,p_7,p_{10}\})
 &=&
 \{\FA{d,f}, \FA{f,d}, \FA{d,e}, \FA{e,d}, \FA{c,f}, \FA{f,c}\}\;.
\end{array}
\]
Also, in Figure~\ref{fi-1}($a,b$), we have the following:
$\enables(\{p_4,p_5,p_8,p_9\})=\enables(\{p_3,p_4,\dots,p_{10}\})$
and $\disables(\{p_4,p_5,p_8,p_9\})=\disables(\{p_3,p_4,\dots,p_{10}\})$.
\eod
\end{example}

We now make precise a concept which is aimed at 
capturing a simple structural notion of a `distributed place'
(set of places)
whose state (marking) --- as the firing progresses --- 
changes monotonically and cyclically 
between 
being empty and being full. 
For example, 
$\PC=\{p_3,p_4,\dots,p_{10}\}$ 
is a distributed place of the marked net in Figure~\ref{fi-1}($a$). 
Initially the places of $\PC$ are empty, 
and firing $a$ and $b$ inserts tokens into its places. 
Note that while being filled, no transition can remove 
tokens from $\PC$. 
Then, \eg transitions $c$ and $d$ can fire and remove all 
the tokens in $\PC$ and, while being emptied, 
no transition can insert tokens into $\PC$.

We start by specifying what is meant by a monotonic change 
of the marking of a set of places $\PC$. 
Note that, in general, firing a single transition is not 
enough to fill with tokens an empty $\PC$, and 
one may need a sequence of transitions 
to do this.
Similarly, firing a single transition is, in 
general, not enough to empty $\PC$ filled with tokens. 

\begin{definition}[in-sequence and out-sequence]
\label{def-kkd}
Let $\PC$ be a non-empty set of places of a marked net.
\begin{enumerate}
 \item 
An \emph{in-sequence of $\PC$} is 
 $\sigma=t_1\dots t_k\in(\In_\PC\cup\Read_\PC)^*$ 
 such that we have, for every $1\leq i \leq k$: 
\[
\begin{array}{lcl}
 t_i\in\In_\PC&\implies&
 \PC\cap \cPOST{t_i} \cap \cPOST{\{t_1,\dots,t_{i-1}\}} =\es
 \\
 t_i\in\Read_\PC&\implies&
 \PREE_\PC(t_i) \subseteq \PC\cap\cPOST{\{t_1,\dots,t_{i-1}\}} \;. 
\end{array}
\] 
Moreover, $\sigma$ is a
\emph{complete in-sequence} if 
$\PC\cap \cPOST{\{t_1,\dots,t_k\}}=\PC$. 
We then denote $\sigma\in\inseq_\PC$ and $\sigma\in\complinseq_\PC$, respectively.
 
 \item 
 An \emph{out-sequence of $\PC$} is 
 $\sigma=t_1\dots t_k\in(\Out_\PC\cup\Read_\PC)^*$ 
 such that we have, for every $1\leq i \leq k$: 
\[
 \PREE_\PC(t_i)\subseteq \PC\setminus \cPRE{\{t_1,\dots,t_{i-1}\}} \;.
\] 
Moreover, $\sigma$ is a \emph{complete out-sequence} if 
$\PC\cap\cPRE{\{t_1,\dots,t_k\}}=\PC$.
We then denote $\sigma\in\outseq_\PC$ 
and $\sigma\in\comploutseq_\PC$, respectively. 
\eod
\end{enumerate} 
\end{definition} 
Intuitively, an in-sequence inserts tokens into
$\PC$ in a monotonic and `safe' way 
(due to 
$t_i\in\In_\PC \implies\PC\cap( \cPOST{t_i} \cap \cPOST{\{t_1,\dots,t_{i-1}\}})=\es$)
whereas an out-sequence removes tokens from $\PC$ in a monotonic way.

\begin{figure}[t!]
\begin{center} 
\begin{tabular}{cc} 

\StandardNet[0.55] 

 \placN{Aa}{2}{9}{0}{p_1}
 \placN{Ab}{4}{9}{0}{p_2} 
 \Whitetran{a}{2}{6}{a}
 \Whitetran{b}{4}{6}{b} 

 \diredge{Aa}{a}\diredge{Ab}{b} 
\end{tikzpicture} 
&
\StandardNet[0.55] 

 \placW{cea}{ 0}{4.5}{0}{p_3\!\!\!} 
 \placW{ceb}{ 2}{4.5}{0}{p_4\!\!\!} 
 \placW{cfa}{ 4}{4.5}{0}{p_5\!\!\!} 
 \placW{cfb}{ 6}{4.5}{0}{p_6\!\!\!} 
 \placE{dea}{ 8}{4.5}{0}{\!\!\!p_7} 
 \placE{deb}{10}{4.5}{0}{\!\!\!p_8} 
 \placE{dfa}{12}{4.5}{0}{\!\!\!p_9} 
 \placE{dfb}{14}{4.5}{0}{\!\!\!p_{10}} 
 
 \Whitetran{a}{ 4}{6}{a}
 \Whitetran{b}{10}{6}{b}
 \Whitetran{c}{ 2}{1.5}{c} 
 \Whitetran{d}{ 4}{1.5}{d}
 \Whitetran{e}{10}{1.5}{e}
 \Whitetran{f}{12}{1.5}{f}
 
 \diredge{a}{cea}\diredge{a}{cfa}\diredge{a}{dea}\diredge{a}{dfa}
 \diredge{b}{ceb}\diredge{b}{cfb}\diredge{b}{deb}\diredge{b}{dfb}
 \diredge{ceb}{c}\diredge{cfb}{c}\diredge{cea}{c}\diredge{cfa}{c}
 \diredge{deb}{d}\diredge{dfb}{d}\diredge{dea}{d}\diredge{dfa}{d}
 \diredge{ceb}{e}\diredge{cfb}{f}\diredge{cea}{e}\diredge{cfa}{f}
 \diredge{deb}{e}\diredge{dfb}{f}\diredge{dea}{e}\diredge{dfa}{f}
\end{tikzpicture} 
\\
($a$) $\PC$
&
($b$) $\PC'$
\\ [4mm]
\StandardNet[0.55] 
 \placW{ceb}{ 2}{4}{0}{p_4} 
 \placW{cfa}{ 4}{4}{0}{p_5} 
 \placE{deb}{ 6}{4}{0}{p_8} 
 \placE{dfa}{ 8}{4}{0}{p_9} 
 
 \Whitetran{a}{4}{6}{a}
 \Whitetran{b}{6}{6}{b}
 \Whitetran{c}{2}{2}{c} 
 \Whitetran{d}{4}{2}{d}
 \Whitetran{e}{6}{2}{e}
 \Whitetran{f}{8}{2}{f} 
 \diredge{a}{cfa} \diredge{a}{dfa}
 \diredge{b}{deb}\diredge{b}{ceb}
 \diredge{ceb}{c} \diredge{cfa}{c}
 \diredge{deb}{d} \diredge{dfa}{d} 
 \diredge{ceb}{e} \diredge{cfa}{f}
 \diredge{deb}{e} \diredge{dfa}{f}
\end{tikzpicture} 
&
\StandardNet[0.55] 
 \placW{cfb}{ 2}{4}{0}{p_6} 
 \placE{dea}{ 4}{4}{0}{p_7} 
 \placE{dfb}{ 6}{4}{0}{p_{10}} 
 
 \Whitetran{a}{4}{6}{a}
 \Whitetran{b}{6}{6}{b}
 \Whitetran{c}{2}{2}{c} 
 \Whitetran{d}{4}{2}{d}
 \Whitetran{e}{6}{2}{e}
 \Whitetran{f}{8}{2}{f}
 
 \diredge{a}{dea}\diredge{b}{cfb}\diredge{b}{dfb}
 \diredge{cfb}{c}\diredge{dfb}{d}\diredge{dea}{d} 
 \diredge{cfb}{f}\diredge{dfb}{f}\diredge{dea}{e} 
\end{tikzpicture} 
\\
($c$) $\PC''$ & ($d$) $\PC'''=\{p_6,p_7,p_{10}\}$
\end{tabular}
\end{center}
\caption {\label{fi-2}
($a,b,c$) Three distributed places together
with surrounding 
transitions for the marked nets in Figure~\ref{fi-1}($a,b$).
($d$) A set of places with surrounding 
transitions which is not a distributed place for the marked
net $\NN_0$ in Figure~\ref{fi-1}($a$).
}
\end{figure}

\begin{definition}[distributed place]
\label{def-distributed-place} 
A \emph{distributed place} of a marked net is a non-empty set of places
$\PC$ such that:
\begin{enumerate}
\item 
 $\PREPOST{\PC}=\In_\PC\cup \Out_\PC \cup\Read_\PC$.
 
 \item
 $\In_\PC\times \Out_\PC \subseteq\enables(\PC)$. 
 
\item 
If $\In_\PC\neq\es$ then, for every $\sigma\in\inseq_\PC$, there is 
$\omega\in\In_\PC^*$ such that $\sigma\circ\omega\in\complinseq_\PC$. 
\\
Equivalently, $\In_\PC\neq\es\implies\inseq_\PC=\pref(\complinseq_\PC)$.

\item
If $\Out_\PC\neq\es$ then, for every $\sigma\in\outseq_\PC$, there is 
$\omega\in\Out_\PC^*$ such that $\sigma\circ\omega\in\comploutseq_\PC$.
\\
Equivalently, $\Out_\PC\neq\es\implies\outseq_\PC=\pref(\comploutseq_\PC)$.
\end{enumerate} 
Also, $\PC$ is a \emph{pure distributed place} if $\Read_\PC=\es$. 
\eod
\end{definition}
Note that Definition~\ref{def-distributed-place} 
is local as it only depends on the 
transitions and arcs 
adjacent to the places making up the distributed place.

As already indicated, a distributed place should behave as a single place.
The idea of the proposed approach is the following: 
the emptiness of a single place corresponds to having no tokens in $\PC$
whereas filling of a single place corresponds to having tokens in 
all the places of $\PC$.
Then, an atomic act of 
filling a single place 
is implemented by a firing sequence in $\complinseq_\PC$ 
filling all the places of $\PC$.
Likewise, 
emptying a single place is implemented by a firing sequence 
in $\comploutseq_\PC$ emptying all the places of $\PC$. 

Directly from Definition~\ref{def-distributed-place} 
we obtain some basic properties of distributed places.

\begin{proposition}
\label{prop-ojdan}
Let $\PC$ be a distributed place of a marked net. Then: 
\begin{equation}
\label{eq-ecqe}
\begin{array}{lll@{~~}c@{~~}l}
\In_\PC=\es
&\implies
&\complinseq_\PC=\es
&\wedge
&\inseq_\PC=\{\lambda\}
\\
\In_\PC\neq\es
&\implies
&\complinseq_\PC\neq\es
&\wedge
&\complinseq_\PC\cap\Read_\PC^*=\es
\\
\Out_\PC = \es
&\implies 
&\comploutseq_\PC = \es
&\wedge
&\outseq_\PC = \Read_\PC^*
\\
\Out_\PC\neq\es
&\implies
&\comploutseq_\PC\neq\es
&\wedge
&\comploutseq_\PC\cap\Read_\PC^*=\es\;.
\end{array} 
\end{equation}
Moreover, 
$\inseq_\PC|_{\In_\PC}\subseteq\inseq_\PC$ and 
$\complinseq_\PC|_{\In_\PC}\subseteq\complinseq_\PC$
as well as 
$\outseq_\PC|_{\Out_\PC}\subseteq\outseq_\PC$ and
$\comploutseq_\PC|_{\Out_\PC}\subseteq\comploutseq_\PC$.
\end{proposition}

\begin{proposition}
\label{prop-oqieqo} 
Let $\PC$ be a distributed place of a marked net $\NN$.
Moreover, let $M\in\reach(\NN)$ and $M\xrightarrow[\NN]{\sigma}M'$
as well as $\sigma|_{\PREPOST{\PC}}=t_1\dots t_k$ ($k\geq 0$)
and $U=\{t_1,\dots,t_k\}$. Then:
\begin{equation} 
\label{eq-croncorn}
\begin{array}{l@{~~}c@{~~}lll}
\PC\cap M=\es
&\wedge
&\sigma|_{\PREPOST{\PC}}\in \inseq_\PC
&\implies
&\PC\cap M'=\PC\cap\cPOST{U} 
\\
\PC\cap M=\PC
&\wedge
&\sigma|_{\PREPOST{\PC}}\in \outseq_\PC
&\implies
&\PC\cap M'=\PC\setminus\cPRE{U}
\\
\PC\cap M=\es
&\wedge
&\sigma|_{\PREPOST{\PC}}\in \complinseq_\PC
&\implies
&\PC\cap M'=\PC 
\\
\PC\cap M=\PC
&\wedge
&\sigma|_{\PREPOST{\PC}}\in \comploutseq_\PC
&\implies
&\PC\cap M'=\es\;.
\end{array}
\end{equation} 
\end{proposition} 

\begin{example}
\label{ex-1}
Figure~\ref{fi-2}($a,b$) depicts two distributed places for the marked 
net $\NN_0$ of Figure~\ref{fi-1}($a$):
$\PC=\{p_1,p_2\}$
and
$\PC'=\{p_3,p_4,p_5,p_6,p_7,p_8,p_9,p_{10}\}$.
Moreover, we have:
$\inseq_{\PC}=\{\lambda\}$,
$\complinseq_{\PC}=\es$,
$\complinseq_{\PC'}=\{ab,ba\}$,
$\outseq_{\PC}=\pref(\{ab,ba\})$, and
$\comploutseq_{\PC'}=\{cd,dc,ef,fe\}$.

We then observe that 
$\PC''=\{p_4,p_5,p_8,p_9\}$ in Figure~\ref{fi-2}($c$) is a distributed place of 
$\NN_0$ with the same 
in-sequences and out-sequences as well as the same 
complete in-sequences and complete out-sequences
as $\PC'$. 
We also observe that the set of places 
$\PC'''=\{p_6,p_7,p_{10}\}$ 
(shown in Figure~\ref{fi-2}($d$)) is not a distributed place 
as $\FA{a,c}\notin\enables(\PC''')$.
\eod
\end{example}

In the case of pure distributed places, all permutations 
of in-sequences (out-sequences) are 
in-sequences (out-sequences).
Such an observation leads to a characterisation of 
distributed places. 

\begin{proposition}
\label{prop-kdkddk} 
A non-empty 
set of places $\PC$ of a marked net is a distributed place iff 
the following hold:
\begin{enumerate} 
\item
$\PREPOST{\PC}=\In_\PC\cup \Out_\PC\cup \Read_\PC$.
 
\item
$\In_\PC\times \Out_\PC \subseteq\enables(\PC)$. 

\item
If $\In_\PC\neq\es$ and $U$ is a maximal subset of $\In_\PC$
such that $\PC\cap\POST{t}\cap\POST{u}=\es$ 
(for all $t\neq u\in U$), then
$\PC\subseteq\POST{U}$. 
 
\item
If $\Out_\PC\neq\es$ and $U$ is a maximal subset of $\Out_\PC$
such that $\PC\cap\PRE{t}\cap\PRE{u}=\es$ (for all $t\neq u\in U$),
then $\PC\subseteq\PRE{U}$.
\end{enumerate} 
\end{proposition} 

\begin{proof}
First we note that parts (1,2) are the 
same as Definition~\ref{def-distributed-place}(1,2). 

($\Longrightarrow$)
To show part (3), let $\In_\PC\neq\es$ and 
$U=\{t_1,\dots,t_k\}$ be as in the formulation
of part (3).
Then, by Definition~\ref{def-kkd}(1) and $\PC\cap\POST{t}\cap\POST{u}=\es$ 
(for all $t\neq u\in U$), we have
$\sigma=t_1\dots t_k\in\inseq_\PC$.

Suppose that $\PC\not\subseteq\POST{U}$.
Then $\sigma\notin\complinseq_\PC$
and so, by Definition~\ref{def-distributed-place}(3), there 
is $\omega=v_1\dots v_n\in\In_\PC^+$ such that 
$\sigma\circ \omega \in\complinseq_\PC$. 
Hence,
$v_1\in\In_\PC$ is such that $\sigma\circ v_1 \in\inseq_\PC$.
We then observe that, by Definition~\ref{def-kkd}(1), 
we have 
$\PC\cap \POST{v_1}\cap\POST{U}=\es$.
Hence, $v_1\notin U$ and $U'=U\cup\{v_1\}\subseteq\In_\PC$ is such that $\PC\cap\POST{t}\cap\POST{u}=\es$ 
(for all $t\neq u\in U'$),
yielding a contradiction with the maximality of $U$. 

To show part (4), 
we proceed in a similar way as above using Definitions~\ref{def-kkd}(2)
and~\ref{def-distributed-place}(4)
rather than Definitions~\ref{def-kkd}(1) and~\ref{def-distributed-place}(3).

($\Longleftarrow$)
To show Definition~\ref{def-distributed-place}(3),
suppose that $\In_\PC\neq\es$ and $\sigma\in\inseq_\PC$.
Then, by Proposition~\ref{prop-ojdan}, $\sigma|_{\In_\PC}=t_1\dots t_k\in\inseq_\PC$.
Hence, $U=\{t_1,\dots,t_k\}\subseteq\In_\PC$ 
is such that, for all $1\leq i<j\leq k$, we have
$t_i\neq t_j$ and $\PC\cap\POST{t_i}\cap\POST{t_j}=\es$. 
If $U$ is maximal, then we have $\PC\subseteq \POST{U}$, and so 
$\sigma\in\complinseq_\PC$.
Otherwise, there is $v\in\In_\PC$ 
such that $v\notin U$ and $\PC\cap\POST{v}\cap\POST{U}=\es$.
Hence $\sigma|_{\In_\PC}\circ v\in\inseq_\PC$. We then can repeat the same argument 
(possibly several, but finitely many, times) until we obtain a complete in-sequence 
$\sigma|_{\In_\PC}\circ \omega\in\complinseq_\PC$. 
Then, by Definition~\ref{def-kkd}(1), $\sigma\circ \omega\in\complinseq_\PC$. 

To show Definition~\ref{def-distributed-place}(4), 
we proceed in a similar way using Definitions~\ref{def-kkd}(2) 
rather than Definitions~\ref{def-kkd}(1).
\end{proof}

The idea behind Proposition~\ref{prop-kdkddk}(3,4) is to provide 
effective tests for Definition~\ref{def-distributed-place}(3,4).
For example, Definition~\ref{def-distributed-place}(3) 
basically means that each in-sequence of $\PC$ which 
cannot be extended with a transition 
belonging to $\In_\PC$ is a complete in-sequence.
In general, a straightforward check that this holds would be 
impractical as there could be infinitely many in-sequences to consider
(note that $\sigma\circ v\in\inseq_\PC$ and $v\in\Read_\PC$ implies
 $\{\sigma\circ v^k\mid k\geq 1\}\subseteq\inseq_\PC$). 
However, as seen in Proposition~\ref{prop-kdkddk}(3),
only (finitely many) in-sequences 
made up of the transitions in $\In_\PC$
need to be considered, and 
a single set $U$ in Proposition~\ref{prop-kdkddk}(3)
represents $|U|!$ of them. 
Hence, Proposition~\ref{prop-kdkddk} could help in identifying 
distributed places contained in marked nets. 

We also have the following immediate result characterising 
singleton distributed places. 

\begin{proposition}
\label{prop-dvndv} 
For every place $p$ of a marked net, $\SSS_p=\{p\}$ is a distributed place
such that:
\begin{enumerate} 
\item 
$\In_{\SSS_p}=\PRE{p}\setminus\POST{p}$, $\Out_{\SSS_p}=\POST{p}\setminus\PRE{p}$, and 
$\Read_{\SSS_p}=\POST{p}\cap\PRE{p}$.
\item 
\makebox[1.7cm][l]
{$\In_{\SSS_p}\neq\es$}
$\implies 
\complinseq_{\SSS_p}=\In_{\SSS_p}\circ \Read_{\SSS_p}^* 
~\wedge~ 
\inseq_{\SSS_p}=\{\lambda\}\cup\In_{\SSS_p}\circ \Read_{\SSS_p}^* $.
\item 
\makebox[1.7cm][l]
{$\In_{\SSS_p}=\es$}
$\implies 
\complinseq_{\SSS_p}=\es 
~\wedge~ 
\inseq_{\SSS_p}=\{\lambda\} $.
\item 
\makebox[1.7cm][l]
{$\Out_{\SSS_p}\neq\es$}
$\implies 
\comploutseq_{\SSS_p}=\Read_{\SSS_p}^*\circ\Out_{\SSS_p}
~\wedge~ 
\outseq_{\SSS_p}= \Read_{\SSS_p}^*
\cup \Read_{\SSS_p}^* \circ \Out_{\SSS_p}$.
\item 
\makebox[1.7cm][l]
{$\Out_{\SSS_p}=\es$}
$\implies 
\comploutseq_{\SSS_p}=\es
~\wedge~ 
\outseq_{\SSS_p}=\Read_{\SSS_p}^*$.
\item 
$\SSS_p$ is pure iff $\POST{p}\cap\PRE{p}=\es$.
\end{enumerate} 
\end{proposition} 
 
Intuitively, a distributed place $\PC$ is 
an implementation of a single place which can be cyclically 
and monotonically filled and emptied.
This is evident in the next result
which asserts that the behaviour of each distributed place of 
safe marked net can be 
understood as execution of alternating in-sequences and out-sequences,
provided that the initial marking is set to `empty' or to `filled'. 
Such a behaviour is captured in the first two lines of 
Eq.\eqref{eq-eri} in the formulation of the next result,
where both $\In_\PC$ and $\Out_\PC$ are non-empty. 
 
\begin{theorem}
\label{prop-cust001}
Let $\PC$ be a distributed place of a safe marked net $\NN=\FA{P,T,\Flow,M_\init}$.
Then, for every $\sigma\in\fseq(\NN)$, we have:
\begin{equation}
\label{eq-eri} 
\sigma|_{\PREPOST{\PC}} \in 
\left\{
\begin{array}{lll@{~~}l@{~~}ll}
 \pref((\complinseq_\PC\circ\comploutseq_\PC)^*) ~~
 & \textit{if} 
 & \In_\PC\neq\es\neq\Out_\PC 
 &\wedge& 
 \PC\cap M_\init=\es 
 \\
 \pref((\comploutseq_\PC\circ\complinseq_\PC)^*)
 & \textit{if}
 & \In_\PC\neq\es\neq\Out_\PC 
 &\wedge& 
 \PC\cap M_\init=\PC 
 \\
 \pref(\complinseq_\PC ) 
 & \textit{if}
 & \In_\PC\neq\es=\Out_\PC 
 &\wedge& 
 \PC\cap M_\init=\es 
 \\
 \pref(\comploutseq_\PC )
 & \textit{if}
 & \In_\PC=\es\neq\Out_\PC 
 &\wedge& 
 \PC\cap M_\init=\PC
\\ 
 \{\lambda\}
 & \textit{if}
 & \In_\PC=\es 
 &\wedge& 
 \PC\cap M_\init=\es
\\ 
 \Read_\PC^*
 & \textit{if}
 & \Out_\PC=\es 
 &\wedge& 
 \PC\cap M_\init=\PC \;. 
\end{array}
\right.
\end{equation} 
\end{theorem}

\begin{proof} 
Let $M_\init\xrightarrow[\NN]{\sigma} M$.
The proof proceeds by induction on the length of $\sigma$.

In the base step, $\sigma=\lambda$ satisfies Eq.\eqref{eq-eri}. 
In the inductive step, we take $\sigma\circ t\in \fseq(\NN)$ 
with $\sigma$ satisfying Eq.\eqref{eq-eri}, and consider 
several cases assuming that $\PC\cap M_\init=\es$ 
(if $\PC\cap M_\init=\PC$
we proceed in a similar way.)
 
\emph{Case 1:} 
 $\In_\PC\neq\es\neq\Out_\PC$ and $\sigma|_{\PREPOST{\PC}} \in
 (\complinseq_\PC\circ\comploutseq_\PC)^*\circ\inseq_\PC$.
 
 Then $\sigma$ can be represented 
 (not necessarily in a unique way due to the presence 
 of transitions in $\Read_\PC$)
 as $\sigma=\kappa\circ\omega$, where:
 \[
 \kappa|_{\PREPOST{\PC}}\in (\complinseq_\PC\circ\comploutseq_\PC)^*
 ~~~\mbox{and}~~~
 \omega|_{\PREPOST{\PC}}\in \inseq_\PC \;.
 \]
 Let $M_\init\xrightarrow[\NN]{\kappa} M'\xrightarrow[\NN]{\omega} M$
 as well as
 $\omega|_{\PREPOST{\PC}}=t_1\dots t_k$ ($k\geq 0$)
 and $U=\{t_1,\dots,t_k\}$.
 Then, by Proposition~\ref{prop-oqieqo}, we have $\PC\cap M'=\es$ and 
 $\PC\cap M=\PC\cap\cPOST{U}$ ($*$).
 
 If $t\notin \PREPOST{\PC}$, then 
 $(\omega\circ t)|_{\PREPOST{\PC}}=\omega|_{\PREPOST{\PC}}\in \inseq_\PC$.
 Otherwise, 
 $(\omega\circ t)|_{\PREPOST{\PC}}=\omega|_{\PREPOST{\PC}} \circ t$ 
 and, by Definition~\ref{def-distributed-place}(1),
exactly one of the following three cases holds.
 
\emph{Case 1.1:} $t\in \Read_\PC$. 

 Then 
 $\omega|_{\PREPOST{\PC}}\circ t\in \inseq_\PC$ follows 
 from 
 $\PREE_\PC(t)\subseteq\PC\cap M=_{\mbox{\small(by ($*$))}}\PC\cap \cPOST{U}$
 and Definition~\ref{def-kkd}(1). 
 
\emph{Case 1.2:} $t\in \In_\PC$. 
 Then, by $\NN$ being safe, $M\cap \cPOST{t}=\es$.
 Hence, $\omega|_{\PREPOST{\PC}}\circ t\in \inseq_\PC$ follows 
 from Definition~\ref{def-kkd}(1) and
 $\PC\cap\cPOST{t} \cap \cPOST{U}
 =_{\mbox{\small(by ($*$))}}\PC\cap M\cap\cPOST{t}\subseteq M\cap \cPOST{t}=\es$. 

\emph{Case 1.3:} $t\in \Out_\PC$. 

 If $\PC\cap M=\PC$ then, by ($*$), 
 $\omega|_{\PREPOST{\PC}}\in \complinseq_\PC$, and 
 so we have
 $\omega|_{\PREPOST{\PC}}\circ t\in \complinseq_\PC\circ\outseq_\PC$. 
 
 If $\PC\cap M\neq\PC$ then, by ($*$), 
 $\omega|_{\PREPOST{\PC}}\notin \complinseq_\PC$.
 Thus, by Definition~\ref{def-distributed-place}(3),
 there is $u\in\In_\PC$ such that 
 $\omega|_{\PREPOST{\PC}}\circ u\in \inseq_\PC $.
 Hence, by Definition~\ref{def-kkd}(1),
 $\PC\cap\cPOST{u}\cap\cPOST{U}=\es$ and so, by ($*$),
 $\PC\cap\cPOST{u}\cap M=\es$.
 On the other hand, $ \PC\cap\cPRE{t}\subseteq \PC\cap M$.
 Hence $\PC\cap\cPOST{u}\cap\cPRE{t}=\es$, and so 
 $\FA{u,t}\notin\enables(\PC)$,
 yielding a contradiction
 with $\In_\PC\times \Out_\PC \subseteq\enables(\PC)$ required by
 Definition~\ref{def-distributed-place}(2). 

\emph{Case 2:}
 $\In_\PC\neq\es\neq\Out_\PC$ and
 $\sigma|_{\PREPOST{\PC}} \in
 (\complinseq_\PC\circ\comploutseq_\PC)^*\circ\complinseq_\PC\circ \outseq_\PC $.
 
 Then $\sigma$ can be represented (not necessarily in a unique way due to the presence 
 of transitions in $\Read_\PC$) 
 as $\sigma=\kappa\circ\omega$,
 where:
 \[
 \kappa|_{\PREPOST{\PC}}\in
 (\complinseq_\PC\circ\comploutseq_\PC)^*\circ\complinseq_\PC
 ~~~\mbox{and}~~~
 \omega|_{\PREPOST{\PC}}\in \outseq_\PC \;.
 \] 
 Let $M_\init\xrightarrow[\NN]{\kappa} M'\xrightarrow[\NN]{\omega} M$
 as well as
 $\omega|_{\PREPOST{\PC}}=t_1\dots t_k$ ($k\geq 0$)
 and $U=\{t_1,\dots,t_k\}$.
 Then, by Proposition~\ref{prop-oqieqo}, we have $\PC\cap M'=\PC$ 
 and $\PC\cap M=\PC\setminus\cPRE{U}$ ($**$).

 If $t\notin \PREPOST{\PC}$, then 
 $(\omega\circ t)|_{\PREPOST{\PC}}=\omega|_{\PREPOST{\PC}}\in \outseq_\PC$.
 Otherwise, 
 $(\omega\circ t)|_{\PREPOST{\PC}}=\omega|_{\PREPOST{\PC}} \circ t$ 
 and, by Definition~\ref{def-distributed-place}(1),
 exactly one of the following three cases holds. 

\emph{Case 2.1:} $t\in \Read_\PC$. 

Then 
 $\omega|_{\PREPOST{\PC}}\circ t\in \outseq_\PC$ follows 
 from 
 $\PREE_\PC(t)\subseteq\PC\cap M=_{\mbox{\small(by ($**$))}}\PC\setminus\cPRE{U}$
 and Definition~\ref{def-kkd}(2). 

\emph{Case 2.2:} $t\in \In_\PC$. 

 If $\PC\cap M=\es$ then, by ($**$), we have 
 $\PC\cap\cPRE{U}=\PC$.
 Hence, $\omega|_{\PREPOST{\PC}}\in \comploutseq_\PC$, and 
 so we have
 $\omega|_{\PREPOST{\PC}}\circ t\in \comploutseq_\PC\circ \inseq_\PC$. 
 
 If $\PC\cap M\neq\es$ then, by ($**$), 
 $\omega|_{\PREPOST{\PC}}\notin \comploutseq_\PC$, and 
 so, by Definition~\ref{def-distributed-place}(4),
 there is $u\in\Out_\PC$ such that 
 $\omega|_{\PREPOST{\PC}}\circ u\in \outseq_\PC $.
 Hence, by Definition~\ref{def-kkd}(2), 
 $\PC\cap\cPRE{u}=\PREE_\PC(u)
 \subseteq \PC\setminus \cPRE{U}=_{\mbox{\small(by ($**$))}} \PC\cap M$. 
 On the other hand, since $\NN$ is safe,
 $\PC\cap M\cap\cPOST{t}=\es$. 
 Hence $\PC\cap \cPRE{u}\cap\cPOST{t} =\es$, and so 
 $\FA{t,u}\notin\enables(\PC)$,
 yielding a contradiction
 with $\In_\PC\times \Out_\PC \subseteq\enables(\PC)$ required by
 Definition~\ref{def-distributed-place}(2). 

\emph{Case 2.3:} $t\in \Out_\PC$. 

 Then $\PREE_\PC(t)\subseteq \PC\cap M
 =_{\mbox{\small(by ($**$))}} \PC\setminus\cPRE{U}$.
 Hence, by Definition~\ref{def-kkd}(2),
 $\omega|_{\PREPOST{\PC}}\circ t\in \outseq_\PC$. 

\emph{Case 3:} 
$\In_\PC\neq\es=\Out_\PC$ and $\sigma|_{\PREPOST{\PC}} \in\inseq_\PC $.

Then 
we proceed similarly as in Case~1, except Case~1.3 which cannot hold.

\emph{Case 4:} 
$\In_\PC=\es$. 

Then, by $\PC\cap M_\init=\es$, we have $\PC\cap M =\es$, for every $M\in\reach(\NN)$.
Hence $\sigma|_{\PREPOST{\PC}}=\lambda$. 
\end{proof} 

The following two crucial properties of a distributed place $\PC$ come into play
when $\PC$ is a part of a safe marked net $\NN$: 
(i) if $\PC$ is currently empty, then it is not possible to `fire' a 
transition $u\in\Out_\PC$ after an incomplete (projected) 
in-sequence of $\PC$; 
and (ii) if $\PC$ is currently filled, then it is not possible to `fire' a 
transition $u\in\In_\PC$ after an incomplete (projected) out-sequence of $\PC$. 
To explain (ii) more precisely, let 
$M_\init\xrightarrow[\NN]{} M'\xrightarrow[\NN]{\omega}M\xrightarrow[\NN]{u}M''$,
where $\PC\subseteq M'$, $u\in\In_\PC$, and
$\omega|_{\PREPOST{\PC}}=t_1\dots t_n$
is an incomplete (projected) out-sequence of $\PC$. 
Clearly, $\PC\cap M = \PC\setminus \cPRE{\{t_1,\dots,t_n\}} $.
Then, by Definition~\ref{def-distributed-place}(4), there is $t\in\Out_\PC$ 
such that $\omega|_{\PREPOST{\PC}}\circ t$ is an out-sequence 
of $\PC$ and so, by Definition~\ref{def-kkd}(2),
$\PC\cap \PRE{t}\subseteq \PC\setminus \cPRE{\{t_1,\dots,t_n\}}=\PC\cap M$.
On the other hand, by $\FA{u,t}\in \In_\PC\times\Out_\PC$ required by 
Definition~\ref{def-distributed-place}(2), we have 
$\FA{u,t}\in\enables(\PC)$, and so
there is $p\in \PC\cap\cPOST{u}\cap\PRE{t}$.
Hence, by $\PC\cap \PRE{t}\subseteq \PC\cap M$, we have 
$p\in M$, and so $u$ cannot be fireable 
at $M$ as $\NN$ is safe.
 
The next result captures a key property of
distributed places from the 
point of view of their application to semantics-preserving net reduction.
In a nutshell, it shows that in safe marked nets covered by 
distributed places
one can apply reductions at local
level (\ie within each distributed place). 

\begin{theorem}
\label{th-main}
Let $\NN=\FA{P,T,\Flow,M_\init}$ be a safe marked net
and, for every place $p\in P$, let $\PC_p$ be a distributed place 
in $\NN$ comprising $p$ such that either
$\PC_p\cap M_\init=\es$ or $\PC_p\cap M_\init=\PC_p$.

Moreover, let $P'\subseteq P$ be a set of places of $\NN$ such that
the following hold, 
for every $p\in P$ 
(below $\PC'_p=\PC_p\cap P'$):
\begin{subequations} 
\begin{equation} 
\label{eq-jjdjd} 
\PRE{(\PC'_p)}=\PRE{(\PC_p)}
~~\mbox{and}~~
\POST{(\PC'_p)}=\POST{(\PC_p)}
\end{equation}
\begin{equation}
\label{eq-jjdjd2} 
\enables(\PC'_p)=\enables(\PC_p)
\end{equation}
\begin{equation}
\label{eq-jjdjd3} 
\disables(\PC'_p)=\disables(\PC_p) \;.
\end{equation}
\end{subequations} 
Then the following hold for the tuple
$\NN'=
\FA{P',T,\Flow',M'_\init}=
\FA{P',T,\Flow|_{T\times P'\cup P'\times T},M_\init\cap P'}$. 
\begin{enumerate}
 \item 
 $\NN'$ is a safe marked net such that $\fseq(\NN')=\fseq(\NN)$. 
 \item
 $\RG_\NN\cong\RG_{\NN'}$ provided that, 
 for all $M\in\reach(\NN)$ and $p\in P$, there is $M'\in\reach(\NN)$ such that: 
 \begin{equation}
 \label{eq-rvri}
 M\xrightarrow[\NN]{} M'~\wedge~ (\PC_p\cap M'=\es~\vee~\PC_p\cap M'=\PC_p)\;.
 \end{equation} 
\end{enumerate} 
\end{theorem}
 
\begin{proof}
(1)
We start by making three observations concerning the 
connectivity between places and transitions
in $\NN$ and $\NN'$ (the first two observations follow immediately from
the definition of $\NN'$):
\begin{itemize}
\item[($*$)] 
For all $p\in P'$ and $t\in T$,
$\FA{t,p}\in\Flow'\iff\FA{t,p}\in\Flow$ 
and
$\FA{p,t}\in\Flow'\iff\FA{p,t}\in\Flow$.
\\
Intuitively, ($*$) means that the connectivity of a place
$p\in P'$ is the same in $\NN$ and $\NN'$.

\item[($**$)] 
For all $p\in P$ and $t\in T$,
$\FA{t,p}\in\Flow'\iff(\FA{t,p}\in\Flow\wedge p\in P')$ 
and
$\FA{p,t}\in\Flow'\iff(\FA{p,t}\in\Flow\wedge p\in P')$.
\\
Intuitively, ($**$) means that the connectivity of 
a transition 
$t\in T$ in $\NN'$ is the connectivity of 
$t$ in $\NN$ restricted to the places in $P'$.

\item[($***$)] 
For all $p\in P\setminus P'$ and $t\in T$,
$t\in\PRE{p}\implies t\in\PRE{(\PC'_p)}$
and
$t\in\POST{p}\implies t\in\POST{(\PC'_p)}$.
\\
Indeed, suppose that $t\in\POST{p}$ (if $t\in\PRE{p}$ we proceed similarly). 
Then, by $p\in\PC_p$ and 
the second part of Eq.\eqref{eq-jjdjd}, 
we have $t\in\POST{(\PC_p)}=\POST{(\PC'_p)}$. 
\\
Intuitively, ($***$) means that the 
connectivity of a place $p\in P\setminus P'$
is `retained' in $\NN'$ through the connectivity of the places in $\PC'_p$.
\end{itemize}
We then immediately conclude that $\FA{P',T,\Flow'}$ is a net as it 
contains no isolated places (since $\FA{P,T,\Flow}$ contains no isolated 
places and ($*$) holds), and each transition has a pre-place 
(since each transition has a pre-place in $\FA{P,T,\Flow}$ and the second 
parts of ($**$) and ($***$) hold).
Hence, $\NN'$ is a marked net.

Next we prove that the marked nets $\NN$ and $\NN'$ generate the 
same firing sequences, \ie $\fseq(\NN')=\fseq(\NN)$.
First, we observe that, by ($*$), we have the following:
\begin{equation}
\label{eq-ierv}
M_\init\xrightarrow[\NN]{\sigma} M 
~~\implies~~
M'_\init \xrightarrow[\NN']{\sigma} M\cap P' \;,
\end{equation} 
which immediately yields $\fseq(\NN)\subseteq\fseq(\NN')$.
To show that the reverse inclusion (\ie $\fseq(\NN')\subseteq\fseq(\NN )$)
also holds, we proceed by induction 
on the length of a firing sequence $\sigma\in\fseq(\NN')$.

In the base step, $\sigma=\lambda\in\fseq(\NN)$.
In the inductive step, we take a firing sequence
$\sigma\circ t\in\fseq(\NN')$ 
such that $\sigma\in\fseq(\NN)$. Suppose 
to the contrary that $\sigma\circ t\notin\fseq(\NN)$. 

Let $M_\init\xrightarrow[\NN]{\sigma} M$. 
Then, by Eq.\eqref{eq-ierv}, we have
$M'_\init\xrightarrow[\NN']{\sigma} M\cap P' $.
Also, since $t\notin\fireable_\NN(M)$, 
there is a place $p\in \PRE{t}\setminus M$.
Clearly, $p\notin P' $ as $t\in\fireable_{\NN'}(M\cap P')$.

Let
$\PC=\PC_p$ and $\PC'=\PC'_p$. 
Then, by $t\in\POST{p}$ and ($***$), we have $t\in\POST{\PC'}$. 

Suppose now that $\PC\cap M_\init=\es$ (if $\PC\cap M_\init=\PC$
the proof is similar).
Taking into account Theorem~\ref{prop-cust001} for $\PC$, we will 
consider four cases. As in each of these cases we will
obtain a contradiction, our 
overall conclusion will be that $\sigma\circ t\in\fseq(\NN)$. 

\emph{Case 1.1:} 
 $\In_\PC\neq\es\neq\Out_\PC$ and $\sigma|_{\PREPOST{\PC}} \in
 (\complinseq_\PC\circ\comploutseq_\PC)^*\circ\inseq_\PC$.
 
 Then $\sigma$ can be represented 
 (not necessarily in a unique way due to the presence 
 of transitions in $\Read_\PC$) 
 as $\sigma=\kappa\circ\omega$,
 where:
\[ 
\kappa|_{\PREPOST{\PC}} \in (\complinseq_\PC\circ\comploutseq_\PC)^*
~~~\mbox{and}~~~
\omega|_{\PREPOST{\PC}} \in \inseq_\PC \;. 
\] 
Let $M_\init\xrightarrow[\NN]{\kappa} M'\xrightarrow[\NN]{\omega} M$.
Then 
$\PC\cap M'=\es$ (by Proposition~\ref{prop-oqieqo}) 
and $\PC\not\subseteq M $ (by $p\in\PC$ and $p\notin M$). 
Hence, by Proposition~\ref{prop-oqieqo},
$\omega|_{\PREPOST{\PC}}\notin\complinseq_\PC$. Thus, by 
Definition~\ref{def-distributed-place}(3), there is $\theta\in\In_\PC^*$
such that $\omega|_{\PREPOST{\PC}}\circ \theta\in \complinseq_\PC$.
Hence, by Definition~\ref{def-kkd}(1),
there is 
$u\in\In_\PC$ occurring in $\theta$ such that 
$p\in\cPOST{u}$
and
$\omega|_{\PREPOST{\PC}}\circ u\in \inseq_\PC$.
Clearly, $\PC\cap\cPOST{u}\subseteq \PC\setminus M$. Moreover,
$\PC\cap \PRE{t} \cap P'\subseteq \PC\cap M$. Hence
$\FA{u,t}\notin \enables(\PC')$, and so by Eq.\eqref{eq-jjdjd2},
we obtain $\FA{u,t}\notin \enables(\PC)$, yielding 
a contradiction with $p\in\PC\cap\cPOST{u}\cap\PRE{t}$.

\emph{Case 1.2:} 
 $\In_\PC\neq\es\neq\Out_\PC$ and
 $\sigma|_{\PREPOST{\PC}} \in
 (\complinseq_\PC\circ\comploutseq_\PC)^*\circ\complinseq_\PC\circ \outseq_\PC $.

Then $\sigma$ can be represented (not necessarily in a unique way due to the presence 
 of transitions in $\Read_\PC$) as $\sigma=\kappa\circ\omega$,
 where:
\[ 
\kappa|_{\PREPOST{\PC}} \in (\complinseq_\PC\circ\comploutseq_\PC)^*\circ
\complinseq_\PC
~~~\mbox{and}~~~
\omega|_{\PREPOST{\PC}} \in \outseq_\PC \;. 
\] 
Let $M_\init\xrightarrow[\NN]{\kappa} M'\xrightarrow[\NN]{\omega} M$.
Then, by Proposition~\ref{prop-oqieqo}, $\PC\cap M' =\PC $. 
Hence, by $p\in\PC$ and $p\notin M$, there is $u\in\Out_\PC$ occurring
in $\omega$ such that $p\in\cPRE{u}$.
Thus $\FA{u,t}\in\disables(\PC)$, and so by Eq.\eqref{eq-jjdjd3}, 
we obtain
$\FA{u,t}\in\disables(\PC')$.
But the latter means that $t\notin\fireable_{\NN'}(M\cap P')$,
yielding a contradiction. 

\emph{Case 1.3:} 
$\In_\PC\neq\es=\Out_\PC$ and
 $\sigma|_{\PREPOST{\PC}} \in \inseq_\PC $. 

Then, by $p\notin M$ and Proposition~\ref{prop-oqieqo},
$\sigma|_{\PREPOST{\PC}}\notin\complinseq_\PC$. Thus, by 
Definition~\ref{def-distributed-place}(3), there is 
$\theta\in\In_\PC^*$ such that 
$\sigma|_{\PREPOST{\PC}}\circ \theta\in \complinseq_\PC$.
Hence, by Definition~\ref{def-kkd}(1),
there is 
$u\in\In_\PC$ occurring in $\theta$ such that 
$p\in\cPOST{u}$
and $\sigma|_{\PREPOST{\PC}}\circ u\in \inseq_\PC$. 
Clearly, $\PC\cap\cPOST{u}\subseteq \PC\setminus M$. Moreover,
$\PC\cap P'\cap\PRE{t}\subseteq \PC\cap M$. Hence
$\FA{u,t}\notin \enables(\PC')$, and so by Eq.\eqref{eq-jjdjd2},
we obtain $\FA{u,t}\notin \enables(\PC)$, yielding 
a contradiction with $p\in\PC\cap(\cPOST{u}\cap\PRE{t})$. 

\emph{Case 1.4:}
$\In_\PC=\es$. 

Then, by $t\in\fireable_{\NN'}(M\cap P')$ and $t\in\POST{\PC'}$, 
we obtain that $\PC\cap \PRE{t}\cap M\cap P'\neq\es$.
Hence $\PC\cap M\neq\es$, yielding a contradiction with 
$\PC\cap M_\init=\es$ and $\In_\PC=\es$. 

Therefore, we have demonstrated that $\fseq(\NN')\subseteq\fseq(\NN)$,
and so $\fseq(\NN')=\fseq(\NN)$.

We then observe that $\NN'$ is safe.
Indeed, suppose that $M'_\init \xrightarrow[\NN']{\sigma}M$
and $t\in\fireable_{\NN'}(M)$. 
Then, by $\sigma\circ t\in\fseq(\NN')=\fseq(\NN)$
and Eq.\eqref{eq-ierv} and $\RG_\NN$ being deterministic, there is 
a marking $\widetilde M$ such that
$M_\init \xrightarrow[\NN]{\sigma}\widetilde M$ and $t\in\fireable_{\NN}(\widetilde M)$
and $M=\widetilde M\cap P'$. Also, since $\NN$ is safe, 
$\cPOST{t}\cap \widetilde M =\es$. Hence, by ($**$) and 
$M\subseteq \widetilde M$, we obtain 
$\{p\in P'\mid \FA{t,p}\in\Flow'\wedge \FA{p,t}\notin\Flow'\}\cap M =\es$.
As a result, $\NN'$ is a safe marked net.

(2)
In the second part of the proof, we show that two firing sequences lead to the same 
marking in $\NN$ iff they lead to the same 
marking in $\NN'$.
This will be sufficient to demonstrate that the reachability 
graphs of $\NN$ and $\NN'$ are isomorphic thanks to 
Proposition~\ref{prop-refrf} and the already 
proved equality $\fseq(\NN)=\fseq(\NN')$. 
In other words, we will show 
that Eq.\eqref{eq-ufufgg} holds with $\NN_0=\NN$ and $\NN_1=\NN'$.

We start by observing that,
by Eq.\eqref{eq-ierv} and $\RG_{\NN'}$ being deterministic, 
if two firing sequences lead to the same 
marking in $\NN$ then they lead to the same 
marking in $\NN'$.

To show that the reverse implication also holds, 
and after taking into account Eq.\eqref{eq-ierv},
suppose to the contrary that $\sigma$ and $\sigma'$
are two firing sequences in $\fseq(\NN)=\fseq(\NN')$ such that: 
\begin{equation}
\label{eq-jdjdj}
M_\init\xrightarrow[\NN]{\sigma} M~~~~~~ 
M_\init\xrightarrow[\NN]{\sigma'} M'~~~~~~ 
M'_\init\xrightarrow[\NN']{\sigma} M\cap P'~~~~~ ~
M'_\init\xrightarrow[\NN']{\sigma'} M'\cap P'~~~~~~ 
M\neq M'~~~~~~ 
M\cap P'=M'\cap P'\;.
\end{equation} 
We first observe that the continuations of 
$\sigma$ and $\sigma'$ in $\NN$ are the same, \ie we have:
\begin{equation}
\label{eq-irirj}
\{\omega\mid \sigma\circ\omega\in\fseq(\NN)\}
=
\{\omega\mid \sigma'\circ\omega\in\fseq(\NN)\}~~~(= \Omega) 
\end{equation}
since the following holds:
\[
\begin{array}{lll}
\{\omega\mid \sigma\circ\omega\in\fseq(\NN)\}
&=_{\mbox{\small(by $\fseq(\NN)=\fseq(\NN')$)}}&
\{\omega\mid \sigma\circ\omega\in\fseq(\NN')\}
\\
&=_{\mbox{\small(by $M\cap P'=M'\cap P'$)}}&
\{\omega\mid \sigma'\circ\omega\in\fseq(\NN')\}
\\
&=_{\mbox{\small(by $\fseq(\NN)=\fseq(\NN')$)}}&
\{\omega\mid \sigma'\circ\omega\in\fseq(\NN)\}\;.
\end{array}
\]
We then observe that,
by Eq.\eqref{eq-jdjdj} and without loss of generality, 
there is a place $p\in M\setminus M'$ which does not belong to $P'$.
Moreover, it turns out that such a place belongs to each marking reachable in $\NN$
from $M$, and does not belong to any marking reachable in $\NN$
from $M'$, \ie
\begin{equation}
\label{eq-eieimf}
M\xrightarrow[\NN]{}\dot M\implies p\in\dot M
~~\mbox{and}~~
M'\xrightarrow[\NN]{}\ddot M\implies p\notin\ddot M\;.
\end{equation}
And such a fact follows immediately from $\Omega |_{\PREPOST{p}}=\{\lambda\}$ 
(see Eq.\eqref{eq-irirj}), which is a 
property that can be shown in the following way.
Suppose to the contrary that there is $\omega=\kappa\circ v\in\Omega$
satisfying $v\in\PREPOST{p}$
and $\kappa|_{\PREPOST{p}}=\lambda$. Let 
\[
M_\init\xrightarrow[\NN]{\sigma} M\xrightarrow[\NN]{\kappa}\overline M 
\xrightarrow[\NN]{v}\overline{\overline M} 
~~\mbox{and}~~ 
M_\init\xrightarrow[\NN]{\sigma'} M'\xrightarrow[\NN]{\kappa}\overline M' 
\xrightarrow[\NN]{v}\overline{\overline M}'\;.
\]
Then, by $p\in M\setminus M'$ and $\kappa|_{\PREPOST{p}}=\lambda$, we
get $p\in \overline M\setminus \overline M'$. Hence, 
if $v\in\POST{p}$ then this contradicts 
$v\in \fireable_\NN(\overline M')$ and $p\notin \overline M'$, 
and if $v\in\PRE{p}\setminus\POST{p}$ 
then this contradicts $v\in \fireable_\NN(\overline M)$, 
$p\in \overline M$ and $\NN$ being safe. 
We have therefore demonstrated that Eq.\eqref{eq-eieimf} holds and are 
ready to move to the last stage of the proof. 

Since $\PC'_p \neq \es$ (which follows from ($***$) and 
the fact that $\NN$ does not contain isolated places), we can find 
$r\in \PC'_p$. Note that 
from the second part of Eq.\eqref{eq-jdjdj} it follows that $r\in M\iff r\in M'$.
Hence it suffices to consider only 
two cases (each leading to a contradiction).

\emph{Case 3.1:} $r\in M\cap M'$.

Then, by the assumption Eq.\eqref{eq-rvri} applied to $\PC_p$
and marking $M'$, there is 
$\theta$
such that $M'\xrightarrow[\NN]{\theta}M''$ and
$\PC_p\cap M''=\es~\vee~\PC_p\cap M''=\PC_p$.
As Eq.\eqref{eq-eieimf} implies $p\notin M''$,
we have $\PC_p\cap M''=\es$, and so $r\notin M''$.
Thus, by Eq.\eqref{eq-irirj}, we have 
$M\xrightarrow[\NN]{\theta} \widetilde M$. 
Moreover, by ($*$), $r\notin \widetilde M$.
Then, by the assumption Eq.\eqref{eq-rvri} applied to $\PC_p$
and marking $\widetilde M$, there is 
$\psi$ such that $\widetilde M\xrightarrow[\NN]{\psi}\widehat M$ and
$\PC_p\cap \widehat M=\es~\vee~\PC_p\cap \widehat M=\PC_p$.
As Eq.\eqref{eq-eieimf} implies $p\in \widehat M$,
we have $\PC_p\cap \widehat M=\PC_p$, and so, by ($*$), $r\in \widehat M$.
We therefore have 
\[
M_\init\xrightarrow[\NN]{\sigma} M
\xrightarrow[\NN]{\theta} \widetilde M\xrightarrow[\NN]{\psi}\widehat M~~~~~
\mbox{as well as}~~~~~
r\in M~~~~~
r\notin \widetilde M~~~~~
r\in \widehat M\;.
\]
Hence, by Theorem~\ref{prop-cust001} for $\PC_p$,
there are $\theta'$ and $\psi'$ such that $\theta\circ\psi=\theta'\circ\psi'$
as well as 
$M_\init\xrightarrow[\NN]{\sigma} M
\xrightarrow[\NN]{\theta'} \widetilde M'\xrightarrow[\NN]{\psi'}\widehat M$
and $\PC_p\cap \widetilde M'=\es$.
This, however, means that $p\notin\widetilde M'$ 
yielding a contradiction with the first part of Eq.\eqref{eq-eieimf}.

\emph{Case 3.2:} $r\notin M\cup M'$. 

Then, by the assumption Eq.\eqref{eq-rvri} applied to $\PC_p$
and marking $M$, there is 
$\theta$
such that $M\xrightarrow[\NN]{\theta}M''$ and
$\PC_p\cap M''=\es~\vee~\PC_p\cap M''=\PC_p$.
As Eq.\eqref{eq-eieimf} implies $p\in M''$,
we have $\PC_p\cap M''=\PC_p$, and so $r\in M''$.
Thus, by Eq.\eqref{eq-irirj}, we have 
$M'\xrightarrow[\NN]{\theta} \widetilde M$. 
Moreover, by ($*$), $r\in \widetilde M$.
Then, by the assumption Eq.\eqref{eq-rvri} applied to $\PC_p$
and marking $\widetilde M$, there is 
$\psi$ such that $\widetilde M\xrightarrow[\NN]{\psi}\widehat M$ and
$\PC_p\cap \widehat M=\es~\vee~\PC_p\cap \widehat M=\PC_p$.
As Eq.\eqref{eq-eieimf} implies $p\notin \widehat M$,
we have $\PC_p\cap \widehat M=\es$, and so, by ($*$), $r\notin \widehat M$.
We therefore have 
\[
M_\init\xrightarrow[\NN]{\sigma} M'
\xrightarrow[\NN]{\theta} \widetilde M\xrightarrow[\NN]{\psi}\widehat M~~~~~
\mbox{as well as}~~~~~
r\notin M~~~~~
r\in \widetilde M~~~~~
r\notin \widehat M\;.
\] 
Hence, by Theorem~\ref{prop-cust001} for $\PC_p$,
there are $\theta'$ and $\psi'$ such that $\theta\circ\psi=\theta'\circ\psi'$
as well as 
$M_\init\xrightarrow[\NN]{\sigma} M'
\xrightarrow[\NN]{\theta'} \widetilde M'\xrightarrow[\NN]{\psi'}\widehat M$
and $\PC_p\cap \widetilde M'=\PC_p$.
This, however, means that $p\in\widetilde M'$ 
yielding a contradiction with
the second part of Eq.\eqref{eq-eieimf}. 
\end{proof}

The last result provides a static condition 
for reducing distributed 
places in safe marked nets. Also, as shown below,
none of the key assumptions in Theorem~\ref{th-main}
can be dropped.
\begin{itemize}
\item 
Assuming that $\NN$ is safe is essential.
Indeed, otherwise we could take any non-safe $\NN$ together with the 
trivial singleton distributed places and $P'=P$, and then
wrongly conclude that $\NN'=\NN$ is safe.

\item 
Assuming Eq.\eqref{eq-jjdjd2} is essential.
Indeed, consider $\NN_1$ in Figure~\ref{fi-3}($a$) with
three distributed places: $\PC_1=\{q_1,q_2\}$, 
$\PC_2=\{q_3,q_4,q_5,q_6\}$, and $\PC_3=\{q_7,q_8\}$.
We can take $P'=\{q_1,q_2,q_3,q_6,q_7,q_8\}$, and the resulting 
reduced marked net $\NN'_1$ in Figure~\ref{fi-3}($b$) 
generates more firing sequences
than $\NN_1$
(\eg $acbd$) even though assumptions other than Eq.\eqref{eq-jjdjd2} are satisfied
(note that in this case
$\PC_2\cap P'=\{q_3,q_6\}$ and 
$\FA{b,d}\in \enables(\PC_2)\setminus \enables(\{q_3,q_6\})$).

\item
Assuming Eq.\eqref{eq-jjdjd3} is essential.
Indeed, consider $\NN_2$ in Figure~\ref{fi-3}($c$) with
two distributed places: $\PC_1=\{r_1,r_2,r_3,r_4\}$ 
 and $\PC_2=\{r_5,r_6,r_7,r_8\}$.
We can take $P'=\{r_1,r_4,r_5,r_6,r_7,r_8\}$, 
and the resulting 
marked net $\NN'_2$ in Figure~\ref{fi-3}($d$) generates more firing sequences
than $\NN_2$
(\eg $ad$) even though assumptions other than Eq.\eqref{eq-jjdjd3} are satisfied
(note that in this case
$\PC_1\cap P'=\{r_1,r_4\}$ and 
$\FA{a,d}\in \disables(\PC_1)\setminus\disables(\{r_1,r_4\})$).
\end{itemize} 

To summarise, the net reduction approach emerging from
the results presented in this section amounts to finding distributed places covering 
a marked net 
(\ie each place belongs to at least one distributed place of the cover)
and a subset $P'$ of net places meeting conditions of Theorem~\ref{th-main}.
In view of Proposition~\ref{prop-dvndv}, one can always choose 
the covering by 
singleton distributed places. But then, due to 
Eq.\eqref{eq-jjdjd}, there is no reduction at all. 
In general, the choice of suitable covering by
distributed places and $P'$ is not unique, 
leaving a scope for developing optimal covers. 
It is also worth noting that to apply Theorem~\ref{th-main} 
in cases when only a partial
cover by distributed places is provided, one can always 
cover the uncovered places by the trivial singleton distributed places. 

\begin{figure}[t!]
\begin{center} 
\begin{tabular}{c@{~~~~~~~~~~~~~}c}
\StandardNet[0.55]

 \placN{Aa}{2}{7.5}{1}{q_1}
 \placN{Ab}{4}{7.5}{1}{q_2}
 
 \placW{ac}{ 0}{4.5}{0}{q_3} 
 \placW{bc}{ 2}{4.5}{0}{q_4} 
 \placE{ad}{ 4}{4.5}{0}{q_5} 
 \placE{bd}{ 6}{4.5}{0}{q_6} 

 \placS{Bc}{ 2}{1.5}{0}{q_7}
 \placS{Bd}{ 4}{1.5}{0}{q_8} 
 
 \Whitetran{a}{2}{6}{a}
 \Whitetran{b}{4}{6}{b}
 \Whitetran{c}{2}{3}{c} 
 \Whitetran{d}{4}{3}{d} 

 \diredge{Aa}{a}\diredge{Ab}{b}
 \diredge{c}{Bc}\diredge{d}{Bd} 
 \diredge{a}{ac}\diredge{a}{ad} 
 \diredge{b}{bc}\diredge{b}{bd}
 \diredge{ac}{c}\diredge{bc}{c} 
 \diredge{ad}{d}\diredge{bd}{d} 
\end{tikzpicture} 
&
\StandardNet[0.55]

 \placN{Aa}{2}{7.5}{1}{q_1}
 \placN{Ab}{4}{7.5}{1}{q_2}
 
 \placW{ac}{ 0}{4.5}{0}{q_3} 
 \placE{bd}{ 6}{4.4}{0}{q_6} 

 \placS{Bc}{ 2}{1.5}{0}{q_7}
 \placS{Bd}{ 4}{1.5}{0}{q_8} 
 
 \Whitetran{a}{2}{6}{a}
 \Whitetran{b}{4}{6}{b}
 \Whitetran{c}{2}{3}{c} 
 \Whitetran{d}{4}{3}{d} 

 \diredge{Aa}{a}\diredge{Ab}{b}
 \diredge{c}{Bc}\diredge{d}{Bd} 
 \diredge{a}{ac} 
 \diredge{b}{bd}
 \diredge{ac}{c} 
 \diredge{bd}{d} 
\end{tikzpicture}
\\
($a$) $\NN_1$
&($b$) $\NN'_1$
\\ [2mm]
\StandardNet[0.55] 
 
 \placN{ac}{ 0}{4}{1}{r_1} 
 \placN{ad}{ 2}{4}{1}{r_2} 
 \placN{bc}{ 4}{4}{1}{r_3} 
 \placN{bd}{ 6}{4}{1}{r_4} 

 \placS{Ba}{ 0}{0}{0}{r_5}
 \placS{Bb}{ 2}{0}{0}{r_6}
 \placS{Bc}{ 4}{0}{0}{r_7}
 \placS{Bd}{ 6}{0}{0}{r_8} 
 
 \Whitetran{a}{0}{2}{a}
 \Whitetran{b}{2}{2}{b}
 \Whitetran{c}{4}{2}{c} 
 \Whitetran{d}{6}{2}{d} 

 \diredge{a}{Ba}\diredge{c}{Ba}
 \diredge{a}{Bb}\diredge{d}{Bb}
 \diredge{c}{Bc}\diredge{b}{Bc} 
 \diredge{b}{Bd}\diredge{d}{Bd} 
 \diredge{ac}{c}\diredge{ac}{a} 
 \diredge{ad}{d}\diredge{ad}{a} 
 \diredge{bc}{b}\diredge{bc}{c}
 \diredge{bd}{b}\diredge{bd}{d} 
\end{tikzpicture} 
&
\StandardNet[0.55] 
 
 \placN{ac}{ 0}{4}{1}{r_1} 
 \placN{bd}{ 6}{4}{1}{r_4} 

 \placS{Ba}{ 0}{0}{0}{r_5}
 \placS{Bb}{ 2}{0}{0}{r_6}
 \placS{Bc}{ 4}{0}{0}{r_7}
 \placS{Bd}{ 6}{0}{0}{r_8} 
 
 \Whitetran{a}{0}{2}{a}
 \Whitetran{b}{2}{2}{b}
 \Whitetran{c}{4}{2}{c} 
 \Whitetran{d}{6}{2}{d} 

 \diredge{a}{Ba}\diredge{c}{Ba}
 \diredge{a}{Bb}\diredge{d}{Bb}
 \diredge{c}{Bc}\diredge{b}{Bc} 
 \diredge{b}{Bd}\diredge{d}{Bd} 
 \diredge{ac}{c}\diredge{ac}{a} 
 \diredge{bd}{b}\diredge{bd}{d} 
\end{tikzpicture} 
\\
($c$) $\NN_2$&($d$) $\NN'_2$
\end{tabular}
\end{center}
\caption {\label{fi-3}
($a,c$) Marked nets $\NN_1$ and $\NN_2$, and ($b,d$) their 
unsuccessful reductions $\NN'_1$ and $\NN'_2$. 
}
\end{figure}

\section{Composing distributed places}
\label{sec-constrdp}

It is possible to derive 
distributed places compositionally. 
Since any $\NN=\FA{P,T,\Flow,M_\init}$ dealt with in the rest of
this paper is such that no two different places have the same 
input transitions and the same output transitions
(\ie $\PRE{p}=\PRE{r}\wedge\POST{p}=\POST{r}$ implies
$p=r$), a place $p$ can be identified by 
the sets $\PRE{p}$ and $\POST{p}$, 
and so $p$ will be denoted as $\pi_Z$, where 
$Z=\{t^\OUT\mid \FA{p,t}\in \Flow\}\cup\{t^\IN\mid \FA{t,p}\in \Flow\}$ is the set 
of tagged input and output transitions of $p$.
We will also use the notation $\PRE{\pi_Z}=\{t\mid t^\IN\in Z\}$
and $\POST{\pi_Z}=\{t\mid t^\OUT\in Z\}$.
In the examples, $Z$ will be denoted as $\pi_{z_1\dots z_k}$ where
$z_1\dots z_k$ is an enumeration of the elements of $Z$.
With such a notation, the transitions and arcs of $\NN$ become 
implicit, as we have:
\begin{equation}
\label{eq-vd}
\begin{array}{lclll} 
T&=&\{t\mid \exists \pi_Z\in P:~t^\IN\in Z \}
&\cup&\{t\mid \exists \pi_Z\in P:~ t^\OUT\in Z\}
\\
\Flow&=&\{\FA{t,\pi_Z}\in T\times P\mid t^\IN\in Z\}&\cup&
\{\FA{\pi_Z,t}\in P\times T\mid t^\OUT\in Z\}\;.
\end{array}
\end{equation}
Hence the structure of $\NN$ can be 
represented by its set of places.
We also denote 
$\INN_t=\SSS_{\pi_{t^\IN}}=\{\pi_{t^\IN}\}$ and 
$\OUTT_t=\SSS_{\pi_{t^\OUT}}=\{\pi_{t^\OUT}\}$, for every transition $t$.
By Proposition~\ref{prop-dvndv}, $\INN_t$ and 
$\OUTT_t$ are singleton
distributed places such that 
$\In_{\INN_t}=\Out_{\OUTT_t}=\{t\}$ and 
$\In_{\OUTT_t}=\Out_{\INN_t}=
\Read_{\INN_t}=\Read_{\OUTT_t}=\es$.
We also have the following immediate characterisation 
of pure distributed places.

\begin{proposition}
\label{pr-rvirbir}
A distributed place $\PC$ is pure iff 
$\{t^\IN,t^\OUT\}\not\subseteq \bigcup\{Z\mid \pi_Z\in \PC\}$,
for each transition~$t$.
\end{proposition} 

What we just introduced is a mere notational convenience
allowing simple additions of new and removal of existing places 
without having to modify the flow relation explicitly. 

\begin{example}
\label{ex-rrf}
Consider $\NN_1$ in Figure~\ref{fi-3}($a$). 
By the above convention, $q_1=\pi_{a^\OUT}$
and $q_3=\pi_{a^\IN c^\OUT}$, and all the places of $\NN_1$ are 
$\pi_{a^\OUT}$,
$\pi_{b^\OUT}$, 
$\pi_{a^\IN c^\OUT }$,
$\pi_{b^\IN c^\OUT }$, 
$\pi_{a^\IN d^\OUT }$,
$\pi_{b^\IN d^\OUT }$, 
$\pi_{c^\IN }$, and
$\pi_{d^\IN }$.
Also, $a\in\PRE{q_3}$ since $q_3=\pi_Z$, with $Z=\{a^\IN,c^\OUT\}$
and $a^\IN\in Z$.
\eod
\end{example} 

When composing two distributed places, $\PC$ and $\PC'$, we 
will assume that they are \emph{separated} which means that
they are connected to disjoint sets of transitions, 
\ie $\PREPOST{\PC}\cap\PREPOST{\PC'}=\es$.

The first way of constructing
new distributed places is simply to take their set union. 
That is, $\PC\oplus\PC'=\PC\cup\PC'$ is the \emph{union} of two 
separated distributed places $\PC$ and $\PC'$.

\begin{theorem}
\label{prop-union}
Let $\PC$ and $\PC'$ be separated distributed places and $\PC''=\PC\oplus\PC'$.

\begin{enumerate}
\item 
$\alpha_{\PC''}=\alpha_\PC\cup\alpha_{\PC'}$ for $\alpha\in\{\In,\Out,\Read\}$. 

\item 
$\In_\PC\neq\es\neq\In_{\PC'}$ implies
$\alpha_{\PC''}=\alpha_\PC |||\alpha_{\PC'}$ 
for $\alpha\in\{\inseq,\complinseq\}$ 
as well as $\inseq_{\PC''}=\pref(\complinseq_{\PC''})$.

\item 
$\Out_\PC\neq\es\neq\Out_{\PC'}$ implies
$\alpha_{\PC''}=\alpha_\PC |||\alpha_{\PC'}$ 
for $\alpha\in\{\outseq,\comploutseq\}$ 
as well as $\outseq_{\PC''}=\pref(\comploutseq_{\PC''})$.

\item
$\PC''$ is a distributed place 
iff 
$\In_\PC=\es=\In_{\PC'}\;\wedge\;\Out_\PC=\es=\Out_{\PC'}$ or
$\In_\PC=\es=\In_{\PC'}\;\wedge\;\Out_\PC\neq\es\neq\Out_{\PC'}$ or
$\In_\PC\neq\es\neq\In_{\PC'}\;\wedge\;\Out_\PC=\es=\Out_{\PC'}$.
Moreover, $\PC''$ is pure iff so are $\PC$ and $\PC'$. 
\end{enumerate}
\end{theorem}
 
\begin{proof} 
Parts (1,2,3) follow directly from the definitions
and an observation that, intuitively, the behaviour 
of $\PC''$ is a concurrent composition of the behaviours
of $\PC$ and $\PC'$.
 
To show the ($\Longrightarrow$) implication in part (4)
we first observe that, by Definition~\ref{def-distributed-place}(2),
$\In_\PC\neq \es \neq\Out_{\PC'}$ and $\Out_\PC\neq \es \neq\In_{\PC'}$
do not hold.
Next, also
$\In_\PC= \es \neq\In_{\PC'}$ and $\In_\PC\neq\es =\In_{\PC'}$
do not hold, as in such cases 
$\In_{\PC''}\neq \es$ and $\complinseq_{\PC''}=\es$, contradicting 
Definition~\ref{def-distributed-place}(3).
Similarly, $\Out_\PC= \es \neq\Out_{\PC'}$ and $\Out_\PC\neq\es =\Out_{\PC'}$
are impossible due to Definition~\ref{def-distributed-place}(4). 

The ($\Longleftarrow$) implication in part (4)
follows from parts (1,2,3) and an observation that 
Definition~\ref{def-distributed-place}(2) holds
since 
$\In_{\PC''}\times\Out_{\PC''}=\es$.

Finally, by $\Read_{\PC''}=\Read_\PC\cup\Read_{\PC'}$, $\PC''$ is 
pure iff both $\PC$ and $\PC'$ are pure. 
\end{proof} 

In the rest of the paper, we will use the following simplified version 
of the last result.

\begin{corollary}
\label{cprop-union}
Let $\PC$ and $\PC'$ be separated pure distributed places 
such that either 
$\In_\PC=\es=\In_{\PC'}\;\wedge\;\Out_\PC\neq\es\neq\Out_{\PC'}$,
or
$\In_\PC\neq\es\neq\In_{\PC'}\;\wedge\;\Out_\PC=\es=\Out_{\PC'}$.
Then $\PC''=\PC\oplus\PC'$ is a pure distributed place 
such that 
$\alpha_{\PC''}=\alpha_\PC\cup\alpha_{\PC'}$ for $\alpha\in\{\In,\Out\}$,
and 
$\alpha_{\PC''}=\alpha_\PC|||\alpha_{\PC'}$ 
for $\alpha\in\{\inseq,\outseq,\complinseq,\comploutseq\}$. 
\eod
\end{corollary}

To support step-wise
construction (or hierarchical analysis) of concurrent systems
a crucial role is often played by some notion of synchronisation 
between components. 
In terms of net construction, a suitable synchronisation between 
different process threads can be 
achieved through the cross-product of 
sets of places.

\begin{figure}[t!]
\begin{center} 
\begin{tabular}{c@{~~~~~~}c@{~~~~~~}c@{~~~~~~}c@{~~~~~~}c@{~~~~~~}c@{~~~~}c}
\StandardNet[0.55]
\placS{Aa}{0}{0}{0}{\pi_{a^\IN}} 
\Whitetran{a}{0}{2}{a} 
\diredge{a}{Aa} 
\end{tikzpicture} 
&
\StandardNet[0.55]
\placS{Aa}{0}{0}{0}{\pi_{b^\IN}} 
\Whitetran{a}{0}{2}{b} 
\diredge{a}{Aa} 
\end{tikzpicture} 
&
\StandardNet[0.55]
\placS{Aa}{0}{0}{0}{\pi_{c^\OUT}} 
\Whitetran{a}{0}{2}{c} 
\diredge{Aa} {a} 
\end{tikzpicture} 
&
\StandardNet[0.55]
\placS{Aa}{0}{0}{0}{\pi_{d^\OUT}} 
\Whitetran{a}{0}{2}{d} 
\diredge{Aa}{a} 
\end{tikzpicture} 
&
\StandardNet[0.55]
\placS{Aa}{0}{0}{0}{\pi_{a^\IN}}
\placS{Ab}{1.5}{0}{0}{\pi_{b^\IN}} 
\Whitetran{a}{0}{2}{a}
\Whitetran{b}{1.5}{2}{b} 
\diredge{a}{Aa}\diredge{b}{Ab}
\end{tikzpicture} 
&
\StandardNet[0.55] 
\placS{Bc}{0}{0}{0}{\pi_{c^\OUT}}
\placS{Bd}{1.5}{0}{0}{\pi_{d^\OUT}} 
\Whitetran{c}{0}{2}{c} 
\Whitetran{d}{1.5}{2}{d} 
\diredge{Bc}{c}\diredge{Bd}{d} 
\end{tikzpicture}
&
\StandardNet[0.55]
 \placS{Aa}{0}{0}{0}{\pi_{a^\IN c^\OUT}}
 \placS{Ab}{2}{0}{0}{\pi_{b^\IN d^\OUT}}
 \placS{Bc}{4}{0}{0}{\pi_{b^\IN c^\OUT}}
 \placS{Bd}{6}{0}{0}{\pi_{a^\IN d^\OUT}} 
 \Whitetran{a}{0}{2}{a}
 \Whitetran{b}{2}{2}{b}
 \Whitetran{c}{4}{2}{c} 
 \Whitetran{d}{6}{2}{d} 
 \diredge{a}{Aa}\diredge{b}{Ab}
 \diredge{Bc}{c}\diredge{Bd}{d} 
 \diredge{Aa}{c}\diredge{b}{Bc}
 \diredge{Ab}{d}\diredge{a}{Bd}
\end{tikzpicture} 
\\
$\INN_a$&$\INN_b$&$\OUTT_c$&$\OUTT_d$&
$\INN_a\oplus\INN_b$&$\OUTT_c\oplus\OUTT_d$&$(\INN_a\oplus\INN_b)\otimes(\OUTT_c\oplus\OUTT_d)$ 
\end{tabular}
\end{center}
\caption {\label{fi-3aaa}
Composing distributed places.
}
\end{figure}

\begin{definition}[cross-product]
\label{def-jdjdj}
The \emph{cross-product} of two non-empty sets of places, 
$Q$ and $R$, 
is 
$Q\otimes R 
=
\{ \pi_{Z\cup Z'}\mid \pi_Z\in Q \wedge \pi_{Z'}\in R \}$.
\eod
\end{definition}

\begin{example}
\label{ex-ewnc}
Figure~\ref{fi-3aaa} depicts four singleton distributed places, 
two distributed places obtained through the union composition,
$\INN_a\oplus\INN_b$ and $\OUTT_c\oplus\OUTT_d$,
and one distributed place obtained through the cross-product composition
$(\INN_a\oplus\INN_b)\otimes(\OUTT_c\oplus\OUTT_d)$. 
The latter is nothing but the 
distributed place 
$\PC_1=\{q_3,q_4,q_5,q_6\}$ of $\NN_1$ in Figure~\ref{fi-3}($a$).
Moreover, for the 
distributed place
$\PC_2=\{r_1,r_2,r_3,r_4\}$ of $\NN_2$ in Figure~\ref{fi-3}($c$), we have 
$
\PC_2=\{\pi_{a^\OUT c^\OUT},
 \pi_{b^\OUT c^\OUT}, 
 \pi_{a^\OUT d^\OUT}, 
 \pi_{b^\OUT d^\OUT} \}
=
\{\pi_{a^\OUT},\pi_{b^\OUT}\}
\otimes
\{\pi_{c^\OUT},\pi_{d^\OUT}\} 
=
(\OUTT_a\oplus\OUTT_b)\otimes(\OUTT_c\oplus\OUTT_d)
$. 
\eod
\end{example} 

\begin{theorem}
\label{th-cpr}
Let $\PC$ and $\PC'$ be separated distributed places. 
Then $\PC''=\PC \otimes\PC'$ is a distributed place 
such that the following hold:
\begin{enumerate}
\item 
$\alpha_{\PC''}=\alpha_\PC\cup\alpha_{\PC'}$ for $\alpha\in\{\In,\Out,\Read\}$. 

\item 
$\inseq_{\PC''}=\inseq_\PC\cup 
\inseq_{\PC'}\cup (\complinseq_\PC 
\cup
\complinseq_{\PC'})\circ \Read_{\PC''}^*$.

\item 
$\complinseq_{\PC''}=(\complinseq_\PC 
\cup
\complinseq_{\PC'})\circ \Read_{\PC''}^*$.

\item 
$\alpha_{\PC''}=\Read_{\PC''}^*\circ(\alpha_\PC \cup \alpha_{\PC'})$
for $\alpha\in\{\outseq,\comploutseq\}$. 
\end{enumerate}
Moreover, $\PC''$ is pure iff so are $\PC$ and $\PC'$.
\end{theorem}

\begin{proof}
We first observe that, since $\PC$ and $\PC'$ are separated,
for every $\pi_{Z''}\in \PC''$ there are 
unique $\pi_Z\in \PC$ and $\pi_{Z'}\in \PC'$ such that $Z''=Z\cup Z'$.
We then observe that, also by $\PC\neq\es\neq\PC'$, for every $x$ of the form $t^\IN$ 
or $t^\OUT$, the following holds:
\begin{equation}
\label{eq-ejvn} 
x \in \bigcup\{Z''\mid\pi_{Z''}\in \PC''\} 
 \iff 
x \in \bigcup\{Z\mid\pi_{Z}\in \PC\}\uplus \bigcup\{Z'\mid\pi_{Z'}\in \PC'\}\;.
\end{equation}
Based on these observations and the definitions, 
for all $t,u \in\PREPOST{\PC}$ and $w,v \in\PREPOST{\PC'}$ and
$\alpha,\beta\in\{\PREE,\POSTT\}$, we have several structural properties
of $\PC''$:
\begin{equation}
\label{eq-ddbd} 
\alpha_{\PC''}(t)
=
\alpha_\PC(t)\otimes \PC'
~~\mbox{and}~~ 
\alpha_{\PC''}(w)
=
\alpha_{\PC'}(w)\otimes \PC \;. 
\end{equation} 
\begin{equation}
\label{eq-ddbwwd}
\begin{array}{lcl@{~}l@{~}l}
\alpha_{\PC''}(t)\cap\beta_{\PC''}(u)
&=&
(\alpha_{\PC}(t)\cap\beta_{\PC}(u))&\otimes& \PC'
\\
\alpha_{\PC''}(w)\cap\beta_{\PC''}(v)
&=&
(\alpha_{\PC'}(w)\cap\beta_{\PC'}(v))&\otimes& \PC 
\\
\alpha_{\PC''}(t)\cup\alpha_{\PC''}(u)
&=&
(\alpha_{\PC}(t)\cup\alpha_{\PC}(u))&\otimes& \PC'
\\
\alpha_{\PC''}(w)\cup\alpha_{\PC''}(v)
&=&
(\alpha_{\PC'}(w)\cup\alpha_{\PC'}(v))&\otimes& \PC \;.
\end{array}
\end{equation}
\begin{equation}
\label{eq-ddrereffrbd} 
\alpha_{\PC''}(t)\cap\beta_{\PC''}(w)
=
\alpha_{\PC}(t)\otimes\beta_{\PC'}(w)\;. 
\end{equation}

The properties captured through 
Eqs.(\ref{eq-ejvn},\ref{eq-ddbd},\ref{eq-ddbwwd},\ref{eq-ddrereffrbd}) 
allow one to prove that $\PC''$ meets the conditions for being 
a distributed place in
Definition~\ref{def-distributed-place}, in the following way.

To show Definition~\ref{def-distributed-place}(1), we need to demonstrate that
$\PREPOST{\PC''}= \In_{\PC''}\cup \Out_{\PC''}\cup\Read_{\PC''}$. 
This equality 
follows from \eqref{eq-ejvn} as well as
$\PREPOST{\mathcal R}= \In_{\mathcal R}\cup \Out_{\mathcal R}\cup\Read_{\mathcal R}$ 
for $\mathcal R\in\{\PC,\PC'\}$. 
Moreover, if $t\in \PREPOST{\PC}$ then, by Eq.\eqref{eq-ddbd}, we have:
\[
\begin{array}{l@{~}l@{~}l@{~}l@{~}l@{~}lll} 
t\in \In_{\PC''}&\iff&
\POSTT_{\PC''}(t)\neq\es=\PREE_{\PC''}(t)&\iff &
\POSTT_{\PC}(t)\otimes \PC'\neq\es=\PREE_{\PC}(t)\otimes \PC' 
\\
&\iff&\POSTT_{\PC}(t)\neq\es=\PREE_{\PC}(t)&\iff&t\in \In_\PC\;.
\end{array} 
\] 
Hence, $\In_{\PC''}=\In_\PC\cup\In_{\PC'}$. In a similar way, we can show that
$\Out_{\PC''}=\Out_\PC\cup\Out_{\PC'}$ and $\Read_{\PC''}=\Read_\PC\cup\Read_{\PC'}$.
As a result, part (1) holds.

To show Definition~\ref{def-distributed-place}(2), we need to demonstrate that
$\In_{\PC''}\times \Out_{\PC''} \subseteq\enables(\PC'')$.
Suppose 
$t\in\In_{\PC''}=\In_\PC\cup\In_{\PC'}$ and 
$u\in\Out_{\PC''}=\Out_\PC\cup\Out_{\PC'}$. Then, without loss of generality, 
we have two cases:
\begin{itemize}

\item 
$t\in\In_\PC$ and $u\in\Out_\PC$.
Then, by \eqref{eq-ddbwwd}, 
$\POSTT_{\PC''}(t)\cap\PREE_{\PC''}(u)=
(\POSTT_{\PC}(t) \cap \PREE_{\PC}(u))\otimes \PC' \neq \es$ 
since $\POSTT_{\PC}(t) \cap \PREE_{\PC}(u)\neq\es$ due to $\FA{t,u}\in\enables(\PC)$.

\item 
$t\in\In_\PC$ and $u\in\Out_{\PC'}$.
Then, by \eqref{eq-ddrereffrbd} and $\POSTT_{\PC}(t)\neq\es\neq\PREE_{\PC'}(u)$, 
we have
$\POSTT_{\PC''}(t)\cap\PREE_{\PC''}(u)=
\POSTT_{\PC}(t)\otimes\PREE_{\PC'}(u)\neq\es$.
\end{itemize}
 
To show Definition~\ref{def-distributed-place}(3,4), we need to demonstrate that 
$\In_{\PC''}\neq\es\implies\inseq_{\PC''}=\pref(\complinseq_{\PC''})$
and 
$\Out_{\PC''}\neq\es\implies\outseq_{\PC''}=\pref(\comploutseq_{\PC''})$.
We observe that 
both implications 
follow 
from parts
(2,3,4) and Definition~\ref{def-distributed-place}(3,4) for $\PC$ and $\PC'$.
We therefore show that parts (2,3,4) do hold.

First, we note that the ($\supseteq$) inclusions in parts (2,3,4) hold by
Eqs.(\ref{eq-ddbd},\ref{eq-ddbwwd}).
We then note that the ($\subseteq$) inclusions in 
parts (2,3,4) hold for the same reasons and the 
following observations which we make for $\inseq_{\PC''}$ 
in part (2) (and the other cases 
are treated similarly).

Suppose that $\sigma=t_1\dots t_k\in\inseq_{\PC''}$ ($k\geq 1$)
and $U=\{t_1,\dots,t_k\}\cap\In_{\PC''}$.
Then, by $\In_{\PC''}=\In_\PC\cup\In_{\PC'}$,
$U\subseteq \In_\PC\cup\In_{\PC'}$.

Suppose that $t\in U\cap\In_{\PC}$
and $u\in U\cap\In_{\PC'}$. Then, by \eqref{eq-ddrereffrbd},
we have $\POSTT_{\PC''}(t)\cap\POSTT_{\PC''}(u)
=\POSTT_{\PC}(t)\otimes\POSTT_{\PC'}(u)\neq\es$ yielding a contradiction.
Hence, without loss of generality, we may assume that $U\subseteq \In_\PC$
and consider two cases.

\emph{Case 1:}
$\{t_1,\dots,t_k\}\cap\Read_{\PC'}=\es$.
Then 
$\{t_1,\dots,t_k\}\subseteq\In_\PC\cup\Read_{\PC}$ and
$\sigma\in\inseq_\PC$ which follows, in particular, from \eqref{eq-ddbwwd}.

\emph{Case 2:}
$\{t_1,\dots,t_k\}\cap\Read_{\PC'}\neq\es$.
Then there is the smallest $i$ such that $t_i\in\Read_{\PC'}$.
Suppose that $t_j\in\In_\PC$.
Then, by \eqref{eq-ddrereffrbd},
$\POSTT_{\PC''}(t_j)\cap\PREE_{\PC''}(t_i)=
\POSTT_\PC(t_j)\otimes\PREE_{\PC'}(t_i)\neq\es$.
Hence, by Definition~\ref{def-kkd}(1), we have $j<i$.
We then observe that 
$t_1\dots t_{i-1}\in\complinseq_\PC$.
Indeed, otherwise there is
$v\in\In_\PC$ such that, by \eqref{eq-ddrereffrbd} and \eqref{eq-ddbwwd},
$\POSTT_{\PC''}(v)\cap\PREE_{\PC''}(t_i)\neq\es$ 
and $\POSTT_{\PC''}(v)\cap \POSTT_{\PC''}(U)=\es$.
Hence, we obtain a contradiction, which means that 
$t_1\dots t_{i-1}\in\complinseq_\PC$,
and so $\sigma\in\complinseq_\PC\circ\Read_{\PC''}^*$.
 
Finally, by $\Read_{\PC''}=\Read_\PC\cup\Read_{\PC'}$, $\PC''$ is 
pure iff both $\PC$ and $\PC'$ are pure.
\end{proof} 

Intuitively, if we ignore transitions
in $\Read_{\PC\otimes\PC'}$, then the distributed places in 
Theorem~\ref{th-cpr}
are such that the empty $\PC\otimes\PC'$ can be filled
either by an 
in-sequence of $\PC$ or an in-sequence of
$\PC'$, and then emptied either by an out-sequence of 
$\PC$ or an out-sequence of $\PC'$. 

\begin{example}
\label{ex-eurvi} 
Coming back to Example~\ref{ex-ewnc}, the decompositions
of sets of places $\PC_1$ and $\PC_2$, together with 
Corollaries~\ref{cprop-union} and~\ref{cth-cpr} as well as 
Proposition~\ref{prop-dvndv},
imply that these two 
sets of places are distributed places.
\eod
\end{example}

In the rest of the paper, we will use the following simplified version 
of the last result.

\begin{corollary}
\label{cth-cpr}
Let $\PC$ and $\PC'$ be separated pure distributed places. 
Then $\PC''=\PC \otimes\PC'$ is a pure distributed place 
such that $\alpha_{\PC''}=\alpha_\PC\cup\alpha_{\PC'}$ 
for $\alpha\in\{\In,\Out,\inseq,\outseq,\complinseq,\comploutseq\}$.
\eod
\end{corollary}

\section{Box algebra and distributed places}
\label{sec-balg}

We consider a fragment of Box Algebra~\cite{BDK-01} 
focusing on Petri nets derived from four control flow operators. 
The syntax of process expressions we will use --- called \emph{box expressions}
--- is as follows:
\begin{equation}
\label{eq-ba} 
 E ~::=~ 
 a ~\mid~
 E\SEQ E ~\mid~
 E\CHOICE E ~\mid~
 E\PAR E ~\mid~
 \ITER{E}{E}{E}
\end{equation}
where $a$ is an atomic action. 
Moreover, since we deal only with control flow aspects and actions 
do not have any special semantical attributes, we assume 
that no action occurs more than once in a 
given box expression, which
simplifies the compositional translation 
from box expressions to marked nets.\footnote{
An alternative would be to associate unique tags with action
occurrences, \eg
using their positions in the syntax tree of a box expression.
}

Intuitively, 
$a$ denotes a process which can execute atomic action $a$ and terminate,
$E\SEQ F$ denotes sequential composition of two processes,
$E\CHOICE F$ denotes choice composition, 
$E\PAR F$ denotes parallel composition, and
$\ITER{E}{F}{G}$ denotes iteration which starts by executing
$E$, then executes $F$ any number of times
(including zero), and terminates by executing~$G$.

\begin{example}
\label{ex-eirrio}
$E_1=\ITER{a\CHOICE b}{c\SEQ(d\PAR e)}{f\PAR g}$
is a box expression specifying a concurrent system which starts 
with the execution of either $a$ or $b$, 
then executes (possibly zero times)
a loop `$c$ followed by concurrent execution of $d$ and $e$', 
and the whole expression terminates by executing $f$ concurrently with~$g$.
Figure~\ref{fi-ee44dd3}($a$) displays the corresponding marked net
which can be derived compositionally, 
and Figure~\ref{fi-ee44dd3}($b$) its reachability graph.
\eod
\end{example}

The syntax in Eq.\eqref{eq-ba} admits box expressions 
which would result in non-safe marked
nets. 
To avoid this, we will work with the following modified 
syntax: 
\begin{equation}
\label{eq-baaaa} 
\begin{array}{lcllllllllllllll}
 E &::=& 
 a &\mid& 
 E\SEQ E &\mid&
 E\CHOICE E &\mid&
 \ITER{E}{H}{E}&\mid&
 E\PAR E&\mid&
 H
 \\
 H &::=& 
 & & 
 E\SEQ E &\mid&
 H\CHOICE H &\mid&
 \ITER{E}{H}{E}
\end{array}
\end{equation} 
The standard translation from box expressions to marked nets adapted
for the notation used in this paper 
is carried out in two steps. 
The translation first associates 
with each box expression $E$ a triple of pairwise disjoint sets of places, 
$ 
\Upsilon_E=\FA{\Upsilon_E^e, \Upsilon_E^x, \Upsilon_E^i} 
$.
The syntax-driven derivation of $\Upsilon_E$ is as follows: 
\begin{equation}
\label{eq-fefw}
\begin{array}{lcl}
 \Upsilon_a
 &=&
 \FA{\OUTT_a, \INN_a, \es}
\\
 \Upsilon_{E\PAR F}
 &=&
 \FA{\Upsilon_E^e\cup \Upsilon_F^e, \Upsilon_E^x\cup \Upsilon_F^x, 
 \Upsilon_E^i\cup \Upsilon_F^i}
 \\
 \Upsilon_{E\SEQ F}
 &=&
 \FA{\Upsilon_E^e, \Upsilon_F^x, 
 \Upsilon_E^i\cup \Upsilon_F^i\cup (\Upsilon_E^x\otimes\Upsilon_F^e )}
 \\
 \Upsilon_{E\CHOICE F}
 &=&
 \FA{\Upsilon_E^e\otimes\Upsilon_F^e, 
 \Upsilon_E^x\otimes\Upsilon_F^x, 
 \Upsilon_E^i\cup \Upsilon_F^i}
 \\
\Upsilon_{\ITER{E}{F}{G}} 
 &=&
 \FA{\Upsilon_E^e, 
 \Upsilon_G^x, 
 \Upsilon_E^i\cup \Upsilon_F^i\cup\Upsilon_G^i\cup ((\Upsilon_E^x\otimes \Upsilon_F^x)
 \otimes(\Upsilon_G^e\otimes\Upsilon_F^e))}
 \end{array}
\end{equation} 
Then the marked net associated with 
$E$ (called in~\cite{BDK-01} a \emph{box}) is given by: 
\[
\BOX(E)=\FA{\PLACES(E), T, \Flow, \Upsilon_E^e }\;,
\]
where $\PLACES(E)=\Upsilon_E^e\cup \Upsilon_E^x \cup \Upsilon_E^i$,
$T$ are the actions occurring in $E$,
and $\Flow$ are the arcs derived as in Eq.\eqref{eq-vd}.
The set $\Upsilon_E^e$ comprises the \emph{entry places} (initially marked), 
$\Upsilon_E^x$ comprises the \emph{exit places} (initially empty), and 
$\Upsilon_E^i$ comprises the \emph{internal places} (initially empty).
Note that the entry and exit places play the role of
active interfaces in the 
compositional translation from
box expressions to marked nets, whereas the internal places are 
carried over to the resulting marked nets without any changes.

\begin{figure}[t!]
\begin{center}
\begin{tabular}{c@{~~~~~~~~~~~~~~~~~~~~~~~~~~}c} 

\StandardNet[0.55]
 \placW{xxx}{3}{9.5}{1}{} 
 \placW{abd}{0}{6}{0}{} 
 \placW{abe}{2}{6}{0}{}
 \placW{acd}{4}{6}{0}{}
 \placW{ace}{6}{6}{0}{}

 \placW{ab}{1}{4}{0}{}
 \placW{ac}{5}{4}{0}{}

 \placW{dd}{-3}{2}{0}{}
 \placW{ee}{9}{2}{0}{}

 \Whitetran{A}{0}{8}{a}
 \Whitetran{B}{6}{8}{b}
 \Whitetran{a}{3}{4}{c}
 \Whitetran{b}{-1}{4}{d}
 \Whitetran{c}{7}{4}{e} 
 \Whitetran{d}{-3}{4}{f} 
 \Whitetran{e}{9}{4}{g}

\diredge{abd}{a}\diredge{abe}{a}\diredge{acd}{a}\diredge{ace}{a}
\diredge{a}{ab}\diredge{ab}{b}\diredge{a}{ac}\diredge{ac}{c}
\diredge{b}{abd}\diredge{b}{abe}\diredge{c}{acd}\diredge{c}{ace}
\diredge{abd}{d}\diredge{abe}{e}\diredge{acd}{d}\diredge{ace}{e}
\diredge{e}{ee}\diredge{d}{dd}
\diredge{xxx}{A}\diredge{xxx}{B}
\diredge{A}{ace}\diredge{A}{acd}\diredge{A}{abe}\diredge{A}{abd}
\diredge{B}{ace}\diredge{B}{acd}\diredge{B}{abe}\diredge{B}{abd}
\end{tikzpicture} 
&
\StandardTS [0.55]
\nodN{a}{3}{6}{M_\init}
\nodE{b}{3}{-1}{} 
\nodE{d}{3}{4}{}
\nodE{e}{0}{2}{}
\nodE{f}{2}{2}{}
\nodE{g}{3}{0}{}
\nodE{h}{4}{2}{}
\nodE{i}{6}{2}{}

\leftbowtext{a}{d}{60}{120}{40}{$b$} 
\rightbowtext{a}{d}{300}{240}{40}{$a$} 
\edgetext{d}{e}{$f$};\edgetext{e}{b}{$g$};
\edgetext{d}{i}{$g$};\edgetext{i}{b}{$f$};
\edgetext{d}{g}{$c$}; 
\edgetext{f}{d}{$d$};\edgetext{g}{f}{$e$};
\edgetext{h}{d}{$e$};\edgetext{g}{h}{$d$};
\end{tikzpicture} 
\\
($a$) $ \BOX(E_1)$
&
($b$) $\RG_{\BOX(E_1)}$ 
\\
\StandardNet[0.55]

 \placN{Aa}{2}{7.5}{1}{q_1}
 \placN{Ab}{4}{7.5}{1}{q_2}
 
 \placW{ac}{ 0}{4.5}{0}{q_3} 
 \placW{bc}{ 2}{4.5}{0}{q_4} 
 \placE{ad}{ 4}{4.5}{0}{q_5} 
 \placE{bd}{ 6}{4.5}{0}{q_6} 

 \placS{Bc}{ 2}{1.5}{0}{q_7}
 \placS{Bd}{ 4}{1.5}{0}{q_8} 
 
 \Whitetran{a}{2}{6}{a}
 \Whitetran{b}{4}{6}{b}
 \Whitetran{c}{2}{3}{c} 
 \Whitetran{d}{4}{3}{d} 

 \diredge{Aa}{a}\diredge{Ab}{b}
 \diredge{c}{Bc}\diredge{d}{Bd} 
 \diredge{a}{ac}\diredge{a}{ad} 
 \diredge{b}{bc}\diredge{b}{bd}
 \diredge{ac}{c}\diredge{bc}{c} 
 \diredge{ad}{d}\diredge{bd}{d} 
\end{tikzpicture} 
&
\StandardNet[0.55] 
 
 \placN{ac}{ 0}{4}{1}{r_1} 
 \placN{ad}{ 2}{4}{1}{r_2} 
 \placN{bc}{ 4}{4}{1}{r_3} 
 \placN{bd}{ 6}{4}{1}{r_4} 

 \placS{Ba}{ 0}{0}{0}{r_5}
 \placS{Bb}{ 2}{0}{0}{r_6}
 \placS{Bc}{ 4}{0}{0}{r_7}
 \placS{Bd}{ 6}{0}{0}{r_8} 
 
 \Whitetran{a}{0}{2}{a}
 \Whitetran{b}{2}{2}{b}
 \Whitetran{c}{4}{2}{c} 
 \Whitetran{d}{6}{2}{d} 

 \diredge{a}{Ba}\diredge{c}{Ba}
 \diredge{a}{Bb}\diredge{d}{Bb}
 \diredge{c}{Bc}\diredge{b}{Bc} 
 \diredge{b}{Bd}\diredge{d}{Bd} 
 \diredge{ac}{c}\diredge{ac}{a} 
 \diredge{ad}{d}\diredge{ad}{a} 
 \diredge{bc}{b}\diredge{bc}{c}
 \diredge{bd}{b}\diredge{bd}{d} 
\end{tikzpicture} 
\\
($c$) $\BOX(E_2)$
&
($d$) $\BOX(E_3)$
\end{tabular}
 
\end{center}
\caption {\label{fi-ee44dd3}
($a$) A net generated for 
$E_1=\ITER{a\CHOICE b}{c\SEQ(d\PAR e)}{f\PAR g}$ and 
($b$) its reachability graph. 
($c$) A marked net generated for 
$E_2=(a\PAR b)\SEQ(c\PAR d)$.
($d$) A marked net generated for 
$E_3=(a\PAR b)\CHOICE(c\PAR d)$.
}
\end{figure}

\begin{example}
\label{ex-vaiv}
For the box expression $E_2=(a\PAR b)\SEQ(c\PAR d)$, we have: 
\[
\begin{array}{lcl}
\Upsilon_{E_2}
&=& 
\FA{\Upsilon_{a\PAR b}^e, 
 \Upsilon_{c\PAR d}^x, 
 \Upsilon_{a\PAR b}^i\cup \Upsilon_{c\PAR d}^i\cup 
 (\Upsilon_{a\PAR b}^x\otimes\Upsilon_{c\PAR d}^e )}
\\
&=&
\FA{\Upsilon^e_a\cup\Upsilon^e_b, \Upsilon^x_c\cup\Upsilon^x_d, 
 \Upsilon^i_a\cup\Upsilon^i_b\cup \Upsilon^i_c\cup\Upsilon^i_d
 \cup 
 ((\Upsilon^x_a\cup\Upsilon^x_b)\otimes(\Upsilon^e_c\cup\Upsilon^e_d)) }
\\
&=&
\FA{\OUTT_a\cup\OUTT_b,~\INN_c\cup\INN_d, 
 \es \cup\es \cup\es \cup\es \cup 
 ((\INN_a\cup\INN_b)\otimes(\OUTT_c\cup\OUTT_d)) }
\\
&=&
\FA{\{\pi_{a^\OUT},\pi_{b^\OUT}\}, 
 \{\pi_{c^\IN},\pi_{d^\IN}\}, 
 \{\pi_{a^\IN c^\OUT},\pi_{b^\IN c^\OUT},
 \pi_{a^\IN d^\OUT},\pi_{b^\IN d^\OUT}\} } \;.
\end{array}
\]
Hence 
$
\PLACES(E_2) =\{\pi_{a^\OUT},
\pi_{b^\OUT},\pi_{c^\IN},\pi_{d^\IN},\pi_{a^\IN c^\OUT}, 
\pi_{b^\IN c^\OUT},\pi_{a^\IN d^\OUT},\pi_{b^\IN d^\OUT}\}$
is the set of places of $\BOX(E_2)$. 
Moreover, $M_\init=\Upsilon_{E_2}^e=\{\pi_{a^\OUT},\pi_{b^\OUT}\}$, and so
$\BOX(E_2 )$ is the marked net 
depicted in Figure~\ref{fi-ee44dd3}($c$),
where: $q_1=\pi_{a^\OUT}$, $q_2=\pi_{b^\OUT}$, $q_3=\pi_{a^\IN c^\OUT}$,
$q_4=\pi_{b^\IN c^\OUT}$, $q_5=\pi_{a^\IN d^\OUT}$, $q_6=\pi_{b^\IN d^\OUT}$,
$q_7=\pi_{c^\IN}$, and $q_8=\pi_{d^\IN}$. 

For the box expression $E_3=(a\PAR b)\CHOICE(c\PAR d)$ the translation 
proceeds as follows:
\[
\begin{array}{lcl}
\Upsilon_{E_3}
&=& 
\FA{\Upsilon_{a\PAR b}^e\otimes \Upsilon_{c\PAR d}^e, 
\Upsilon_{a\PAR b}^x \otimes\Upsilon_{c\PAR d}^x, 
 \Upsilon_{a\PAR b}^i\cup \Upsilon_{c\PAR d}^i }
 \\
 &=& 
\FA{(\OUTT_a\cup\OUTT_b)\otimes 
 (\OUTT_c\cup\OUTT_d),
 (\INN_a\cup\INN_b)\otimes 
 (\INN_c\cup\INN_d),
 \es\cup\es }
\\
&=&
\FA{\{\pi_{a^\OUT c^\OUT},\pi_{a^\OUT d^\OUT},
 \pi_{b^\OUT c^\OUT},\pi_{b^\OUT d^\OUT}\}, 
 \{\pi_{a^\IN c^\IN},\pi_{a^\IN d^\IN},
 \pi_{b^\IN c^\IN},\pi_{b^\IN d^\IN}\}, 
 \es }\;.
\end{array}
\]
Hence 
$\PLACES(E_3)=
\{\pi_{a^\OUT c^\OUT},\pi_{a^\OUT d^\OUT},\pi_{b^\OUT c^\OUT},\pi_{b^\OUT d^\OUT},
\pi_{a^\IN c^\IN},\pi_{a^\IN d^\IN},\pi_{b^\IN c^\IN},\pi_{b^\IN d^\IN}\}$
is the set of places of $\BOX(E_3)$.
Moreover, 
$
M_\init=\Upsilon_{E_3}^e=\{\pi_{a^\OUT c^\OUT},\pi_{a^\OUT d^\OUT},
\pi_{b^\OUT c^\OUT},\pi_{b^\OUT d^\OUT}\} 
$, and so $\BOX(E_3)$ is the marked net 
of Figure~\ref{fi-ee44dd3}($d$),
where:
$r_1=\pi_{a^\OUT c^\OUT}$, 
$r_2=\pi_{a^\OUT d^\OUT}$,
$r_3=\pi_{b^\OUT c^\OUT}$,
$r_4=\pi_{b^\OUT d^\OUT}$,
$r_5=\pi_{a^\IN c^\IN}$,
$r_6=\pi_{a^\IN d^\IN}$,
$r_7=\pi_{b^\IN c^\IN}$, and
$r_8=\pi_{b^\IN d^\IN}$. 
\eod
\end{example} 

The translation from box expressions generates 
marked nets in
which one can identify distributed places needed for the proposed net-reduction approach based on Theorem~\ref{th-main}.

For every box expression $E$, the set of \emph{distributed places} of $E$
is defined as $\DP(E)=\Delta_E^e\cup\Delta_E^x\cup\Delta_E^i$, where
$\Delta_E=\FA{\Delta_E^e,\Delta_E^x,\Delta_E^i}$ is a triple of 
sets of sets of the entry, exit, and internal 
places of $\BOX(E)$, respectively.
In the syntax-driven definition of $\Delta_E$ given below,
we denote
$\oplus\Delta=D_1\oplus\dots\oplus D_k$
for any set $\Delta=\{D_1,\dots,D_k\}$: 

\begin{equation}
\label{eq-fefwdd}
\begin{array}{lcl}
 \Delta_a
 &=&
 \FA{\{\{\OUTT_a\}\},\{\{\INN_a\}\},\es}
\\
 \Delta_{E\PAR F}
 &=&
 \FA{\Delta_E^e\cup \Delta_F^e, \Delta_E^x\cup \Delta_F^x, 
 \Delta_E^i\cup \Delta_F^i}
 \\
 \Delta_{E\SEQ F}
 &=&
 \FA{\Delta_E^e, \Delta_F^x, 
 \Delta_E^i\cup \Delta_F^i\cup 
 \{(\oplus\Delta_E^x)\otimes(\oplus\Delta^e_F )\}}
 \\
 \Delta_{E\CHOICE F}
 &=&
 \FA{
 (\oplus\Delta_E^e)\otimes(\oplus\Delta_F^e),
 (\oplus\Delta_E^x)\otimes(\oplus\Delta_F^x),
 \Delta_E^i\cup \Delta_F^i} 
 \\
\Delta_{\ITER{E}{F}{G}} 
 &=&
 \FA{\Delta_E^e, 
 \Delta_G^x, 
 \Delta_E^i\cup \Delta_F^i\cup\Delta_G^i\cup 
 \{((\oplus\Delta_E^x)\otimes (\oplus\Delta_F^x))
 \otimes
 ((\oplus\Delta_F^e)\otimes(\oplus\Delta_G^e))\}
 }\;.
 \end{array}
\end{equation} 

\begin{example}
\label{ex-erhe} 
For the box expressions 
$E_2=(a\PAR b)\SEQ(c\PAR d)$ and 
$E_3=(a\PAR b)\CHOICE(c\PAR d)$ 
with $\BOX(E_2)$ and 
$\BOX(E_3)$ depicted in Figure~\ref{fi-ee44dd3}($c,d$), and
the box expression $E_4=(a\PAR b);((c\PAR d)\CHOICE(e\PAR f))$ 
with $\BOX(E_4)$ depicted in Figure~\ref{fi-1aee}($a$), we have:
\[
\begin{array}{rcl}
\Delta_{E_2}
&=&
\FA{\{\{q_1\},\{q_2\}\},\{\{q_7\},\{q_8\}\},\{\{q_3,q_4,q_5,q_6\}\}} 
\\
\DP(E_2)
&=&
\{\{q_1\},\{q_2\},\{q_7\},\{q_8\},\{q_3,q_4,q_5,q_6\}\}
\\
&=&
 \{\{\pi_{a^\OUT}\},\{\pi_{b^\OUT}\}, 
 \{\pi_{c^\IN}\},\{\pi_{d^\IN}\},
 \{\pi_{a^\IN c^\OUT},\pi_{b^\IN c^\OUT},
 \pi_{a^\IN d^\OUT},\pi_{b^\IN d^\OUT}\} \} 
\\
\Delta_{E_3}
&=&\FA{\{\{r_1,r_2,r_3,r_4\}\},\es,\{\{r_5,r_6,r_7,r_8\}\}}
\\
\DP(E_3)
&=&\{\{r_1,r_2,r_3,r_4\},\{r_5,r_6,r_7,r_8\}\}
\\
&=&
\{\{\pi_{a^\OUT c^\OUT},\pi_{a^\OUT d^\OUT} ,
\pi_{b^\OUT c^\OUT},\pi_{b^\OUT d^\OUT} \}, 
\{\pi_{a^\IN c^\IN},\pi_{a^\IN d^\IN},
\pi_{b^\IN c^\IN},\pi_{b^\IN d^\IN}\}\}
\\
\Delta_{E_4}
&=&
\FA{\{\{p_1\},\{p_2\}\},\{\{p_{11},p_{12},p_{13},p_{14}\}\},
\{\{p_3,p_4,p_5,p_6,p_7,p_8,p_9,p_{10}\}\}}
\\
\DP(E_4)
&=&
\{\{p_1\},\{p_2\},\{p_{11},p_{12},p_{13},p_{14}\},
\{p_3,p_4,p_5,p_6,p_7,p_8,p_9,p_{10}\}\} 
\;.
\end{array}
\]
Note that $\BOX(E_2)$ was discussed as 
$N_1$ after the proof of Theorem~\ref{th-main} 
with a different set of distributed places covering its places:
$\{q_1,q_2\}$, $\{q_7,q_8\}$, and $\{q_3,q_4,q_5,q_6\}$.
\eod
\end{example} 

The next result subsumes a number of basic properties of marked nets 
translated from box expressions
as well as the corresponding sets of distributed places. 

\begin{proposition}
\label{prop-ewceiu}
Let $E$ be a box expression defined by 
the syntax in Eq.\eqref{eq-baaaa} 
and $M$ be a reachable marking of $\BOX(E)=\FA{P,T,\Flow,M_\init}$.
\begin{enumerate}

\item 
$\BOX(E)$ is a safe marked net.

\item
$\Upsilon_E^e=M_\init=\{p\in P\mid \PRE{p}=\es\}\neq\es$
and 
$\Upsilon_E^x=\{p\in P\mid \POST{p}=\es\}\neq\es$.
Moreover, if $E$ is generated by the second line of the syntax 
in Eq.\eqref{eq-baaaa}, then 
$\POST{(\Upsilon_E^e)}\cap\PRE{(\Upsilon_E^x)}=\es$.

\item 
$\Upsilon_E^x$ is a marking reachable from $M$.

\item 
$(M\subseteq \Upsilon_E^e\;\vee\;\Upsilon_E^e\subseteq M)
\implies M=\Upsilon_E^e$ and 
$(M\subseteq\Upsilon_E^x\;\vee\;\Upsilon_E^x\subseteq M)\implies M=\Upsilon_E^x$. 

\item 
$\DP(E)$ is a set of pairwise disjoint pure distributed places 
covering $P$. 

\item
If $\PC\neq\PC'\in\DP(E)$ then 
$\PRE{\PC}\cap\PRE{\PC'}=\POST{\PC}\cap\POST{\PC'}=\es$. 
\end{enumerate}
\end{proposition}

\begin{proof}
Parts (1,2,3,4) are results proven in~\cite{BDK-01,DBLP:journals/iandc/BestDK02}.
Parts (5,6) follow by simultaneous 
induction on the structure of $E$
and the syntax-driven definitions 
Eqs.(\ref{eq-baaaa},\ref{eq-fefw},\ref{eq-fefwdd}), using part (2),
Corollaries~\ref{cprop-union} and~\ref{cth-cpr},
and the assumption that 
no action is used more than once in a given box expression.
\end{proof}

Intuitively, Proposition~\ref{prop-ewceiu}(4) means that 
the initial marking $\Upsilon_E^e$ of $\BOX(E)$ is monotonically emptied,
and that the final marking $\Upsilon_E^x$ is monotonically filled (and no token is
`left behind' when all the exit places are filled).

\begin{figure}[t!]
\begin{center}
\begin{tabular}{cc} 
\StandardNet[0.55] 
 \placN{Aa}{ 4}{8}{1}{p_1}
 \placN{Ab}{10}{8}{1}{p_2}
 
 \placW{cea}{ 0}{4.5}{0}{p_3\!\!} 
 \placW{ceb}{ 2}{4.5}{0}{p_4\!\!} 
 \placW{cfa}{ 4}{4.5}{0}{p_5\!\!} 
 \placW{cfb}{ 6}{4.5}{0}{p_6\!\!} 
 \placE{dea}{ 8}{4.5}{0}{\!\!p_7} 
 \placE{deb}{10}{4.5}{0}{\!\!p_8} 
 \placE{dfa}{12}{4.5}{0}{\!\!p_9} 
 \placE{dfb}{14}{4.5}{0}{\!\!p_{10}}

 \placS{cebx}{ 2}{-0.5}{0}{p_{11}} 
 \placS{cfax}{ 4}{-0.5}{0}{p_{12}} 
 \placS{debx}{10}{-0.5}{0}{p_{13}} 
 \placS{dfax}{12}{-0.5}{0}{p_{14}} 
 
 \Whitetran{a}{ 4}{6.5}{a}
 \Whitetran{b}{10}{6.5}{b}
 \Whitetran{c}{ 2}{2}{c} 
 \Whitetran{d}{ 4}{2}{d}
 \Whitetran{e}{10}{2}{e}
 \Whitetran{f}{12}{2}{f}

 \diredge{Aa}{a}\diredge{Ab}{b} 
 
 \diredge{a}{cea}\diredge{a}{cfa}\diredge{a}{dea}\diredge{a}{dfa}
 \diredge{b}{ceb}\diredge{b}{cfb}\diredge{b}{deb}\diredge{b}{dfb}
 \diredge{ceb}{c}\diredge{cfb}{c}\diredge{cea}{c}\diredge{cfa}{c}
 \diredge{deb}{d}\diredge{dfb}{d}\diredge{dea}{d}\diredge{dfa}{d}
 \diredge{ceb}{e}\diredge{cfb}{f}\diredge{cea}{e}\diredge{cfa}{f}
 \diredge{deb}{e}\diredge{dfb}{f}\diredge{dea}{e}\diredge{dfa}{f}
 
 \diredge{c}{cebx} \diredge{c}{cfax}
 \diredge{d}{debx} \diredge{d}{dfax}
 \diredge{e}{cebx} \diredge{f}{cfax}
 \diredge{e}{debx} \diredge{f}{dfax}
\end{tikzpicture}
&
\StandardNet[0.55] 
 \placN{Aa}{ 4}{8}{1}{p_1}
 \placN{Ab}{6}{8}{1}{p_2}
 
 \placW{ceb}{ 2}{4.5}{0}{p_4\!\!} 
 \placW{cfa}{ 4}{4.5}{0}{p_5\!\!} 
 \placE{deb}{6}{4.5}{0}{\!\!p_8} 
 \placE{dfa}{8}{4.5}{0}{\!\!p_9} 
 
 \placS{cebx}{ 2}{-.5}{0}{p_{11}} 
 \placS{dfax}{8}{-.5}{0}{p_{14}} 
 
 \Whitetran{a}{4}{6.5}{a}
 \Whitetran{b}{6}{6.5}{b}
 \Whitetran{c}{2}{2}{c} 
 \Whitetran{d}{4}{2}{d}
 \Whitetran{e}{6}{2}{e}
 \Whitetran{f}{8}{2}{f}

 \diredge{Aa}{a}\diredge{Ab}{b} 
 
 \diredge{a}{cfa} \diredge{a}{dfa}
 \diredge{b}{deb}\diredge{b}{ceb}
 \diredge{ceb}{c} \diredge{cfa}{c}
 \diredge{deb}{d} \diredge{dfa}{d} 
 \diredge{ceb}{e} \diredge{cfa}{f}
 \diredge{deb}{e} \diredge{dfa}{f}

 \diredge{c}{cebx} 
 \diredge{d}{dfax}
 \diredge{e}{cebx} 
 \diredge{f}{dfax} 
\end{tikzpicture} 
\\[-3mm]
($a$) $\NN_3$ &($b$) $\NN_3'$
\end{tabular} 
\end{center}
\caption {\label{fi-1aee}
($a$) A marked net $\NN_3=\BOX((a\PAR b);((c\PAR d)\CHOICE(e\PAR f)))$ and 
($b$) its reduced version $\NN_3'$. 
Note that 
$p_1=\pi_{a^\OUT}$,
$p_2=\pi_{b^\OUT}$,
$p_3=\pi_{a^\IN c^\OUT e^\OUT}$,
$p_4=\pi_{b^\IN c^\OUT e^\OUT}$,
$p_5=\pi_{a^\IN c^\OUT f^\OUT}$,
$p_6=\pi_{b^\IN c^\OUT f^\OUT}$,
$p_7=\pi_{a^\IN d^\OUT e^\OUT}$,
$p_8=\pi_{b^\IN d^\OUT e^\OUT}$,
$p_9=\pi_{a^\IN d^\OUT f^\OUT}$,
$p_{10}=\pi_{b^\IN d^\OUT f^\OUT}$,
$p_{11}=\pi_{c^\IN e^\IN}$,
$p_{12}=\pi_{c^\IN f^\IN}$,
$p_{13}=\pi_{d^\IN e^\IN}$, 
$p_{14}=\pi_{d^\IN f^\IN}$.
}
\end{figure}

\section{Connection graphs and distributed places}
\label{sec-cg} 

According to Proposition~\ref{prop-ewceiu}($5$), the set of places of 
a marked net derived from a box expression
can be partitioned into distributed places. It might therefore seem reasonable
to apply our reduction result, Theorem~\ref{th-main},
to derive an effective strategy. 
Such an approach is of course possible, though
it suffers from a crucial drawback 
stemming from the fact that the number of places of $\BOX(E)$ 
can be exponential in the size of $E$,
and so constructing $\BOX(E)$ explicitly may be computationally too expensive.
We therefore propose a different 
way of generating a reduced version of $\BOX(E)$ based on Theorem~\ref{th-main}, 
which
employs a novel (linear) representation of the set of places of $\BOX(E)$.

To carry out effective reduction based on distributed places,
we will employ \emph{connection graphs}
which are undirected graphs $G=\FA{V,A}$ without self-loops, 
where $V$ is a finite set of vertices of the 
form $a^\IN$ and $a^\OUT$, and 
$A$ is a set of edges. 
The empty connection graph $\FA{\es,\es}$ is denoted by $\es$, 
and a singleton connection graph $\FA{\{v\},\es}$ by $v$. 
We also have the following:
\begin{itemize} 

\item The \emph{union} of two disjoint
connection graphs, $G$ and $G'$, is denoted by $G\PLUS G'$. 
Moreover, for a set $\Gamma=\{G_1,\dots,G_k\}$ of pairwise 
disjoint connection graphs, we use
$\PLUS\Gamma$ to denote the connection graph $G_1\PLUS\dots\PLUS G_k$.

\item
The \emph{join} of two disjoint connection graphs, 
$G $ and $G'$, is the
graph $G\join G'$ constructed from their union by 
adding an edge between every vertex of $G$ and every vertex of $G'$.

\item
A \emph{clique} in a connection graph $G$ is a non-empty set of vertices $\clq$
which are pairwise connected by edges.
Moreover, $\clq$ is a \emph{maximal clique} 
if it is not a proper subset of any other clique in $G$
(in particular, each isolated vertex is a maximal clique). 
We denote this by 
$\clq\in\maxCLIQUE(G)$.

\item 
Each maximal clique
$\clq =\{a_1^\IN,\dots,a_m^\IN,b_1^\OUT,\dots,b_n^\OUT\}$ 
\emph{generates} a place
$\pi_\clq =\pi_{a_1^\IN \dots a_m^\IN b_1^\OUT \dots b_n^\OUT}$.
Moreover,
$\Upsilon_G=\{\pi_\clq \mid \clq \in\maxCLIQUE(G)\}$
are the \emph{places generated} by a connection graph $G$.
\end{itemize}

Intuitively, the edges of a connection graph express three kinds of relationships
between actions occurring in a box expression: 
(i) an edge between $a^\OUT$ and $b^\OUT$ means that $a$ and $b$ are in 
\emph{forward conflict};
(ii) an edge between $a^\IN$ and $b^\IN$ means that $a$ and $b$ are in 
\emph{backward conflict}; and 
(iii) an edge between $a^\IN$ and $b^\OUT$ means that $b$ is
\emph{causally dependent} on $a$. Hence,
a clique $\clq =\{a_1^\IN,\dots,a_k^\IN,b_1^\OUT,\dots,b_l^\OUT\}$
represents $(k+l)(k+l-1)/2$ relationships which can all be succinctly 
implemented by a single place with
input transitions $a_1,\dots,a_k$ and output transitions $b_1,\dots,b_l$,
\ie by $\pi_\clq$ defined above. 

In this paper, the maximal cliques of connection graphs 
will provide implicit representations of places
in marked nets derived from box expressions.
Moreover, the maximal cliques of \emph{connected} connection graphs 
will provide implicit representations of distributed places
in marked nets derived from box expressions. 

The next result captures a crucial link between the
union and cross-product operations on 
separated distributed places,
and the union and join operations on disjoint connection graphs.

\begin{proposition}
\label{prop-comp1xddd}
Let $G$ and $G'$ be two connection graphs
such that $\Upsilon_G$ and $\Upsilon_{G'}$ 
are separated pure distributed places.
Then:
\begin{enumerate}
\item 
$\Upsilon_{G\PLUS G'}$ is a pure distributed place satisfying 
$\Upsilon_{G\PLUS G'}=\Upsilon_G\oplus\Upsilon_{G'}$
provided that $\PRE{(\Upsilon_G)}=\PRE{(\Upsilon_{G'})}=\es$ or
$\POST{(\Upsilon_G)}=\POST{(\Upsilon_{G'})}=\es$.
\item 
$\Upsilon_{G{\join}G'}$ is a pure distributed place satisfying 
$\Upsilon_{G{\join}G'}=\Upsilon_G\otimes\Upsilon_{G'}$.
\end{enumerate} 
\end{proposition} 

\begin{proof}
It follows from Corollaries~\ref{cprop-union} and~\ref{cth-cpr}
as well as the following two immediate properties: 
\[
\begin{array}{lcl}
\maxCLIQUE(K\PLUS K') &=&
\maxCLIQUE(K)\cup \maxCLIQUE(K')
\\
\maxCLIQUE(K\join K') &=&
\{\clq\cup \clq'\mid \clq\in\maxCLIQUE(K) \wedge \clq'\in\maxCLIQUE(K')\}\;,
\end{array} 
\] 
where $K$ and $K'$ are disjoint connection graphs.
\end{proof} 

As a consequence of this result, 
the distributed places in $\DP(E)$
obtained by successive applications
of the cross-product and union can be
generated by connection graphs obtained by successive applications
of the join and union operations.
The connection graphs are constructed by first associating 
with each box expression $E$ a triple of sets of connection graphs, 
$\Gamma_E=\FA{\Gamma_E^e,\Gamma_E^x, \Gamma_E^i}$, 
following the syntax of box expressions: 
\begin{equation}
\label{eq-jewe}
\begin{array}{lcl}
\Gamma_a
&=&
 \FA{\{a^\OUT\},\{a^\IN\}, \es}
 \\
 \Gamma_{E\PAR F}
 &=&
 \FA{\Gamma_E^e\cup \Gamma_F^e, 
 \Gamma_E^x\cup \Gamma_F^x, 
 \Gamma_E^i\cup \Gamma_F^i}
\\
\Gamma_{E\SEQ F}
&=&
 \FA{\Gamma_E^e, \Gamma_F^x, 
 \Gamma_E^i\cup \Gamma_F^i\cup 
 \{\PLUS\Gamma_E^x{\join}\PLUS\Gamma_F^e\}}
 \\
 \Gamma_{E\CHOICE F}
 &=&
 \FA{\PLUS\Gamma_E^e{\join}\PLUS\Gamma_F^e, 
 \PLUS\Gamma_E^x{\join}\PLUS\Gamma_F^x, 
 \Gamma_E^i\cup \Gamma_F^i}
 \\
 \Gamma_{\ITER{E}{F}{G}}
 &=& 
 \FA{\Gamma_E^e, 
 \Gamma_G^x, 
 \Gamma_E^i\cup \Gamma_F^i\cup\Gamma_G^i\cup
 \{(\PLUS\Gamma_E^x{\join}\PLUS\Gamma_F^x){\join}
 (\PLUS\Gamma_F^e{\join}\PLUS\Gamma_G^e)\} 
 }
\;.
\end{array}
\end{equation}
We then define the \emph{connection graph} of $E$
as $\CG(E)=(\PLUS\Gamma_E^e)\PLUS (\PLUS\Gamma_E^x)\PLUS (\PLUS\Gamma_E^i)$.

\begin{proposition} 
\label{pr-jrvenj}
Let $E$ be a box expression defined by 
the syntax in Eq.\eqref{eq-baaaa}.
\begin{enumerate}
\item 
$\Gamma_E^e$, $\Gamma_E^x$, and $\Gamma_E^i$ are 
pairwise disjoint sets of non-empty connected
connection graphs.
\item
$\Gamma_E^e\cup\Gamma_E^x\cup\Gamma_E^i$
is a set of pairwise disjoint connection graphs.
\item 
$\Delta_E^\alpha=\{\Upsilon_G\mid G\in\Gamma_E^\alpha\}$
and $\Upsilon_E^\alpha=\Upsilon_{\PLUS\Gamma_E^\alpha}$ (for $\alpha\in\{e,x,i\}$).
\item 
$\DP(E)
=\{\Upsilon_G\mid G\in\Gamma_E^e\cup\Gamma_E^x\cup\Gamma_E^i\}$.
\item 
$\PLACES(E)
=\bigcup\{\Upsilon_G\mid G\in\Gamma_E^e\cup\Gamma_E^x\cup\Gamma_E^i\}
=\Upsilon_{\CG(E)}$.
\end{enumerate} 
\end{proposition} 

\begin{proof}
The result
follows by simultaneous induction on the structure 
of $E$ and the syntax-driven definitions 
Eqs.(\ref{eq-baaaa},\ref{eq-fefw},\ref{eq-fefwdd},\ref{eq-jewe}), using Proposition~\ref{prop-comp1xddd} 
and the assumption that 
no action is used more than once in a given box expression.
Note that the distributed places composed 
when deriving $\Delta_E$ are separated 
since, in particular, 
$\Upsilon^e_E$ and $\Upsilon^x_E$ are separated distributed places
for $E$ 
generated by the second line of the syntax in Eq.\eqref{eq-baaaa}
(see Proposition~\ref{prop-ewceiu}(2)). 
Note also that each new connection graph constructed in 
Eq.(\ref{eq-jewe}) is connected since the join operation applied to 
two non-empty connection graphs yields a connected connection graph,
and all the component connection graphs involved in 
the constructions in Eq.(\ref{eq-jewe})
are non-empty due to Proposition~\ref{prop-ewceiu}(2).
\end{proof} 
Note that the last result implies that each distributed 
place in $\DP(E)$ is generated by a connected component 
of the connection graph $\CG(E)$.

The connection graph of a box expression $E$ 
can be represented as an expression built following 
Eq.\eqref{eq-jewe} from singleton 
and empty connection graphs using the union and join operators. 
The size of such an expression is linear in the size of $E$. 

\section{Reduction based on connection graphs }
\label{sec-cgss} 

Having defined the connection graph $\CG(E)$ of a 
box expression $E$ which 
can be used to generate 
all the distributed places in $\DP(E)$,
we now present a method to retain sufficiently
many maximal cliques of $\CG(E)$ to make Theorem~\ref{th-main} work.
The method is based on a solution of the edge clique cover problem for $\CG(E)$. 
\begin{quote}
The \emph{edge clique cover problem} (\textsc{eccp}) 
for a connection graph 
consists in finding the smallest number 
of maximal cliques such that: (i) for every edge, there is at least one clique 
that contains both endpoints of this edge; and (ii) 
each isolated vertex is contained in a singleton maximal clique. 
Moreover, a 
\emph{partial} \textsc{eccp} (\textsc{peccp})
is a variant of \textsc{eccp}, 
where some of the edges are assumed to be already covered, though
the maximal cliques used in the cover may still use such edges. 
\end{quote}
We then formulate a 
three-stage procedure for reducing the marked net translated from a box expression 
$E$ defined by the syntax in Eq.\eqref{eq-baaaa}.
 
\begin{itemize}
\item []\textsc{Stage i:} 
Derive $\CG(E)$ using Eq.\eqref{eq-jewe}.

\item []\textsc{Stage ii:} 
Apply any algorithm for finding a solution 
$\mathcal C$ of \textsc{eccp} for $\CG(E)$. 

\item []\textsc{Stage iii:} 
Construct a tuple
$N'=\FA{P',T',\Flow',M'_\init}$,
where:
\begin{equation}
\label{eq-ocoernc}
\begin{array}{lcl}
P'&=&\{\pi_\clq\mid \clq\in \mathcal C\}\\
T'&=&\bigcup\{\PRE{\pi_\clq}\cup\POST{\pi_\clq}\mid \clq\in \mathcal C\}
\\
\Flow'&=&\bigcup\{\PRE{\pi_\clq}\times\{\pi_\clq\}
\mid \clq\in \mathcal C\}
~\cup~
\bigcup\{\{\pi_\clq\}\times\POST{\pi_\clq}\mid \clq\in \mathcal C\}
\\
M'_\init&=&\{\pi_\clq \mid\clq\in \mathcal C \wedge \PRE{\pi_\clq}=\es\}\;.
\end{array}
\end{equation}
\end{itemize} 

The soundness of the above procedure is established in 
the next result which demonstrates that the reachability 
graph of the reduced safe marked net 
is isomorphic to the reachability graph of 
the original marked net generated by the box expression.

\begin{theorem}
\label{th-hddshh}
Let $E$ be a box expression 
defined by the syntax in Eq.\eqref{eq-baaaa}, and 
$\NN'$ be a tuple defined in Eq.\eqref{eq-ocoernc}.
Then $\NN'$ is a safe marked net such that 
$\RG_{\NN'}\cong\RG_{\BOX(E)}$. 
\end{theorem} 

\begin{proof} 
This proof consists of a single application of Theorem~\ref{th-main}
to the marked net generated from $E$ and the marked net constructed in 
\textsc{Stage~i--Stage~iii}.

Let $\NN=\BOX(E)=\FA{P,T,\Flow,M_\init}$. 
Moreover, let
$\PC_p=\PC$ and $\PC'_p=\PC_p\cap P'=\{\pi_\clq\in \PC_p\mid \clq\in \mathcal C\}$, for every distributed place 
$\PC\in\DP(E)$ and every place $p\in \PC$.

Below we show that all the assumptions needed to apply 
Theorem~\ref{th-main} for $\NN$ and $\NN'$ are satisfied.
Note that, throughout the proof, 
$\NN$, $\NN'$, $\PC_p$, $\PC'_p$, $M_\init'$, and $P'$ 
play the same roles as in the formulation of
Theorem~\ref{th-main}.
\begin{itemize} 
\item
\emph{$\NN$ is a safe marked net.}

This holds thanks to Proposition~\ref{prop-ewceiu}(1). 

\item 
\emph{For every $p\in P$, $\PC_p$ is a distributed place 
in $\NN$ comprising $p$.}

By Proposition~\ref{prop-ewceiu}(5), $\DP(E)$ 
is a set of pairwise disjoint distributed places 
covering the places of $P$. Hence, $\PC_p$ is defined and
unique, for every $p\in P$. Clearly, $p\in\PC_p$. 

\item
\emph{For every $p\in P$,
either $\PC_p\cap M_\init=\es$ or $\PC_p\cap M_\init=\PC_p$.}

This holds 
since $M_\init=\Upsilon_E^e=\bigcup\Delta_E^e$ and $\Delta_E^e\subseteq\DP(E)$ and, 
by Proposition~\ref{prop-ewceiu}(5), 
the distributed places in $\DP(E)$ are pairwise disjoint.

\item 
\emph{$P'\subseteq P$ is a set of places of $\NN$.}

This follows from Proposition~\ref{pr-jrvenj}(5).

\item
$\NN'=\FA{P',T',\Flow',M'_\init}$
is a marked net such that
$\NN'=\FA{P',T,\Flow|_{T\times P'\cup P'\times T},M_\init\cap P'}$ ($*$). 
Indeed, we have the following:
\begin{itemize}
\item 
$T'=T$ since $\mathcal C$ covers all the vertices of $\CG(E)$, and
so for every $t\in T$ there is $\clq\in\mathcal C$ such that 
$t^\OUT\in\clq$. Hence $t\in\POST{\pi_\clq}\subseteq T'$, and so $T\subseteq T'$.
Moreover, $t$ has at least one pre-place in $\NN'$.
\item
$\PRE{\pi_\clq}\cup\POST{\pi_\clq}\neq\es$ (for each $\pi_\clq\in P'$)
since $\clq\neq\es$, and so $\pi_\clq$ is not isolated.
\item
$\Flow'=\Flow|_{T\times P'\cup P'\times T}$ due to $T'=T$ and 
Eqs.(\ref{eq-vd},\ref{eq-ocoernc}).
\end{itemize} 

\item 
\emph{For every $p\in P$, 
$\PRE{(\PC'_p)}=\PRE{(\PC_p)}$
and 
$\POST{(\PC'_p)}=\POST{(\PC_p)}$, \ie Eq.\eqref{eq-jjdjd} holds.}

Clearly, 
$\PRE{(\PC'_p)}\subseteq\PRE{(\PC_p)}$ and 
$\POST{(\PC'_p)}\subseteq\POST{(\PC_p)}$ hold 
as $\PC'_p\subseteq\PC_p$.
The reverse inclusion also holds 
since, for every $p\in P$ and $t\in\PREPOST{(\PC_p)}$:
\begin{itemize}
\item 
$t\in\PRE{(\PC_p)}$ implies that there is $\pi_\clq\in\PC'_p$ 
such that $t^\IN\in\clq$, and so $t\in\PRE{\pi_\clq}$.

\item 
$t\in\POST{(\PC_p)}$ implies that there is $\pi_\clq\in\PC'_p$ 
such that $t^\OUT\in\clq$, and so $t\in\POST{\pi_\clq}$.
\end{itemize}
Indeed, suppose $t\in\PRE{(\PC_p)}$. By Proposition~\ref{pr-jrvenj},
there is a connected component $G$ of $\CG(E)$ such that 
$\PC_p=\Upsilon_G$. Hence, $t^\IN$ is a vertex of $G$. As a consequence, since 
$G$ is a connected component of $\CG(E)$ and 
$\mathcal C$ is a solution of \textsc{eccp} for $\CG(E)$,
there is $\clq\in\maxCLIQUE(G)\cap\mathcal C$ such that $t^\IN\in\clq$.
We therefore have $\pi_\clq\in P'$, and so $\pi_\clq\in\PC'_p$.
 
If $t\in\POST{(\PC_p)}$ we proceed in a similar way.

\item
\emph{For every $p\in P$,
$\enables(\PC'_p)=\enables(\PC_p)$, \ie Eq.\eqref{eq-jjdjd2} holds.} 

Note that 
$
 \enables(\PC'_p)=\{\FA{t,u}\in T\times T\mid 
 \PC'_p\cap(\POST{t}\cap\PRE{u})\neq\es \}$
and
$\enables(\PC_p)
 =
 \{\FA{t,u}\in T\times T\mid 
 \PC_p\cap(\POST{t}\cap\PRE{u})\neq\es \} 
$
since $\PC_p$ is pure and ($*$) holds.
 
Clearly, $\enables(\PC'_p)\subseteq\enables(\PC_p)$ holds
as $\PC'_p\subseteq\PC_p$.
To prove the reverse inclusion, 
suppose that $\FA{t,u}\in\enables(\PC_p)$.
By Proposition~\ref{pr-jrvenj},
there is a connected component $G$ of $\CG(E)$ such that 
$\PC_p=\Upsilon_G$.
Hence there is $\clq\in\maxCLIQUE(G)$ such that $t^\IN,u^\OUT\in\clq$,
and so there is an edge between $t^\IN$ and $u^\OUT$ in $G$.
As a consequence, since 
$G$ is a connected component of $\CG(E)$ and 
$\mathcal C$ is a solution of \textsc{eccp} for $\CG(E)$,
there is $\clq'\in\maxCLIQUE(G)\cap\mathcal C$ such that $t^\IN,u^\OUT\in\clq'$. 
We therefore have $\pi_{\clq'}\in P'$, and so $\pi_{\clq'}\in\PC'_p$.
Moreover, $\pi_{\clq'}\in \POST{t}\cap\PRE{u}$, and so $\FA{t,u}\in\enables(\PC'_p)$. 

\item
\emph{For every $p\in P$,
$\disables(\PC'_p)=\disables(\PC_p)$, \ie Eq.\eqref{eq-jjdjd3} holds.} 

This can be shown 
in a similar way as above using the coverage of edges between the relevant vertices 
of the form $t^\OUT$.

\item 
\emph{For every $M\in\reach(\NN)$ and $p\in P$, 
there is $M'\in\reach(\NN)$ such that 
$M\xrightarrow[\NN]{} M'$ and $(\PC_p\cap M'=\es~\vee~\PC_p\cap M'=\PC_p)$.}

This
is satisfied due to Proposition~\ref{prop-ewceiu}(3,4) and 
the definition of $\DP(E)$.
\end{itemize}
As a result Theorem~\ref{th-main} can be applied, and 
so $\NN'$ is a safe marked net such that $\RG_\NN\cong\RG_{\NN'}$.
\end{proof}

It is also possible to apply the following two optimisations to 
the procedure proved sound in the last result.

\begin{itemize}
\item[]\textsc{Optimisation~i:} 
 At the end of \textsc{Stage~i}, mark all the edges between pairs of vertices of the 
 form $t^\IN$ as already covered. Then, in \textsc{Stage~ii}, use any 
 algorithm for solving \textsc{peccp}. 
 
\item []\textsc{Optimisation~ii:} 
 At the end of \textsc{Stage~iii},
 delete all the places in $\{p\in P'\mid \POST{p}=\es\}$.
\end{itemize} 

To begin with, 
\textsc{Optimisation~i}
does not change the conclusions of Theorem~\ref{th-hddshh}
as the coverage of edges
between vertices of the form $t^\IN$ was not needed in the proof
of Theorem~\ref{th-hddshh}.
To validate \textsc{Optimisation~ii}, we assume the notation 
as in the proof
of Theorem~\ref{th-hddshh} and additionally denote:
\[
P''=\{p\in P'\mid \POST{p}\neq\es\}
~~\mbox{and}~~
\NN''=\FA{P'',T,\Flow'',M''_\init}=
\FA{P'',T,\Flow|_{T\times P''\cup P''\times T},M_\init\cap P''}\;.
\]
Note that $P'\setminus P''\subseteq \Upsilon^x_E$ 
(see Proposition~\ref{prop-ewceiu}(2)).
Clearly, $\NN''$ is a marked net as 
$\POST{(P'\setminus P'')}=\es$ and so each transition 
has a pre-place in $P''$.
We will show that $\RG_{\NN'}\cong\RG_{\NN''}$. 
 
We first observe that $\fseq(\NN')=\fseq(\NN'')$ which follows from 
$\POST{(P'\setminus P'')}=\es$. 
Hence, by
Proposition~\ref{prop-refrf} and $\fseq(\NN)=\fseq(\NN')$, it suffices 
to show that two firing sequences $\sigma$ and $\sigma'$ lead to the same 
marking in $\NN'$ iff they lead to the same 
marking in $\NN''$
(\ie that Eq.\eqref{eq-ufufgg} holds 
with $\NN_0=\NN'$ and $\NN_1=\NN''$). 
We immediately observe that the ($\Longrightarrow$) implication holds
as $P''\subseteq P'$.
To show that the ($\Longleftarrow$) implication also holds, suppose that
$\sigma$ and $\sigma'$ lead in $\NN''$ to the same marking, 
but in $\NN'$ to two different markings, $M$ and $M'$, respectively.

Suppose that $p\in M\setminus M'$. 
Then we have $p\in\Upsilon_E^x$ as
$\sigma$ and $\sigma'$ lead to the same marking in $\NN''$. 
By $\fseq(\NN)=\fseq(\NN')$ demonstrated in the proof of Theorem~\ref{th-hddshh}
and Proposition~\ref{prop-ewceiu}(3), 
there is 
$\sigma''=t_1\dots t_k$ such that
$M_\init\xrightarrow[\NN]{\sigma'}M''
\xrightarrow[\NN]{\sigma''}\Upsilon_E^x$. 
As $p\notin M'$ and $P'\subseteq P$, we also have $p\notin M''$.
Hence, since $p\in\Upsilon_E^x$, we obtain 
$\PRE{p}\cap \{t_1,\dots,t_k\}\neq \es$.

We then observe that, by $\fseq(\NN)=\fseq(\NN')=\fseq(\NN'')$,
we have $\sigma'\circ\sigma''\in\fseq(\NN'')$.
Moreover, since $\sigma$ and $\sigma'$ lead to the same marking in $\NN''$, 
$\sigma\circ\sigma''\in\fseq(\NN'')=\fseq(\NN')$.
Hence, we have 
$M'_\init\xrightarrow[\NN']{\sigma}M
\xrightarrow[\NN']{t_1}M_1
\dots \xrightarrow[\NN']{t_k}M_k$.
This, together with $p\in M$ and $\POST{p}=\es$,
means that $p\in M_1\cap\dots\cap M_k$.
On the other hand, 
$\PRE{p}\cap \{t_1,\dots,t_k\}\neq \es$ yielding 
a contradiction with $\NN'$ being safe. 
We therefore conclude that $\RG_\NN\cong\RG_{\NN'}\cong\RG_{\NN''}$,
and so \textsc{Optimisation~ii} has been validated. 

We now move to evaluating effectiveness of the reduction method based 
on connection graphs, starting with an immediate observation.

As the number of vertices of $\CG(E)$ is $2n$, where $n$ is 
the number of actions occurring in $E$, one needs at
most $2n(2n-1)/2=n(2n-1)$ maximal cliques to form a suitable 
$\mathcal C$ (note that one can separately cover each edge by any maximal clique), 
and so there are at most $n(2n-1)$ places in the resulting marked
net. 
Sometimes, the number of places can be much smaller, \eg 
$O(\log n)$, as illustrated by the example in Figure~\ref{fi-ee3}($c$)
which is derived from the box expression 
$(a_1\PAR b_1)\CHOICE\dots\CHOICE(a_{10}\PAR b_{10})$.
This clearly compares favourably with the 
result of the original translation in~\cite{BDK-01,DBLP:journals/iandc/BestDK02} 
which can yield $\BOX(E)$ of exponential size. 

Crucially, connection graphs are 
the well-known \emph{complement-reducible graphs (cographs)} \cite{L-71}.
The \textsc{peccp} is trivially in \textsc{np} and many existing 
\textsc{eccp} algorithms
can be adapted to solve \textsc{peccp}, \eg \cite{GGHN-09}. 
Moreover, since \textsc{eccp}
is \textsc{np}-complete for general
graphs and is a special case of \textsc{peccp}, 
it follows that \textsc{peccp} is \textsc{np}-complete for general
graphs. However, for connection graphs
derived for box expressions --- which are cographs,
\ie a very restricted subclass of graphs --- 
many \textsc{np}-complete problems 
become polynomial. Having said that, to our knowledge, the question 
whether \textsc{eccp} or \textsc{peccp} is \textsc{np}-complete
on cographs is still open. 

Note also that for modelling 
control flows the optimality is not
required and the minimality of $\mathcal C$ is not needed to prove Theorem~\ref{th-hddshh}, 
so fast heuristic algorithms computing small 
but not necessarily smallest covers by maximal cliques
would be sufficient. 

\subsection{Examples} 

\begin{figure}[t!]
\begin{center} 
 
\begin{tabular}{c@{~~~~~~~~~~~~}c@{~~~~~~~~~~~~}c}
\Relation[1]
\nodW{ai}{1}{1}{a^\IN}
\nodW{ao}{1}{1.5}{a^\OUT}
\nodE{bi}{2}{1}{b^\IN}
\nodE{bo}{2}{1.5}{b^\OUT}
\nodW{ci}{1}{0}{c^\IN}
\nodW{co}{1}{0.5}{c^\OUT}
\nodE{di}{2}{0}{d^\IN}
\nodE{do}{2}{0.5}{d^\OUT}
\path(ai)edge [-](co);
\path(bi)edge [-](do);
\path(ai)edge [-](do);
\path(bi)edge [-](co);
\end{tikzpicture}
&
\Relation[1]
\nodW{ai}{0}{0.5}{a^\IN}
\nodW{bi}{0}{0}{b^\IN}
\nodE{ci}{1}{0.5}{c^\IN}
\nodE{di}{1}{0}{d^\IN}

\nodW{ao}{0}{1.5}{a^\OUT}
\nodW{bo}{0}{1}{b^\OUT}
\nodE{co}{1}{1.5}{c^\OUT}
\nodE{do}{1}{1}{d^\OUT}

\path(ai)edge [-](ci);
\path(ai)edge [-](di);
\path(bi)edge [-](ci);
\path(bi)edge [-](di);
\path(ao)edge [-](co);
\path(ao)edge [-](do);
\path(bo)edge [-](co);
\path(bo)edge [-](do);
\end{tikzpicture}
&
\Relation[1]
\nodW{ai}{0}{0.5}{a^\IN}
\nodW{bi}{0}{0}{b^\IN}
\nodE{ci}{1}{0.5}{c^\IN}
\nodE{di}{1}{0}{d^\IN}

\nodW{ao}{0}{1.5}{a^\OUT}
\nodW{bo}{0}{1}{b^\OUT}
\nodE{co}{1}{1.5}{c^\OUT}
\nodE{do}{1}{1}{d^\OUT}

\path(ai)edge [dotted](ci);
\path(ai)edge [dotted](di);
\path(bi)edge [dotted](ci);
\path(bi)edge [dotted](di);
\path(ao)edge [-](co);
\path(ao)edge [-](do);
\path(bo)edge [-](co);
\path(bo)edge [-](do);
\end{tikzpicture}
\\ [-3mm]
($a$) $\CG(E_2)$
&
($b$) $\CG(E_3)$
&
($c$) 
\\
\StandardNet[0.55]
 \placN{Aa}{2}{7.5}{1}{}%{q_1}
 \placN{Ab}{4}{7.5}{1}{}%{q_2}
 
 \placW{ac}{ 0}{4.5}{0}{}%{q_3} 
 \placW{bc}{ 2}{4.5}{0}{}%{q_4} 
 \placE{ad}{ 4}{4.5}{0}{}%{q_5} 
 \placE{bd}{ 6}{4.5}{0}{}%{q_6} 

 \placS{Bc}{ 2}{1.5}{0}{}%{q_7}
 \placS{Bd}{ 4}{1.5}{0}{}%{q_8} 
 
 \Whitetran{a}{2}{6}{a}
 \Whitetran{b}{4}{6}{b}
 \Whitetran{c}{2}{3}{c} 
 \Whitetran{d}{4}{3}{d} 

 \diredge{Aa}{a}\diredge{Ab}{b}
 \diredge{c}{Bc}\diredge{d}{Bd} 
 \diredge{a}{ac}\diredge{a}{ad} 
 \diredge{b}{bc}\diredge{b}{bd}
 \diredge{ac}{c}\diredge{bc}{c} 
 \diredge{ad}{d}\diredge{bd}{d} 
\end{tikzpicture} 
& 
\StandardNet[0.55] 
 \placN{ac}{ 0}{4}{1}{}%{r_1} 
 \placN{ad}{ 2}{4}{1}{}%{r_2} 
 \placN{bc}{ 4}{4}{1}{}%{r_3} 
 \placN{bd}{ 6}{4}{1}{}%{r_4} 

 \placS{Ba}{ 0}{0}{0}{}%{r_5}
 \placS{Bb}{ 2}{0}{0}{}%{r_6}
 \placS{Bc}{ 4}{0}{0}{}%{r_7}
 \placS{Bd}{ 6}{0}{0}{}%{r_8} 
 
 \Whitetran{a}{0}{2}{a}
 \Whitetran{b}{2}{2}{b}
 \Whitetran{c}{4}{2}{c} 
 \Whitetran{d}{6}{2}{d} 

 \diredge{a}{Ba}\diredge{c}{Ba}
 \diredge{a}{Bb}\diredge{d}{Bb}
 \diredge{c}{Bc}\diredge{b}{Bc} 
 \diredge{b}{Bd}\diredge{d}{Bd} 
 \diredge{ac}{c}\diredge{ac}{a} 
 \diredge{ad}{d}\diredge{ad}{a} 
 \diredge{bc}{b}\diredge{bc}{c}
 \diredge{bd}{b}\diredge{bd}{d} 
\end{tikzpicture} 
&
\StandardNet[0.55] 
 \placN{ac}{ 0}{4}{1}{}%{r_1} 
 \placN{ad}{ 2}{4}{1}{}%{r_2} 
 \placN{bc}{ 4}{4}{1}{}%{r_3} 
 \placN{bd}{ 6}{4}{1}{}%{r_4} 

 \placS{Ba}{ 0}{0}{0}{}%{r_5} 
 \placS{Bd}{ 6}{0}{0}{}%{r_8} 
 
 \Whitetran{a}{0}{2}{a}
 \Whitetran{b}{2}{2}{b}
 \Whitetran{c}{4}{2}{c} 
 \Whitetran{d}{6}{2}{d} 

 \diredge{a}{Ba}\diredge{b}{Bd}
 \diredge{c}{Ba}\diredge{d}{Bd} 
 
 \diredge{ac}{c}\diredge{ac}{a} 
 \diredge{ad}{d}\diredge{ad}{a} 
 \diredge{bc}{b}\diredge{bc}{c}
 \diredge{bd}{b}\diredge{bd}{d} 
\end{tikzpicture} 
\\[-3mm]
($d$) $\BOX(E_2)$&
($e$) $\BOX(E_3)$&
($f$) 
\end{tabular} 
\end{center}
\caption{
\label{fig-773rtr}
($a$) The 
connection graph and ($d$) marked net
generated for $E_2=(a\PAR b)\SEQ(c\PAR d)$.
($b$) The 
connection graph and ($e$) marked net
generated for $E_3=(a\PAR b)\CHOICE(c\PAR d)$.
($c$) The connection graph for $E_3$ with dotted edges 
already covered and ($f$) the resulting reduced marked net.
}
\end{figure} 

Note: Except for the last two examples,
we do not apply \textsc{Optimisation~ii} which removes
exit places in order to facilitate comparison with the standard construction.

For the box expression 
$E_2=(a\PAR b)\SEQ (c\PAR d)$ 
with $\BOX(E_2)$
depicted in Figure~\ref{fig-773rtr}($d$), we have:
\[ 
\begin{array}{lcl}
 \Gamma_{E_2}
 &=&
 \FA{\Gamma_{a\PAR b}^e, 
 \Gamma_{c\PAR d}^x, 
 \Gamma_{a\PAR b}^i\cup\Gamma_{c\PAR d}^i\cup
 \{\PLUS\Gamma_{a\PAR b}^x{\join}\PLUS\Gamma_{c\PAR d}^e\}}
 \\
 &=&
 \FA{\Gamma_a^e\cup\Gamma_b^e, 
 \Gamma_c^x\cup\Gamma_d^x, 
 \Gamma_a^i\cup\Gamma_b^i\cup\Gamma_c^i\cup\Gamma_d^i\cup
 \{(\PLUS(\Gamma_a^x\cup\Gamma_b^x){\join}\PLUS(\Gamma_c^e\cup\Gamma_d^e)\}} 
\\ 
 &=&
 \FA{\{a^\OUT\}\cup\{ b^\OUT\}, \{c^\IN\}\cup\{d^\IN\}, 
 \es\cup\es\cup\es\cup\es\cup
 \{\PLUS\{a^\IN, b^\IN\}{\join}\PLUS\{c^\OUT,d^\OUT\}\}} 
\\ 
 &=&
 \FA{\{a^\OUT,b^\OUT\}, \{c^\IN,d^\IN\}, 
 \{(a^\IN\PLUS b^\IN){\join}(c^\OUT\PLUS d^\OUT)\}} 
 \;. 
\end{array}
\]
Hence
$\CG(E_2)=
 a^\OUT\PLUS b^\OUT\PLUS c^\IN\PLUS d^\IN\PLUS
 ((a^\IN\PLUS b^\IN){\join}(c^\OUT\PLUS d^\OUT))$
 as shown in Figure~\ref{fig-773rtr}($a$). 
The only solution of 
\textsc{eccp} yields the following cover: 
$\{a^\OUT\}$,
$\{b^\OUT\}$,
$\{c^\IN\}$,
$\{d^\IN\}$,
$\{a^\IN,c^\OUT\}$,
$\{a^\IN,d^\OUT\}$,
$\{b^\IN,c^\OUT\}$, and
$\{b^\IN,d^\OUT\}$.
The resulting marked net is in this case the same 
as the marked net derived using the 
standard translation shown in Figure~\ref{fig-773rtr}($d$).

Figure~\ref{fig-773rtr}($b$) depicts the 
connection graph for the box expression 
$E_3=(a\PAR b)\CHOICE(c\PAR d)$.
The only outcome of solving 
\textsc{eccp} yields the following cover: 
$\{a^\OUT,c^\OUT\}$,
$\{a^\OUT,d^\OUT\}$,
$\{b^\OUT,c^\OUT\}$,
$\{b^\OUT,d^\OUT\}$,
$\{a^\IN,c^\IN\}$,
$\{a^\IN,d^\IN\}$,
$\{b^\IN,c^\IN\}$, and
$\{b^\IN,d^\IN\}$.
The resulting marked net is in this case the same 
as that derived using the 
standard translation shown in Figure~\ref{fig-773rtr}($e$).
Moreover, Figure~\ref{fig-773rtr}($c$) 
indicates the connection graph of $E_3$ after applying 
\textsc{Optimisation~i}. A possible outcome of solving 
\textsc{peccp} yields the following cover:
$\{a^\OUT,c^\OUT\}$,
$\{a^\OUT,d^\OUT\}$,
$\{b^\OUT,c^\OUT\}$,
$\{b^\OUT,d^\OUT\}$,
$\{a^\IN,c^\IN\}$, and
$\{b^\IN,d^\IN\}$.
The resulting marked net is depicted in Figure~\ref{fig-773rtr}($f$).

\begin{figure}[ht!]
\begin{center} 

\begin{tabular}{c@{~}c}
\Relation[1]
\nodW{ao}{0}{-0.5}{a^\OUT}
\nodE{bo}{3}{-0.5}{b^\OUT}
\nodW{ai}{0}{2}{a^\IN}
\nodE{bi}{3}{2}{b^\IN}
\nodW{co}{0}{1}{c^\OUT} 
\nodW{do}{1}{0}{d^\OUT} 
\nodE{eo}{2}{0}{e^\OUT} 
\nodE{fo}{3}{1}{f^\OUT} 

\nodW{ci}{0}{-1}{c^\IN} 
\nodW{di}{0}{-1.5 }{d^\IN} 
\nodE{ei}{3}{-1.5 }{e^\IN} 
\nodE{fi}{3}{-1}{f^\IN}

\path(ai)edge [-](co);
\path(ai)edge [-](do);
\path(ai)edge [-](eo);
\path(ai)edge [-](fo); 
\path(bi)edge [-](co);
\path(bi)edge [-](do);
\path(bi)edge [-](eo);
\path(bi)edge [-](fo); 

\path(co)edge [-](fo); 
\path(co)edge [-](eo); 
\path(do)edge [-](fo); 
\path(do)edge [-](eo); 

\path(ci)edge [-](fi); 
\path(ci)edge [-](ei); 
\path(di)edge [-](fi); 
\path(di)edge [-](ei);
\end{tikzpicture}
&
\Relation[1]
\nodW{ao}{0}{-0.5}{a^\OUT}
\nodE{bo}{3}{-0.5}{b^\OUT}
\nodW{ai}{0}{2}{a^\IN}
\nodE{bi}{3}{2}{b^\IN}
\nodW{co}{0}{1}{c^\OUT} 
\nodW{do}{1}{0}{d^\OUT} 
\nodE{eo}{2}{0}{e^\OUT} 
\nodE{fo}{3}{1}{f^\OUT} 

\nodW{ci}{0}{-1}{c^\IN} 
\nodW{di}{0}{-1.5 }{d^\IN} 
\nodE{ei}{3}{-1.5}{e^\IN} 
\nodE{fi}{3}{-1}{f^\IN}

\path(ai)edge [-](co);
\path(ai)edge [-](do);
\path(ai)edge [-](eo);
\path(ai)edge [-](fo); 
\path(bi)edge [-](co);
\path(bi)edge [-](do);
\path(bi)edge [-](eo);
\path(bi)edge [-](fo); 

\path(co)edge [-](fo); 
\path(co)edge [-](eo); 
\path(do)edge [-](fo); 
\path(do)edge [-](eo); 

\path(ci)edge [dotted](fi); 
\path(ci)edge [dotted](ei); 
\path(di)edge [dotted](fi); 
\path(di)edge [dotted](ei);
\end{tikzpicture}
\\ 
($a$) $\CG(E_4)$ & 
($b$) 
\\
\StandardNet[0.55] 
 \placN{Aa}{ 4}{8}{1}{}%{p_1}
 \placN{Ab}{10}{8}{1}{}%{p_2}
 
 \placW{cea}{ 0}{4.5}{0}{}%{p_3\!\!} 
 \placW{ceb}{ 2}{4.5}{0}{}%{p_4\!\!} 
 \placW{cfa}{ 4}{4.5}{0}{}%{p_5\!\!} 
 \placW{cfb}{ 6}{4.5}{0}{}%{p_6\!\!} 
 \placE{dea}{ 8}{4.5}{0}{}%{\!\!p_7} 
 \placE{deb}{10}{4.5}{0}{}%{\!\!p_8} 
 \placE{dfa}{12}{4.5}{0}{}%{\!\!p_9} 
 \placE{dfb}{14}{4.5}{0}{}%{\!\!p_{10}} 

 \placS{cebx}{ 2}{-0.5}{0}{}%{p_{12}} 
 \placS{cfax}{ 4}{-0.5}{0}{}%{p_{13}} 
 \placS{debx}{10}{-0.5}{0}{}%{p_{16}} 
 \placS{dfax}{12}{-0.5}{0}{}%{p_{17}} 

 \Whitetran{a}{ 4}{6.5}{a}
 \Whitetran{b}{10}{6.5}{b}
 \Whitetran{c}{ 2}{2}{c} 
 \Whitetran{d}{ 4}{2}{d}
 \Whitetran{e}{10}{2}{e}
 \Whitetran{f}{12}{2}{f}

 \diredge{Aa}{a}\diredge{Ab}{b} 
 
 \diredge{a}{cea}\diredge{a}{cfa}\diredge{a}{dea}\diredge{a}{dfa}
 \diredge{b}{ceb}\diredge{b}{cfb}\diredge{b}{deb}\diredge{b}{dfb}
 \diredge{ceb}{c}\diredge{cfb}{c}\diredge{cea}{c}\diredge{cfa}{c}
 \diredge{deb}{d}\diredge{dfb}{d}\diredge{dea}{d}\diredge{dfa}{d}
 \diredge{ceb}{e}\diredge{cfb}{f}\diredge{cea}{e}\diredge{cfa}{f}
 \diredge{deb}{e}\diredge{dfb}{f}\diredge{dea}{e}\diredge{dfa}{f}
 
 \diredge{c}{cebx} \diredge{c}{cfax}
 \diredge{d}{debx} \diredge{d}{dfax}
 \diredge{e}{cebx} \diredge{f}{cfax}
 \diredge{e}{debx} \diredge{f}{dfax}
\end{tikzpicture}
&
\StandardNet[0.55] 
 \placN{Aa}{ 4}{8}{1}{}%{p_1}
 \placN{Ab}{6}{8}{1}{}%{p_2}
 
 \placW{ceb}{ 2}{4.5}{0}{}%{p_4\!\!} 
 \placW{cfa}{ 4}{4.5}{0}{}%{p_5\!\!} 
 \placE{deb}{6}{4.5}{0}{}%{\!\!p_8} 
 \placE{dfa}{8}{4.5}{0}{}%{\!\!p_9} 
 
 \placS{cebx}{ 2}{-.5}{0}{}%{p_{12}} 
 \placS{dfax}{8}{-.5}{0}{}%{p_{17}} 
 
 \Whitetran{a}{4}{6.5}{a}
 \Whitetran{b}{6}{6.5}{b}
 \Whitetran{c}{2}{2}{c} 
 \Whitetran{d}{4}{2}{d}
 \Whitetran{e}{6}{2}{e}
 \Whitetran{f}{8}{2}{f}

 \diredge{Aa}{a}\diredge{Ab}{b} 
 
 \diredge{a}{cfa} \diredge{a}{dfa}
 \diredge{b}{deb}\diredge{b}{ceb}
 \diredge{ceb}{c} \diredge{cfa}{c}
 \diredge{deb}{d} \diredge{dfa}{d} 
 \diredge{ceb}{e} \diredge{cfa}{f}
 \diredge{deb}{e} \diredge{dfa}{f}

 \diredge{c}{cebx} 
 \diredge{d}{dfax}
 \diredge{e}{cebx} 
 \diredge{f}{dfax} 
\end{tikzpicture} 
\\[-3mm]
($c$) $\BOX(E_4)$&
($d$)
\\[4mm]
\Relation[1] 
\nodN{ai}{0.5}{2}{a^\IN}
\nodN{ei}{1.5}{2}{e^\IN}
\nodN{bi}{2.5}{2}{b^\IN}

\nodN{ci}{0.5}{3.2}{c^\IN}
\nodN{eo}{1.5}{3.2}{e^\OUT}
\nodN{di}{2.5}{3.2}{d^\IN}

\nodW{co}{0.5}{0}{c^\OUT} 
\nodW{do}{0}{1}{d^\OUT} 
\nodE{fo}{3}{1}{f^\OUT} 
\nodE{go}{2.5}{0}{g^\OUT} 

\nodS{ao}{0}{-0.5}{a^\OUT} 
\nodS{bo}{0.7}{-0.5}{b^\OUT} 
\nodS{fi}{2.3}{-0.5}{f^\IN}
\nodS{gi}{3}{-0.5}{g^\IN}

\path(ci)edge [-](eo);\path(di)edge [-](eo); 

\path(bi)edge [dotted](ei);\path(ai)edge [dotted](ei); 
\path(co)edge [-](fo);\path(do)edge [-](fo);
\path(co)edge [-](go);\path(do)edge [-](go);
\path(ai)edge [-](co);\path(ai)edge [-](do);\path(ai)edge [-](fo);\path(ai)edge [-](go);
\path(ei)edge [-](co);\path(ei)edge [-](do);\path(ei)edge [-](fo);\path(ei)edge [-](go);
\path(bi)edge [-](co);\path(bi)edge [-](do);\path(bi)edge [-](fo);\path(bi)edge [-](go);
\end{tikzpicture}
&
\StandardNet[0.55] 
 \placW{xxx}{0}{6.5}{1}{} 
 \placW{yyy}{6}{6.5}{1}{} 
 \placW{abd}{0}{3}{0}{} 
 \placW{abe}{2}{3}{0}{}
 \placW{acd}{4}{3}{0}{}
 \placW{ace}{6}{3}{0}{}

 \placW{ab}{1}{1}{0}{}
 \placW{ac}{5}{1}{0}{} 

 \Whitetran{A}{0} {5}{a}
 \Whitetran{B}{6} {5}{b}
 \Whitetran{a}{3} {1}{e}
 \Whitetran{b}{-1}{0}{c}
 \Whitetran{c}{7} {0}{d} 
 \Whitetran{d}{-2}{1}{f} 
 \Whitetran{e}{8} {1}{g}

\diredge{a}{abd}\diredge{a}{abe}\diredge{a}{acd}
\diredge{ab}{a}\diredge{ac}{a}\diredge{a}{ace}

\diredge{c}{ac}\diredge{acd}{c}\diredge{ace}{c}
\diredge{b}{ab}\diredge{abd}{b}\diredge{abe}{b}

\diredge{abe}{e}\diredge{acd}{d}\diredge{ace}{e}\diredge{abd}{d}

\diredge{xxx}{A}\diredge{yyy}{B}
\diredge{A}{ace} \diredge{A}{abd}
\diredge{B}{acd}\diredge{B}{abe}
\end{tikzpicture} 
\\
($e$) $\CG(E_5)$&($f$)
\end{tabular}
\end{center}
\caption{
\label{fig-773}
($a$) The 
connection graph and ($c$) marked net
generated for $E_4=(a\PAR b);((c\PAR d)\CHOICE(e\PAR f))$.
($b$) The connection graph for $E_4$ with dotted edges 
already covered, and ($d$) the resulting reduced marked net.
($e$) The connection graph 
and 
($f$) marked net generated for 
$E_5=\ITER{a\PAR b}{(c\PAR d)\SEQ e}{f\PAR g}$. 
}
\end{figure}
 
Figure~\ref{fig-773}($a$) shows the connection graph for 
$E_4=(a\PAR b);((c\PAR d)\CHOICE(e\PAR f))$,
and Figure~\ref{fig-773}($c$) shows the corresponding marked net. 
Figure~\ref{fig-773}($b$) shows the same connection 
graph after designating some edges as already covered
(\textsc{Optimisation~i}). A possible outcome of solving 
\textsc{peccp} yields the following cover:
$\{a^\OUT\}$,
$\{b^\OUT\}$,
$\{c^\IN,e^\IN\}$,
$\{d^\IN,f^\IN\}$,
$\{a^\IN,c^\OUT,f^\OUT\}$,
$\{a^\IN,d^\OUT,f^\OUT\}$,
$\{b^\IN,c^\OUT,e^\OUT\}$, and
$\{b^\IN,d^\OUT,e^\OUT\}$.
The resulting marked net is shown in Figure~\ref{fig-773}($d$). 

For the box expression 
$E_5=\ITER{a\PAR b}{(c\PAR d)\SEQ e}{f\PAR g}$ we have the following
after noting that 
$\Gamma_{w\PAR z}=\FA{\{w^\OUT,z^\OUT\},\{w^\IN, z^\IN\},\es}$, for all
actions $w$ and $z$:
\[
\begin{array}{lcl}
 \Gamma_{E_5} 
 &=&
 \FA{\Gamma_{a\PAR b}^e, 
 \Gamma_{f\PAR g}^x, 
 \Gamma_{a\PAR b}^i\cup
 \Gamma_{(c\PAR d)\SEQ e}^i\cup
 \Gamma_{f\PAR g}^i\cup
 \{
 (\PLUS\Gamma_{a\PAR b}^x{\join}\PLUS\Gamma_{(c\PAR d)\SEQ e}^x)
 {\join}
 (\PLUS\Gamma_{f\PAR g}^e{\join}\PLUS\Gamma_{(c\PAR d)\SEQ e}^e)
 \}}
 \\
 &=&
 \FA{
 \{a^\OUT,b^\OUT\}, 
 \{f^\IN,g^\IN\}, 
 \{\PLUS\Gamma_{c\PAR d}^x{\join}\PLUS\Gamma_e^e, 
 ((a^\IN\PLUS b^\IN){\join}\PLUS\Gamma_e^x)
 {\join}
 ((f^\OUT\PLUS g^\OUT){\join}\PLUS\Gamma_{c\PAR d}^e))\}}
 \\
 &=&
 \FA{
 \{a^\OUT,b^\OUT\}, 
 \{f^\IN,g^\IN\}, 
 \{(c^\IN\PLUS d^\IN){\join}e^\OUT, 
 ((a^\IN\PLUS b^\IN){\join} e^\IN)
 {\join}
 (( f^\OUT\PLUS g^\OUT){\join}(c^\OUT\PLUS d^\OUT))\}}
 \;.
\end{array}
\]
Hence, the connection graph of $E_5$ is as follows:
\[ 
 a^\OUT\PLUS b^\OUT\PLUS
 f^\IN\PLUS g^\IN\PLUS 
 ((c^\IN\PLUS d^\IN){\join}e^\OUT)\PLUS 
 ((a^\IN\PLUS b^\IN){\join} e^\IN)
 {\join}
 (((f^\OUT\PLUS g^\OUT){\join}(c^\OUT\PLUS d^\OUT))))\;. 
\]
It is shown in Figure~\ref{fig-773}($e$) with dotted lines indicating 
edges already covered. 
This cograph has 14 maximal cliques, and the following 10 
provide a minimal solution of \textsc{peccp}:
$\{a^\OUT\}$, 
$\{b^\OUT\}$, 
$\{f^\IN\}$, 
$\{g^\IN\}$, 
$\{c^\IN,e^\OUT\}$,
$\{d^\IN,e^\OUT\}$,
$\{a^\IN, e^\IN, f^\OUT,c^\OUT\}$, 
$\{a^\IN, e^\IN, d^\OUT,g^\OUT\}$,
$\{b^\IN, e^\IN, g^\OUT,c^\OUT\}$, and
$\{b^\IN, e^\IN, d^\OUT,f^\OUT\}$.
Then, after \textsc{Optimisation~ii}, 
we obtain the marked net in Figure~\ref{fig-773}($f$).

Figure~\ref{fig-7reewr}($a$) shows the connection graph for 
$E_6=((a\PAR b)\CHOICE(c\PAR d))\SEQ e$.
The only outcome of solving 
\textsc{eccp} yields the following cover:
$\{e^\IN\}$,
$\{a^\OUT,c^\OUT\}$,
$\{b^\OUT,c^\OUT\}$,
$\{a^\OUT,d^\OUT\}$,
$\{b^\OUT,d^\OUT\}$,
$\{a^\IN,c^\IN,e^\OUT\}$,
$\{b^\IN,c^\IN,e^\OUT\}$,
$\{a^\IN,d^\IN,e^\OUT\}$, and
$\{b^\IN,d^\IN,e^\OUT\}$. 
Figure~\ref{fig-7reewr}($c$) shows the derived marked net.
Then, Figure~\ref{fig-7reewr}($b$) shows 
$\CG(E_6)$ after designating some edges as already covered. 
A possible outcome of solving 
\textsc{peccp} yields the following cover:
$\{e^\IN\}$,
$\{a^\OUT,c^\OUT\}$,
$\{b^\OUT,c^\OUT\}$,
$\{a^\OUT,d^\OUT\}$,
$\{b^\OUT,d^\OUT\}$,
$\{a^\IN,c^\IN,e^\OUT\}$, and
$\{b^\IN,d^\IN,e^\OUT\}$.
The marked net resulting from \textsc{Optimisation~i}
is depicted in Figure~\ref{fig-7reewr}($d$), 
and after \textsc{Optimisation~ii}, 
we obtain the marked net in Figure~\ref{fig-7reewr}($e$).

\begin{figure}[ht!]
\begin{center} 
\begin{tabular}{c@{~~~~~~~~~~~~}c@{~~~~~~~~~~~~}c}
\Relation[1]
\nodW{ai}{0}{0.5}{a^\IN}
\nodW{bi}{0}{0}{b^\IN}
\nodE{ci}{1}{0.5}{c^\IN}
\nodE{di}{1}{0}{d^\IN}
\nodE{eo}{0.5}{-0.5}{e^\OUT}

\nodW{ao}{0}{1.5}{a^\OUT}
\nodW{bo}{0}{1}{b^\OUT}
\nodE{co}{1}{1.5}{c^\OUT}
\nodE{do}{1}{1}{d^\OUT}
\nodE{ei}{0.5}{2}{e^\IN}

\path(ai)edge (ci);
\path(ai)edge (di);
\path(bi)edge (ci);
\path(bi)edge (di);
\path(ao)edge [-](co);
\path(ao)edge [-](do);
\path(bo)edge [-](co);
\path(bo)edge [-](do);
\path(ai)edge [-](eo);
\path(bi)edge [-](eo);
\path(ci)edge [-](eo);
\path(di)edge [-](eo);
\end{tikzpicture}
&
\Relation[1]
\nodW{ai}{0}{0.5}{a^\IN}
\nodW{bi}{0}{0}{b^\IN}
\nodE{ci}{1}{0.5}{c^\IN}
\nodE{di}{1}{0}{d^\IN}
\nodE{eo}{0.5}{-0.5}{e^\OUT}

\nodW{ao}{0}{1.5}{a^\OUT}
\nodW{bo}{0}{1}{b^\OUT}
\nodE{co}{1}{1.5}{c^\OUT}
\nodE{do}{1}{1}{d^\OUT}
\nodE{ei}{0.5}{2}{e^\IN}

\path(ai)edge [dotted](ci);
\path(ai)edge [dotted](di);
\path(bi)edge [dotted](ci);
\path(bi)edge [dotted](di);
\path(ao)edge [-](co);
\path(ao)edge [-](do);
\path(bo)edge [-](co);
\path(bo)edge [-](do);
\path(ai)edge [-](eo);
\path(bi)edge [-](eo);
\path(ci)edge [-](eo);
\path(di)edge [-](eo);
\end{tikzpicture}
\\ 
($a$) $\CG(E_6)$
&
($b$) 
 \\
\StandardNet[0.55] 
 \placN{ac}{ 0}{4}{1}{}%{r_1} 
 \placN{ad}{ 2}{4}{1}{}%{r_2} 
 \placN{bc}{ 4}{4}{1}{}%{r_3} 
 \placN{bd}{ 6}{4}{1}{}%{r_4} 

 \placS{Ba}{ 0}{0}{0}{}%{r_5}
 \placS{Bb}{ 2}{0}{0}{}%{r_6}
 \placS{Bc}{ 4}{0}{0}{}%{r_7}
 \placS{Bd}{ 6}{0}{0}{}%{r_8} 
 \placS{Be}{ 5}{-2}{0}{}%{r_8} 
 
 \Whitetran{a}{0}{2}{a}
 \Whitetran{b}{2}{2}{b}
 \Whitetran{c}{4}{2}{c} 
 \Whitetran{d}{6}{2}{d} 
 \Whitetran{e}{3}{-2}{e}

 \diredge{a}{Ba}\diredge{c}{Ba}
 \diredge{a}{Bb}\diredge{d}{Bb}
 \diredge{c}{Bc}\diredge{b}{Bc} 
 \diredge{b}{Bd}\diredge{d}{Bd} 
 \diredge{ac}{c}\diredge{ac}{a} 
 \diredge{ad}{d}\diredge{ad}{a} 
 \diredge{bc}{b}\diredge{bc}{c}
 \diredge{bd}{b}\diredge{bd}{d} 
 \diredge{Ba}{e}\diredge{Bb}{e}
 \diredge{Bc}{e}\diredge{Bd}{e}
 \diredge{e}{Be}
\end{tikzpicture} 
&
\StandardNet[0.55] 
 \placN{ac}{ 0}{4}{1}{}%{r_1} 
 \placN{ad}{ 2}{4}{1}{}%{r_2} 
 \placN{bc}{ 4}{4}{1}{}%{r_3} 
 \placN{bd}{ 6}{4}{1}{}%{r_4} 

 \placS{Ba}{ 0}{0}{0}{}%{r_5} 
 \placS{Bd}{ 6}{0}{0}{}%{r_8} 
 \placS{Be}{ 5}{-2}{0}{}%{r_8} 
 
 \Whitetran{a}{0}{2}{a}
 \Whitetran{b}{2}{2}{b}
 \Whitetran{c}{4}{2}{c} 
 \Whitetran{d}{6}{2}{d} 
 \Whitetran{e}{3}{-2}{e}

 \diredge{a}{Ba}\diredge{c}{Ba} 
 \diredge{b}{Bd}\diredge{d}{Bd} 
 \diredge{ac}{c}\diredge{ac}{a} 
 \diredge{ad}{d}\diredge{ad}{a} 
 \diredge{bc}{b}\diredge{bc}{c}
 \diredge{bd}{b}\diredge{bd}{d} 
 \diredge{Ba}{e} 
 \diredge{Bd}{e}
 \diredge{e}{Be}
\end{tikzpicture} 
&
\StandardNet[0.55] 
 \placN{ac}{ 0}{4}{1}{}%{r_1} 
 \placN{ad}{ 2}{4}{1}{}%{r_2} 
 \placN{bc}{ 4}{4}{1}{}%{r_3} 
 \placN{bd}{ 6}{4}{1}{}%{r_4} 

 \placS{Ba}{ 0}{0}{0}{}%{r_5} 
 \placS{Bd}{ 6}{0}{0}{}%{r_8} 
 
 \Whitetran{a}{0}{2}{a}
 \Whitetran{b}{2}{2}{b}
 \Whitetran{c}{4}{2}{c} 
 \Whitetran{d}{6}{2}{d} 
 \Whitetran{e}{3}{-2}{e}

 \diredge{a}{Ba}\diredge{c}{Ba} 
 \diredge{b}{Bd}\diredge{d}{Bd} 
 \diredge{ac}{c}\diredge{ac}{a} 
 \diredge{ad}{d}\diredge{ad}{a} 
 \diredge{bc}{b}\diredge{bc}{c}
 \diredge{bd}{b}\diredge{bd}{d} 
 \diredge{Ba}{e} 
 \diredge{Bd}{e} 
\end{tikzpicture} 
\\ 
($c$) $\BOX(E_6)$&
($d$) &
($e$) 
\end{tabular} 
\end{center}
\caption{
\label{fig-7reewr}
($a$) The 
connection graph and ($c$) marked net
generated for $E_6=((a\PAR b)\CHOICE(c\PAR d))\SEQ e$.
($b$) The connection graph for $E_6$ with dotted edges 
already covered. ($d$) The result of applying
\textsc{Optimisation~i} and ($e$) of applying
\textsc{Optimisation~ii}.
}
\end{figure}

\section{Concluding remarks}
 
In future work we plan to extend the results obtained
in this paper to 
box expressions employing other, practically relevant, 
operators. Moreover, we are interested in developing 
effective algorithms for identifying distributed places 
in general safe marked nets.

The work presented in this paper may contribute to the future
development of tools for constructing formally verifiable and
interpretable models for AI, such as Tsetlin Machines, Binary Neural
Networks, and Logic-Gate Networks.

\bibliographystyle{fundam} 
\bibliography{mybibliography} 

@inproceedings{DBLP:conf/dfg/PadbergU03,
  author       = {Julia Padberg and
                  Milan Urb{\'{a}}sek},
  editor       = {Hartmut Ehrig and
                  Wolfgang Reisig and
                  Grzegorz Rozenberg and
                  Herbert Weber},
  title        = {Rule-Based Refinement of Petri Nets: {A} Survey},
  booktitle    = {Petri Net Technology for Communication-Based Systems - Advances in
                  Petri Nets},
  series       = {Lecture Notes in Computer Science},
  volume       = {2472},
  pages        = {161--196},
  publisher    = {Springer},
  year         = {2003} 
}

@article{ehrig2006transformations,
  title={Transformations of Petri nets},
  author={Ehrig, Hartmut and Hoffmann, Kathrin and Padberg, Julia},
  journal={Electronic notes in theoretical computer science},
  volume={148},
  number={1},
  pages={151--172},
  year={2006},
  publisher={Elsevier}
}

@inproceedings{DBLP:conf/apn/EsparzaS90,
  author       = {Javier Esparza and
                  Manuel Silva Su{\'{a}}rez},
  editor       = {Grzegorz Rozenberg},
  title        = {Top-down synthesis of live and bounded free choice nets},
  booktitle    = {Advances in Petri Nets 1991, Papers from the 11th International Conference
                  on Applications and Theory of Petri Nets, Paris, France, June 1990},
  series       = {Lecture Notes in Computer Science},
  volume       = {524},
  pages        = {118--139},
  publisher    = {Springer},
  year         = {1990} 
}

@inproceedings{DBLP:conf/apn/ChanTSY26,
  author       = {Alex Chan and
                  Mohamed Tarraf and
                  Rishad A. Shafik and
                  Alex Yakovlev},
  editor       = {J{\"{o}}rg Desel and
                  Anna Kalenkova},
  title        = {{ANIMATE:} Automated Framework for Scalable Design of Tsetlin Machines
                  Using 1-Safe Petri Nets},
  booktitle    = {Application and Theory of Petri Nets and Concurrency - 47th International
                  Conference, {PETRI} {NETS} 2026, Hamburg, Germany, June 22-26, 2026,
                  Proceedings},
  series       = {Lecture Notes in Computer Science},
  volume       = {16567},
  pages        = {130--153},
  publisher    = {Springer},
  year         = {2026} 
}

@book{Desel_Esparza_1995, 
 place={Cambridge}, 
 series={Cambridge Tracts in Theoretical Computer Science}, 
 title={Free Choice Petri Nets}, 
 publisher={Cambridge University Press}, 
 author={Desel, Jorg and Esparza, Javier}, 
 year={1995} }

@ARTICLE{Lee1985,
  author={Lee, Kwang-Hyung and Favrel, Joël},
  journal={IEEE Transactions on Systems, Man, and Cybernetics}, 
  title={Hierarchical reduction method for analysis and decomposition of Petri nets}, 
  year={1985},
  volume={SMC-15},
  number={2},
  pages={272-280}}

@article{murata1989petri,
  title={Petri nets: Properties, analysis and applications},
  author={Murata, Tadao},
  journal={Proceedings of the IEEE},
  volume={77},
  number={4},
  pages={541--580},
  year={1989},
  publisher={IEEE}
}

@article{DBLP:journals/jcss/VerbeekWAH10,
  author       = {H. M. W. Verbeek and
                  Moe Thandar Wynn and
                  Wil M. P. van der Aalst and
                  Arthur H. M. ter Hofstede},
  title        = {Reduction rules for reset/inhibitor nets},
  journal      = {J. Comput. Syst. Sci.},
  volume       = {76},
  number       = {2},
  pages        = {125--143},
  year         = {2010} 
}

@inproceedings{DBLP:conf/mfcs/BerthelotR76,
  author       = {G{\'{e}}rard Berthelot and
                  G{\'{e}}rard Roucairol},
  editor       = {Antoni W. Mazurkiewicz},
  title        = {Reduction of Petri-Nets},
  booktitle    = {Mathematical Foundations of Computer Science 1976, 5th Symposium,
                  Gdansk, Poland, September 6-10, 1976, Proceedings},
  series       = {Lecture Notes in Computer Science},
  volume       = {45},
  pages        = {202--209},
  publisher    = {Springer},
  year         = {1976}
}

@inproceedings{DBLP:conf/concur/Desel90,
  author       = {J{\"{o}}rg Desel},
  editor       = {Jos C. M. Baeten and
                  Jan Willem Klop},
  title        = {Reduction and Design of Well-behaved Concurrent Systems},
  booktitle    = {{CONCUR} '90, Theories of Concurrency: Unification and Extension,
                  Amsterdam, The Netherlands, August 27-30, 1990, Proceedings},
  series       = {Lecture Notes in Computer Science},
  volume       = {458},
  pages        = {166--181},
  publisher    = {Springer},
  year         = {1990}
}

@article{DBLP:journals/iandc/Esparza94,
  author       = {Javier Esparza},
  title        = {Reduction and Synthesis of Live and Bounded Free Choice Petri Nets},
  journal      = {Inf. Comput.},
  volume       = {114},
  number       = {1},
  pages        = {50--87},
  year         = {1994}
}

@inproceedings{DBLP:conf/fase/EsparzaH16,
  author       = {Javier Esparza and
                  Philipp Hoffmann},
  editor       = {Perdita Stevens and
                  Andrzej Wasowski},
  title        = {Reduction Rules for Colored Workflow Nets},
  booktitle    = {Fundamental Approaches to Software Engineering - 19th International
                  Conference, {FASE} 2016, Held as Part of the European Joint Conferences
                  on Theory and Practice of Software, {ETAPS} 2016, Eindhoven, The Netherlands,
                  April 2-8, 2016, Proceedings},
  series       = {Lecture Notes in Computer Science},
  volume       = {9633},
  pages        = {342--358},
  publisher    = {Springer},
  year         = {2016}
}

@article{DBLP:journals/jcss/SuzukiM83,
  author       = {Ichiro Suzuki and
                  Tadao Murata},
  title        = {A Method for Stepwise Refinement and Abstraction of Petri Nets},
  journal      = {J. Comput. Syst. Sci.},
  volume       = {27},
  number       = {1},
  pages        = {51--76},
  year         = {1983}
}

@article{DBLP:journals/topnoc/BernardinelloLNP22,
  author       = {Luca Bernardinello and
                  Irina A. Lomazova and
                  Roman Nesterov and
                  Lucia Pomello},
  title        = {Property-Preserving Transformations of Elementary Net Systems Based
                  on Morphisms},
  journal      = {Trans. Petri Nets Other Model. Concurr.},
  volume       = {16},
  pages        = {1--23},
  year         = {2022}
}

@inproceedings{DBLP:conf/apn/SchnoebelenS00,
  author       = {Philippe Schnoebelen and
                  Natalia Sidorova},
  editor       = {Mogens Nielsen and
                  Dan Simpson},
  title        = {Bisimulation and the Reduction of Petri Nets},
  booktitle    = {Application and Theory of Petri Nets 2000, 21st International Conference,
                  {ICATPN} 2000, Aarhus, Denmark, June 26-30, 2000, Proceedings},
  series       = {Lecture Notes in Computer Science},
  volume       = {1825},
  pages        = {409--423},
  publisher    = {Springer},
  year         = {2000}
}

@inproceedings{DBLP:conf/ac/Berthelot86,
  author       = {G{\'{e}}rard Berthelot},
  editor       = {Wilfried Brauer and
                  Wolfgang Reisig and
                  Grzegorz Rozenberg},
  title        = {Transformations and Decompositions of Nets},
  booktitle    = {Petri Nets: Central Models and Their Properties, Advances in Petri
                  Nets 1986, Part I, Proceedings of an Advanced Course, Bad Honnef,
                  Germany, 8-19 September 1986},
  series       = {Lecture Notes in Computer Science},
  volume       = {254},
  pages        = {359--376},
  publisher    = {Springer},
  year         = {1986} 
}

@inproceedings{DBLP:conf/spin/BerthomieuBD18,
  author       = {Bernard Berthomieu and
                  Didier Le Botlan and
                  Silvano Dal{-}Zilio},
  editor       = {Mar{\'{\i}}a{-}del{-}Mar Gallardo and
                  Pedro Merino},
  title        = {Petri Net Reductions for Counting Markings},
  booktitle    = {Model Checking Software - 25th International Symposium, {SPIN} 2018,
                  Malaga, Spain, June 20-22, 2018, Proceedings},
  series       = {Lecture Notes in Computer Science},
  volume       = {10869},
  pages        = {65--84},
  publisher    = {Springer},
  year         = {2018} 
}

@book{BDK-01,
  author    = {Eike Best and
               Raymond R. Devillers and
               Maciej Koutny},
  title     = {{P}etri net algebra},
  series    = {Monographs in Theoretical Computer Science. An {EATCS} Series},
  publisher = {Springer},
  year      = {2001}
}

@ARTICLE{CSKLY-21,
  author={Chan, Alex and Sokolov, Danil and Khomenko, Victor and Lloyd, David and Yakovlev, Alex},
  journal={IEEE Transactions on Computer-Aided Design of Integrated Circuits and Systems},
  title={Burst Automaton: Framework for Speed-Independent Synthesis Using Burst-Mode Specifications},
  year={2023},
  volume={42},
  number={5},
  pages={1560-1573} }

@book{CKKLY-02,
  author    = {Jordi Cortadella
  and
  Michael Kishinevsky and
  Alex Kondratyev and
  Luciano Lavagno and Alex Yakovlev},
  title     = {Logic Synthesis for Asynchronous Controllers and Interfaces},
  publisher = {Springer},
  year      = {2002}
}

@article{GGHN-09,
    author = {Gramm, Jens and Guo, Jiong and H\"{u}ffner, Falk and Niedermeier, Rolf},
    title = {Data Reduction and Exact Algorithms for Clique Cover},
    year = {2009},
    publisher = {ACM},
    volume = {13},
    issn = {1084-6654},
    journal = {ACM J. Experimental Algorithmics},
    articleno = {2},
    numpages = {15},
}

@inproceedings{KKY-22, 
  author       = {Victor Khomenko and
                  Maciej Koutny and
                  Alex Yakovlev},
  editor       = {Luca Bernardinello and
                  Laure Petrucci},
  title        = {Avoiding Exponential Explosion in Petri Net Models of Control Flows},
  booktitle    = {Application and Theory of Petri Nets and Concurrency - 43rd International
                  Conference, {PETRI} {NETS} 2022, Bergen, Norway, June 19-24, 2022,
                  Proceedings},
  series       = {Lecture Notes in Computer Science},
  volume       = {13288},
  pages        = {261--277},
  publisher    = {Springer},
  year         = {2022} 
}

@inproceedings{V-98,
  author    = {Antti Valmari},
  editor    = {Wolfgang Reisig and
               Grzegorz Rozenberg},
  title     = {The State Explosion Problem},
  booktitle = {Lectures on {P}etri Nets {I:} Basic Models, Advances in {P}etri Nets,
               the volumes are based on the Advanced Course on {P}etri Nets, held in
               Dagstuhl, September 1996},
  series    = {Lecture Notes in Computer Science},
  volume    = {1491},
  pages     = {429--528},
  publisher = {Springer},
  year      = {1996}
}

@book{ccs,
  author    = {Robin Milner},
  title     = {A Calculus of Communicating Systems}, 
  publisher = {Springer},
  year      = {1980}
}

@inproceedings{old,
  author    = {Ernst{-}R{\"{u}}diger Olderog},
  editor    = {Grzegorz Rozenberg},
  title     = {Operational {P}etri net semantics for {CCSP}},
  booktitle = {Advances in {P}etri Nets 1987, covers the 7th European Workshop on Applications
               and Theory of {P}etri Nets, Oxford, UK, June 1986},
  series    = {Lecture Notes in Computer Science},
  volume    = {266},
  pages     = {196--223},
  publisher = {Springer},
  year      = {1986} 
}

@article{kmh,
  author    = {Victor Khomenko and
               Roland Meyer and
               Reiner H{\"{u}}chting},
  title     = {A Polynomial Translation of pi-calculus {FCP}s to Safe {P}etri Nets},
  journal   = {Logical Methods in Computer Science},
  volume    = {9},
  number    = {3},
  year      = {2013} 
 }

@book{L-71,
  author    = {H. Lerchs},
  title     = {On cliques and kernels}, 
  publisher = {Tech. Report, Dept. of Comp. Sci., Univ. of Toronto},
  year      = {1971}
}

@inproceedings{gm,
  author    = {Ursula Goltz and
               Alan Mycroft},
  editor    = {Jan Paredaens},
  title     = {On the Relationship of {CCS} and {P}etri Nets},
  booktitle = {Automata, Languages and Programming, 11th Colloquium, Antwerp, Belgium,
               July 16-20, 1984, Proceedings},
  series    = {Lecture Notes in Computer Science},
  volume    = {172},
  pages     = {196--208},
  publisher = {Springer},
  year      = {1984}
}

@book{csp,
  author    = {C.A.R. Hoare},
  title     = {Communicating Sequential Processes}, 
  publisher = {Prentice-Hall},
  year      = {1985}
}

@article{bbd,
  author    = {{\'{E}}ric Badouel and
               Luca Bernardinello and
               Philippe Darondeau},
  title     = {The Synthesis Problem for Elementary Net Systems is {NP}-Complete},
  journal   = {Theoretical Computer Science},
  volume    = {186},
  number    = {1-2},
  pages     = {107--134},
  year      = {1997} 
}

@inproceedings{DBLP:conf/rex/GlabbeekG89,
  author    = {Rob J. van Glabbeek and
               Ursula Goltz},
  editor    = {J. W. de Bakker and
               Willem P. de Roever and
               Grzegorz Rozenberg},
  title     = {Refinement of Actions in Causality Based Models},
  booktitle = {Stepwise Refinement of Distributed Systems, Models, Formalisms, Correctness,
               {REX} Workshop, Mook, The Netherlands, May 29 - June 2, 1989, Proceedings},
  series    = {Lecture Notes in Computer Science},
  volume    = {430},
  pages     = {267--300},
  publisher = {Springer},
  year      = {1989} 
}

@incollection{DBLP:conf/apn/BestDH92,
  author    = {Eike Best and
               Raymond R. Devillers and
               Jon G. Hall},
  editor    = {Grzegorz Rozenberg},
  title     = {The box calculus: {A} new causal algebra with multi-label communication},
  booktitle = {Advances in {P}etri Nets 1992, The {DEMON} Project},
  series    = {Lecture Notes in Computer Science},
  volume    = {609},
  pages     = {21--69},
  publisher = {Springer},
  year      = {1992} 
}

@article{DBLP:journals/iandc/BestDK02,
  author       = {Eike Best and
                  Raymond R. Devillers and
                  Maciej Koutny},
  title        = {The Box Algebra = {P}etri Nets + Process Expressions},
  journal      = {Information and Computation},
  volume       = {178},
  number       = {1},
  pages        = {44--100},
  year         = {2002} }

@inproceedings{DBLP:conf/concur/KhomenkoKY22,
  author       = {Victor Khomenko and
                  Maciej Koutny and
                  Alex Yakovlev},
  editor       = {Bartek Klin and
                  Slawomir Lasota and
                  Anca Muscholl},
  title        = {Slimming down {P}etri Boxes: Compact {P}etri Net Models of Control Flows},
  booktitle    = {33rd International Conference on Concurrency Theory, {CONCUR} 2022,
                  September 12-16, 2022, Warsaw, Poland},
  series       = {LIPIcs},
  volume       = {243},
  pages        = {8:1--8:16},
  publisher    = {Schloss Dagstuhl - Leibniz-Zentrum f{\"{u}}r Informatik},
  year         = {2022} 
}

@book{BBD-book,
  title =	"Petri Net Synthesis",
  author =	"{\'E}ric Badouel and Luca Bernardinello and Philippe
		 Darondeau",
  publisher =	"Springer",
  year = 	"2015",
  pages =	"1--325",
  series =	"Texts in Theoretical Computer Science. An EATCS
		 Series",
}

@inproceedings{DBLP:conf/concur/Pietkiewicz-Koutny98,
  author       = {Marta Pietkiewicz{-}Koutny},
  editor       = {Davide Sangiorgi and
                  Robert de Simone},
  title        = {Synthesis of {ENI}-systems Using Minimal Regions},
  booktitle    = {{CONCUR} '98: Concurrency Theory, 9th International Conference, Nice,
                  France, September 8-11, 1998, Proceedings},
  series       = {Lecture Notes in Computer Science},
  volume       = {1466},
  pages        = {565--580},
  publisher    = {Springer},
  year         = {1998} 
}

@inproceedings{DBLP:conf/apn/KhomenkoKY25,
  author       = {Victor Khomenko and
                  Maciej Koutny and
                  Alex Yakovlev},
  editor       = {Elvio Gilberto Amparore and
                  Lukasz Mikulski},
  title        = {Distributed Places and Safe Net Reduction},
  booktitle    = {Application and Theory of Petri Nets and Concurrency - 46th International
                  Conference, {PETRI} {NETS} 2025, Paris, France, June 22-27, 2025,
                  Proceedings},
  series       = {Lecture Notes in Computer Science},
  volume       = {15714},
  pages        = {265--286},
  publisher    = {Springer},
  year         = {2025}
}

\end{document}